\documentclass[10pt,preprint]{aastex}

\newcommand{\kms}{km~s$^{-1}$}
\begin{document}

\voffset -0.1in


\shorttitle{$^{12}$CO, $^{13}$CO, and C$_2$}
\shortauthors{Sonnentrucker et al.}

\title{Abundances and Behavior of $^{12}$CO, $^{13}$CO, and C$_2$ in Translucent Sight Lines.\footnotemark}

\author{P. Sonnentrucker\altaffilmark{2}, D. E. Welty\altaffilmark{3}, J. A. Thorburn\altaffilmark{4}, and D. G. York\altaffilmark{3}}

\altaffiltext{1}{Based on observations made with the NASA/ESA Hubble Space Telescope, obtained from the Data Archive at the Space Telescope Science Institute, which is operated by the Association of Universities for Research in Astronomy, Inc., under NASA contract NAS 5-26555. 
Based in part on observations obtained with the Apache Point Observatory 3.5m telescope, which is owned and operated by the Astrophysical Research Consortium.}
\altaffiltext{2}{Department of Physics and Astronomy, Johns Hopkins University, 3400 North Charles Street, Baltimore, MD 21218; sonnentr@pha.jhu.edu}
\altaffiltext{3}{Department of Astronomy and Astrophysics, University of Chicago, 5640 South Ellis Avenue, Chicago, IL 60637; welty@oddjob.uchicago.edu, don@oddjob.uchicago.edu}
\altaffiltext{4}{Yerkes Observatory, University of Chicago, Williams Bay, WI 53191; thorburn@yerkes.uchicago.edu}

\begin{abstract}

Using UV spectra obtained with {\it FUSE}, {\it HST}, and/or {\it IUE} together with higher resolution optical spectra, we determine interstellar column densities of $^{12}$CO, $^{13}$CO, and/or C$_2$ for ten Galactic sight lines with $E(B-V)$ ranging from 0.37 to 0.72. 
The $N$(CO)/$N$(H$_{2}$) ratio varies over a factor of 100 in this sample, due primarily to differences in $N$(CO). 
For a given $N$(H$_2$), published models of diffuse and translucent clouds predict less CO than is observed.
The $J$ = 1--3 rotational levels of $^{12}$CO are sub-thermally populated in these sight lines, with $T_{ex}$ typically between 3 and 7 K. 
In general, there appears to be no significant difference between the excitation temperatures of $^{12}$CO and $^{13}$CO.
Fits to the higher resolution CO line profiles suggest that CO (like CN) is concentrated in relatively cold, dense gas.

We obtain C$_2$ column densities from the F-X (0-0) band at 1341 \AA\ (three sight lines; $J$ = 0--12), the F-X (1-0) band at 1314 \AA\ (one sight line; $J$ = 0--12), the D-X (0-0) band at 2313 \AA\ (four sight lines; $J$ = 0--18), and the A-X (3-0) and (2-0) bands at 7719 and 8757 \AA\ (seven sight lines; $J$ = 0--12). 
Comparisons among those column densities yield a set of mutually consistent band $f$-values for the UV and optical C$_2$ bands, but also reveal some apparent anomalies within the F-X (0-0) band.
Both the kinetic temperature $T_{\rm k}$ inferred from the C$_2$ rotational populations and the excitation temperature $T_{02}$(C$_2$) are generally smaller than the corresponding $T_{01}$(H$_2$) --- suggesting that C$_2$ is concentrated in colder, denser gas than H$_2$. 

Incorporating additional column density data for \ion{K}{1}, HD, CH, C$_2$, C$_3$, CN, and CO from the literature (for a total sample of 74 sight lines), we find that 
(1) CO is most tightly correlated with CN;
(2) the ratios $^{12}$CO/H$_2$ and $^{13}$CO/H$_2$ both are fairly tightly correlated with the density indicator CN/CH (but C$_2$/H$_2$ is not); and
(3) the ratio $^{12}$CO/$^{13}$CO is somewhat anti-correlated with both CN/CH and $N$(CO).  
Sight lines with $^{12}$CO/$^{13}$CO below the average Galactic value of $^{12}$C/$^{13}$C appear to sample colder, denser gas in which isotope exchange reactions have enhanced $^{13}$CO, relative to $^{12}$CO. 

\end{abstract}


\section{Introduction}
\label{sec-intro}

Apart from H$_2$ and (in some cases) HD, CO is the most abundant interstellar molecule.  
Determinations of the abundance and rotational excitation of the dominant $^{12}$C$^{16}$O and its isotopomers thus can provide significant information about the physical conditions in Galactic and extragalactic molecular clouds.
The CO/H$_{2}$ abundance ratio is often assumed to be roughly constant in dense molecular gas, with a ``canonical'' value of order 10$^{-4}$ derived primarily from millimetric and radio observations of CO and inferred abundances of H$_{2}$ in Galactic molecular clouds (Young \& Scoville 1991). 
Under that assumption, measurements of CO are used to estimate the total H$_2$ content when no data are available for H$_2$. 
In more diffuse gas, both H$_2$ and CO were detected directly via their absorption bands in the far-UV using spectra from the {\it Copernicus} satellite; longer wavelength UV bands of CO were observed with {\it IUE}. 
Analysis of the {\it Copernicus} spectra showed that the CO/H$_{2}$ ratio varied by two orders of magnitude (from about 5 $\times$ 10$^{-8}$ to 5 $\times$ 10$^{-6}$) (Federman et al. 1980) --- so that CO is not a good predictor for H$_{2}$ in more diffuse material. 
Theoretical models of cloud chemistry suggested that such large variations in the CO/H$_{2}$ ratio could be expected in diffuse ($A_{\rm V}$ $\la$ 1 mag) and translucent (1 mag $\la$ $A_{\rm V}$ $\la$ 5 mag) regions still permeated by dissociating photons (e.g., van Dishoeck \& Black 1986a, 1988) as well as in regions perturbed by the recent passage of dissociative shocks (Neufeld \& Dalgarno 1989). 
The relatively low sensitivity and resolution of {\it Copernicus} and {\it IUE}, however, allowed only limited investigation of more heavily reddened sight lines, where the CO/H$_2$ ratio might be expected to approach the dense cloud value. 

Determining observationally the population distribution in the individual rotational levels of any molecular species can yield constraints on the physical properties characterizing the molecular gas (e.g., van Dishoeck et al. 1991; Federman et al. 1997b; Wannier et al. 1997). 
Observations of the rotational excitation of C$_2$, for example, have yielded estimates for the temperatures and densities in diffuse molecular clouds (e.g., van Dishoeck \& de Zeeuw 1984); data for CN have been used both to constrain the temperature of the cosmic microwave background (e.g., Roth \& Meyer 1995) and to estimate local hydrogen densities (e.g., Black \& van Dishoeck 1991).
The pervasive abundance of the CO and H$_{2}$ molecules in the interstellar gas make those species especially important tools for investigations of the gas properties. 
Unfortunately, however, the spectrographs on {\it Copernicus} and {\it IUE} could not resolve the rotational structure of the CO bands.

Isotopic ratio measurements provide another means to investigate the chemical processes occurring in the interstellar medium. 
Carbon, whose isotopes $^{12}$C and $^{13}$C are products of primary and secondary nucleosynthesis processes (Langer \& Penzias 1990), respectively, is well suited for such studies. 
Within about 1--2 kpc of the Sun, the current average $^{12}$C/$^{13}$C ratio is about 70 (e.g., Wannier et al. 1982; Stahl \& Wilson 1992; Lucas \& Liszt 1998), with some variations; there also is an apparent systematic decrease in the ratio toward the Galactic center (e.g., Wilson 1999; Milam et al. 2005).
When determined from molecules unaffected by chemical fractionation (e.g., CH$^{+}$), the $^{12}$C/$^{13}$C ratio is an indicator of Galactic stellar evolution. 
The (optical) absorption lines from $^{13}$CH$^+$, however, are typically very weak.
Because chemical fractionation can occur for CO, the $^{12}$CO/$^{13}$CO ratio gives additional information on chemical processes in the molecular gas (e.g., Bally \& Langer 1982; Sheffer et al. 1992). 
Again, however, the limited sensitivity of {\it Copernicus} and {\it IUE} made it difficult to obtain reliable measures of the generally weak $^{13}$CO absorption.

The higher far-UV sensitivity of the {\it FUSE} instrument (Moos et al. 2000), together with the higher sensitivity and resolution of the {\it HST} GHRS and STIS echelle modes in the near-UV, however, has allowed us to revisit these issues. 
The {\it FUSE} spectra contain numerous absorption lines from the Lyman and Werner bands of molecular hydrogen, from which accurate total H$_{2}$ column densities can be derived even toward fainter, more heavily reddened stars (e.g., Rachford et al. 2002). 
While the B-X (0-0) (1150 \AA), C-X (0-0) (1087 \AA), and E-X (0-0) (1076 \AA) bands arising from the $X ^{1}\Sigma^{+}$ ground electronic state of CO are also present within the {\it FUSE} wavelength coverage, an accurate determination of the CO column densities from the {\it FUSE} data alone is often hindered by line blending, saturation, and unresolved structure in the absorption profiles. 
With GHRS and STIS, however, one could observe weaker bands in the CO A-X system between 1240 and 1550 \AA\ at high enough resolution and S/N to discern and measure the individual rotational ($J$) levels of the bands --- allowing accurate estimates of their individual column densities.  
Moreover, some of the weaker CO spin-forbidden (triplet-singlet) intersystem bands between 1300 and 1550 \AA\ can be detected once the permitted bands become saturated (Morton \& Noreau 1994) --- enabling more reliable determinations of higher CO column densities.  
Several recent studies, for example, have used UV spectra from {\it HST} and/or {\it FUSE} to determine accurate $N$(CO) toward the moderately reddened stars $\zeta$ Oph (Lambert et al. 1994), $\rho$ Oph and $\chi$ Oph (Federman et al. 2003), X Per (Sheffer, Federman, \& Lambert 2002a; Sheffer, Lambert, \& Federman 2002b), and HD 203374A (Sheffer, Federman, \& Andersson 2003); Pan et al. (2005) have reported $N$(H$_2$) and $N$(CO) toward a number of stars in Cep OB2 and Cep OB3.
High-resolution GHRS spectra of the C$_2$ D-X (Mulliken) (0-0) band toward $\zeta$ Oph have enabled measurements of the C$_2$ rotational levels up to $J$ = 24 (Lambert, Sheffer, \& Federman 1995) --- higher $J$ than had been obtainable from the weaker A-X (Phillips) bands in the optical and near-IR.

In this paper, we present a study of the $^{12}$CO, $^{13}$CO, and/or C$_{2}$ molecules toward ten moderately reddened [$E(B-V)$ = 0.37--0.72] Galactic stars, based on UV spectra obtained with {\it FUSE}, {\it IUE}, and/or {\it HST}; on higher resolution optical spectra of CH, CN, and \ion{K}{1}; and on very high S/N optical spectra of C$_2$.
This combination of UV and optical data has allowed us to perform detailed analyses of the CO and C$_{2}$ profiles and to derive column densities for the individual rotational levels of the ground electronic and vibrational states --- up to $J$ = 3 for $^{12}$CO, $J$ = 2 for $^{13}$CO, and $J$ = 18 for C$_2$. 
In \S~\ref{sec-obsred}, we discuss the observational data and the data reduction procedures. 
In \S~\ref{sec-analysis}, we describe the methods used to determine column densities from the spectra. 
In \S~\ref{sec-phys}, we combine our new results with existing literature data for H$_2$, HD, CH, C$_2$, C$_3$, CN, and CO (for a total sample of 74 stars) and discuss the physical conditions characterizing the diffuse molecular gas using this expanded star sample. 
Column densities of H$_2$, CH, C$_2$, CN, $^{12}$CO, and $^{13}$CO for that total sample are tabulated in the Appendix.
In \S~\ref{sec-summ}, we summarize the main results of this study.

\section{Observations and Data Reduction}
\label{sec-obsred}

In order to investigate the behavior of the CO/H$_2$ and $^{12}$CO/$^{13}$CO ratios at higher overall extinctions and column densities, we selected sight lines according to the following general criteria:
(1) inclusion in the {\it FUSE} survey of translucent sight lines (Rachford et al. 2002; for H$_2$ and CO);
(2) availability of higher resolution {\it HST} GHRS or STIS spectra of the CO A-X bands (for $^{12}$CO, $^{13}$CO, and the CO rotational structure); and
(3) availability of high-resolution optical spectra of \ion{K}{1}, CH, and CN (for information on component structure).  
For the sight line toward HD~73882 (Snow et al. 2000), which was not observed with {\it HST}, and for several other sight lines with {\it HST} data, we also have analyzed archival high-dispersion {\it IUE} spectra of the CO A-X and intersystem bands.  
The ten sight lines in the resulting sample (Table~\ref{tab:stars}) have $E(B-V)$ between 0.37 and 0.72 mag, visual extinction $A_{\rm V}$ between 0.83 and 2.36 mag, $N$(H$_2$) from 3--13 $\times$ 10$^{20}$ cm$^{-2}$, and molecular fractions $f$(H$_2$) between 0.14 and 0.76. 
The ten sight lines also probe a variety of regions in the Galactic interstellar medium --- with two (HD~24534, HD~27778) in Taurus/Perseus, one (HD~147888) in Sco-Oph, three (HD~206267, HD~207198, HD~210839) in Cep OB2, and one (HD~210121) through a high-latitude molecular cloud.

\subsection{{\it FUSE} Data}
\label{sec-fuse}

\begin{figure}
\epsscale{1.0}
\plotone{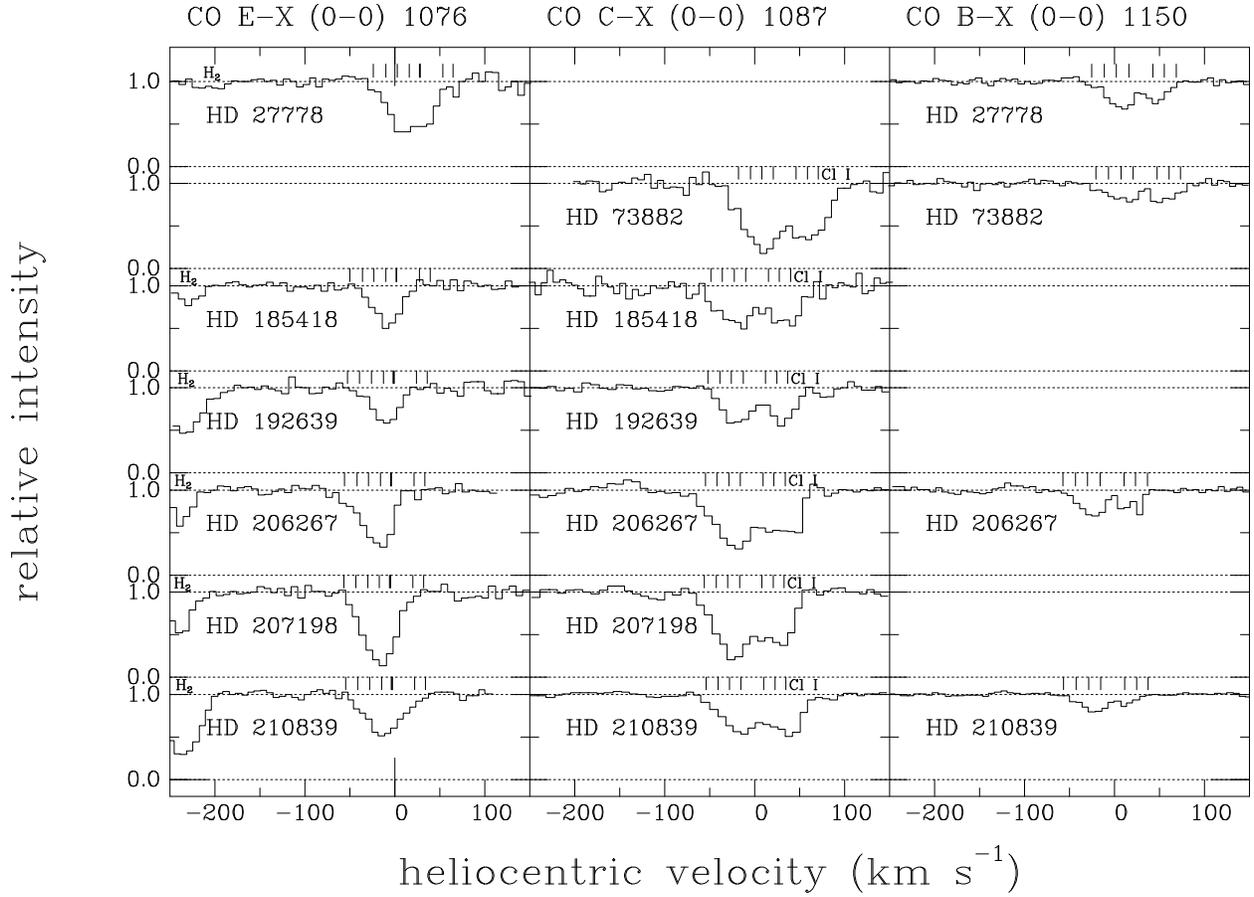}
\caption{{\it FUSE} spectra of $^{12}$CO E-X (0-0) (1076 \AA), C-X (0-0) (1087 \AA), and B-X (0-0) (1150 \AA) bands toward seven stars. 
Tick marks denote the individual rotational lines in each band (R, Q, and P branches for E-X; R and P branches for C-X and B-X).
The C-X (0-0) P branch is blended with a line of Cl~I.
Velocities are heliocentric.} 
\label{fig:fuse}
\end{figure}

{\it FUSE} observed our sample stars under programs P116 and X021 (PI: T. P. Snow), which were dedicated to the study of translucent sight lines (Rachford et al. 2002). 
The exposures used for each sight line are listed in Table~\ref{tab:uvdata}.
The observations were obtained in time tag mode through the low-resolution aperture (LWRS) and were processed with version 2.0.5 of the CALFUSE pipeline (Kruk \& Murphy 2001; Sahnow et al. 2000). 
Eight detector segments cover sections of the total wavelength range from 912 to 1187 \AA, with most spectral regions covered by two to four segments. 
The resolution is $\sim$0.062 \AA\ (FWHM), corresponding to about $\sim$18 \kms\, or $\sim$9 pixels. 
For each segment, the exposures were co-added after cross-correlating and shifting the spectra (by at most 8 pixels) with respect to the brightest exposure. 
The spectra were binned by 4 pixels (slightly less than one-half resolution element) in order to increase the signal-to-noise (S/N) ratio without degrading the resolution delivered by the spectrograph optics. 
The co-added spectra typically have S/N of about 10 per resolution element in the LiF1A spectrum and up to 15 per resolution element in the LiF2A spectrum. 
The LiF1B spectrum was excluded from the data analysis because of the presence of a well-known detector artifact which causes an artificial flux deficiency in this segment (Kruk \& Murphy 2001). 
Figure~\ref{fig:fuse} presents the profiles of the CO E-X (0-0) (1076 \AA), C-X (0-0) (1087 \AA), and B-X (0-0) (1150 \AA) bands, as functions of heliocentric velocity, detected in the far-UV spectra of seven of the ten sight lines. 

\subsection{{\it HST} Data}
\label{sec-hst}

The calibrated GHRS or STIS spectra listed in Table~\ref{tab:uvdata} were retrieved from the MAST archive located at the Space Telescope Science Institute. 
All the STIS data were automatically reduced with the latest version of the CALSTIS pipeline. 
The STIS spectra were obtained using either the FUV MAMA detector and E140H echelle mode or the NUV detector and E230H echelle mode, at resolutions of either 2.75 \kms\, (for the 0.''2 $\times$ 0.''2 or 0.''2 $\times$ 0.''09 slits) or 1.5 \kms\, (for the 0.''1 $\times$ 0.''03 slit).
The E140H spectra generally cover the wavelength range from about 1150 to 1370 \AA\ --- including the A-X (7-0) through (12-0) bands of $^{12}$CO and $^{13}$CO (between 1240 and 1350 \AA) and the F-X (0-0) and (1-0) bands of C$_2$ at 1341 \AA\ and 1314 \AA\ (as well as many atomic absorption lines).  
Unfortunately, the E140H spectra generally do not include the strongest CO intersystem bands. 
The more extensive E140H spectra of HD~24534, which do include those intersystem bands, are described by Sheffer et al. (2002a, 2002b).
The E230H spectra cover the range from about 2130 to 2400 \AA\ --- including the D-X (0-0) band of C$_2$ at 2313 \AA\ (and various atomic lines).
The extracted spectral orders were corrected for scattered light contamination using the procedures described by Howk \& Sembach (2000) or Valenti et al. (2002). 
For the spectra of HD~206267 and HD~210839, which were obtained through the narrowest slit, additional processing was performed by E. Jenkins (private communication) in order to account for unbalancing of the even and odd pixels on the detector (see Jenkins \& Tripp 2001 for details). 
The continua were normalized to unity using low-order polynomials. 
In most cases, the S/N ratios (estimated from the fluctuations in the continuum regions) range from 16 to 36 per resolution element; the C$_2$ F-X band spectra for HD~24534, however, have S/N $\sim$ 130, and the C$_2$ D-X band spectra (for three stars) have S/N $\sim$ 85--120. 
Figures~\ref{fig:stis1}--\ref{fig:stis3} show the profiles of some of the $^{12}$CO A-X bands toward six of the stars; Figure~\ref{fig:13co} shows the $^{13}$CO A-X (7-0) (1347 \AA) band profiles toward four of the stars and the $^{13}$CO A-X (4-0) (1421 \AA) band toward HD~24534. 
The individual rotational levels ($J$ = 0, 1, 2, 3) of the ground vibrational state resolved in each CO band are numbered above the spectra; for $^{12}$CO, the R, Q, and P branches are left-to-right in each case.   

The GHRS ECH-A spectra of HD~210121 cover the $^{12}$CO A-X (2-0) (1477 \AA) and (4-0) (1419 \AA) bands, the corresponding $^{13}$CO bands (1478 and 1421 \AA), and the relatively strong $^{12}$CO a'14 intersystem band (1419 \AA), at a resolution of about 3.5 \kms.
The GHRS ECH-B spectra of HD~24534 cover the C$_2$ D-X (0-0) band at 2313 \AA.
Because the spectra were obtained at several fp-split offsets, the STSDAS routines POFFSETS, DOPOFF, and SPECALIGN were used to align and combine the individual spectra. 
The S/N values range from 15 to 20 per resolution element for HD~210121 and are about 100 per resolution element for HD~24534.

\subsection{{\it IUE} Data}
\label{sec-iue}

High-dispersion (FWHM $\sim$ 25 \kms) {\it IUE} spectra of HD~27778, HD~73882, and HD~207198 were retrieved from the archive (Table~\ref{tab:uvdata}).  
For all three stars, multiple spectra had been obtained at several different offsets within the large aperture, in order to reduce the effects of fixed-pattern noise when the spectra are combined.  
For these three stars, the S/Ns range from about 15--45 per resolution element in the aligned, co-added spectra, for wavelengths between about 1320 and 1550 \AA\ [covering the CO A-X bands (0-0) through (8-0)].
Figure~\ref{fig:iue} shows some of the stronger $^{12}$CO A-X bands toward HD~27778 and HD~73882 (see also Figure~7 of Morton \& Noreau 1994); the corresponding $^{13}$CO A-X bands and some of the intersystem bands of $^{12}$CO can also be discerned.  
While the permitted A-X bands are saturated toward both of those stars, the stronger intersystem bands toward HD~73882 indicate a higher $N$(CO) in that line of sight.
The {\it IUE} data for HD~210121 were described by Welty \& Fowler (1992).
 
\subsection{Optical Data}
\label{sec-opt}

High-resolution (FWHM $\sim$ 0.6--3.6 km s$^{-1}$) spectra of \ion{K}{1} (7698 \AA), CH (4300 \AA), and CN (3874 \AA) were obtained with the Kitt Peak coud\'{e} feed telescope and the McDonald 2.7m telescope. 
A detailed discussion of the reduction and analysis of the \ion{K}{1} spectra has been given by Welty \& Hobbs (2001); the CH and CN spectra, processed in similar fashion, will be discussed by Welty, Snow \& Morton (in preparation). 
Profiles of the optical lines for six of the sight lines discussed in this paper are included in Figures~\ref{fig:stis1}--\ref{fig:stis3}; some have also been shown by Pan et al. (2004).

Medium resolution (FWHM $\sim$ 8 km s$^{-1}$) but very high S/N ($\ga$ 1000) optical spectra were acquired for nine of the stars with the 3.6m telescope and the ARC echelle spectrograph at Apache Point Observatory as part of an observing program designed to investigate the nature of the diffuse interstellar bands. 
The details on the reduction and analysis of these spectra and, in particular, of the C$_{2}$ A-X (Phillips) bands detected in seven of the sight lines studied here, are described in Thorburn et al. (2003). 

\section{Data Analysis}
\label{sec-analysis}

Several factors have made it difficult to derive accurate CO column densities for more heavily reddened sight lines.
The lower resolution UV spectra from {\it Copernicus}, {\it IUE}, and {\it FUSE} yield equivalent widths for a number of the CO bands, but do not resolve either the (often complex) component structure or the rotational structure of the bands.
Blends with stellar lines can also complicate the interpretation of the spectra, especially when the S/N and/or $vsini$ are relatively low.
In addition, for these higher column density sight lines, many of the permitted CO bands that can be detected are significantly saturated.  

Fortunately, the higher resolution (and often higher S/N) spectra obtained with the {\it HST} GHRS and STIS echelle modes at least partially resolve the CO rotational structure and enable reliable measurement of weaker CO A-X and intersystem bands even toward fainter, more heavily reddened targets. 
Agreement between several recent observational and theoretical studies suggests that the $f$-values now available for the intersystem bands are reliable (Sheffer et al. 2002a; Eidelsberg \& Rostas 2003). 
Finally, still higher resolution optical spectra of \ion{K}{1}, CH, and (especially) CN can provide useful information as to the velocity component structure --- enabling more accurate modeling of the observed CO line profiles.

Three methods were used to estimate the interstellar CO column densities.  
First, empirical curves of growth were constructed using the measured equivalent widths of the CO bands.
The empirical curves were then compared with theoretical curves generated for single components with a range in the Doppler broadening parameter ($b$), for two representative CO rotational population distributions characterized by different excitation temperatures.  
Second, the Apparent Optical Depth method (Hobbs 1971; Savage \& Sembach 1991) was applied to the observed CO line profiles.
Third, the absorption-line profiles were fitted using multi-component models derived from higher resolution optical spectra of \ion{K}{1}, CH, and CN.  
Profile fitting was used to determine column densities from the UV and optical lines of C$_2$.
Because the individual gas components along these sight lines are at best partially resolved in the UV spectra, determination of the column densities for individual components is often uncertain with STIS data and essentially impossible with the lower resolution {\it FUSE} data. 
We will therefore only discuss total sight line column densities in the present work. 

\subsection{Curves of Growth}
\label{sec-cog}

In order to construct the empirical curves of growth (COG), total equivalent widths of the $^{12}$CO and $^{13}$CO bands seen in the normalized UV spectra were measured over velocity ranges which include all the detected rotational lines. 
Table~\ref{tab:ew1} lists the CO equivalent widths found for the A-X, B-X, C-X, and E-X bands from {\it FUSE} and STIS data (eight sight lines); Table~\ref{tab:ew2} gives the values for the A-X and intersystem bands found from GHRS, STIS, and/or {\it IUE} data (six sight lines). 
The CO equivalent widths for the well-studied sight lines toward X Per (HD 24534) and $\zeta$ Oph (HD 149757) are included in Table~\ref{tab:ew2} for comparison.
In all cases, the listed 1-$\sigma$ uncertainties include contributions from both photon noise and continuum-fitting uncertainty, added in quadrature; upper limits are 3-$\sigma$.
Because of the wide velocity range covered by the ensemble of rotational lines, the continuum-fitting uncertainties generally are dominant.
Inspection of the equivalent widths for HD~27778 and HD~207198 indicates that there is generally good agreement between the values determined from {\it IUE} and STIS spectra for the CO A-X (7-0) and (8-0) bands at 1322 and 1344 \AA.
The equivalent widths for HD~185418 and HD~192639 are consistent with those listed by Sonnentrucker et al. (2002, 2003).
Because the CO C-X (0-0) band at 1087 \AA\ is often blended with an adjacent \ion{Cl}{1} line (see Fig.~\ref{fig:fuse}) and, in the B stars in our sample, with a stellar absorption feature, it was not used in the construction of the curves of growth. 
An estimate of its equivalent width is, nevertheless, given for the two stars toward which the blending was less severe.
The E-X (0-0) band at 1076 \AA\ is in the wing of the very strong H$_2$ (2-0) R(0) line at 1077 \AA, making the appropriate continuum difficult to estimate in some cases.  

The rest wavelengths and oscillator strengths listed in Tables~\ref{tab:ew1} and \ref{tab:ew2} were taken from Morton \& Noreau (1994) for the CO A-X bands and from Federman et al. (2001) for the far-UV B-X (0-0) (1150 \AA), C-X (0-0) (1087 \AA), and E-X (0-0) (1076 \AA) bands. 
For the CO intersystem bands, we have adopted the rest wavelengths of Eidelsberg \& Rostas (2003) and (where possible) the empirical $f$-values of Sheffer et al. (2002a), which yield smoother curves of growth for the equivalent widths measured toward both HD~24534 (X Per) and HD~149757 ($\zeta$ Oph) than do the corresponding $f$-values of Eidelsberg \& Rostas.
For the $^{12}$CO bands with both {\it FUSE} and STIS data, the range in $f\lambda$ can span nearly three orders of magnitude from the weakest A-X (12-0) (1246 \AA) to the strongest E-X (0-0) (1076 \AA) band; a similar range in $f\lambda$ is covered for sight lines with both STIS and {\it IUE} data. 

In principle, the CO curve of growth depends on (1) the rotational structure of the individual bands, (2) the excitation temperatures characterizing the various rotational levels, and (3) the interstellar component structure (e.g., Black \& van Dishoeck 1988).
Theoretical curves of growth were generated for single interstellar components with a range of $b$-values (0.3--5.0 \kms\ --- as an approximation to more complex component structures), using the CO rotational structure of the A-X (6-0) band (taken as representative of all the A-X bands).  
Separate curves were generated for the far-UV E-X, C-X, and B-X bands, for which the rotational structure is somewhat different.  
For $^{12}$CO, curves were generated for two sets of relative populations in the $J$ = 0--3 rotational levels.  
The first had level populations 0/1/2/3 = 1.0/1.0/0.2/0.05, which are similar to the average ratios found for the sight lines in our small sample.
The relative populations in the $J$ = 0--2 levels, for example, correspond to an excitation temperature of about 5 K.
The second had level populations 0/1/2/3 = 1.0/0.5/0.02/0.0001, corresponding to an excitation temperature of about 3 K (typical of some other sight lines reported in the literature).
For $^{13}$CO, equal populations were assumed for $J$ = 0 and 1 (the only levels generally detected in the STIS spectra).
The sums of the $f$-values for the various individual transitions (R, Q, and P branches) for each $J$ are all equal to the $f$-value of the band as a whole, and the COG analysis of the integrated band equivalent widths yields the total column density over all $J$.

Columns (2) and (7) in Table~\ref{tab:nco} list the total $^{12}$CO and $^{13}$CO column densities (respectively) derived by comparing the empirical curves of growth with the theoretical single-component curves for $T_{\rm ex}$ $\sim$ 3 K and/or for $T_{\rm ex}$ $\sim$ 5 K; column (3) lists the best-fit ``effective'' $b$-value. 
In each case, the uncertainty given for the column density corresponds to the range in $N$(CO) giving rms deviations less than 1.5 times that of the best ($N$,$b$) pair, for that best-fit $b$-value.
Figure~\ref{fig:cog} shows the $^{12}$CO curves of growth for the sight lines toward HD~27778 and HD~207198, with the adopted values for $N$ and $b$. 
In the figure, the solid lines show the $T_{\rm ex}$ $\sim$ 5 K theoretical curves for the A-X bands (which in most cases should be adequate for the intersystem, B-X, and C-X bands); the dotted lines show the theoretical curves for the E-X (0-0) band at 1076 \AA, displaced downward by 1.0 for clarity. 
The data point for the observed E-X (0-0) band is plotted against both sets of theoretical curves to show that the curves for the E-X (0-0) band lie slightly below those for the A-X bands.

For most of the sight lines, at least two CO bands lie on or near the linear part of the curve of growth. 
Accurate total column densities can be derived from such optically thin bands with minimal assumptions regarding the gas velocity distribution. 
When additional, stronger bands are measured, the COG analysis can (in principle) also yield some information regarding the overall gas distribution, as long as (1) the $f$-value range for the stronger bands is large enough that different degrees of saturation are observed and (2) the CO excitation temperature is reasonably well determined. 
For eight of the nine sight lines analyzed, the effective $b$-values for the empirical curves, obtained by comparison with the theoretical curves, range from about 0.5 to 2.0 \kms. 
The very small effective $b$-values for HD~147888 and HD~210839 (0.5 km~s$^{-1}$) likely reflect the widths of the single dominant narrow components seen in CN in those two sight lines.
The higher values for HD~27778 and HD~206267 (both $\sim$ 1 \kms) and for HD~207198 ($\sim$ 2.0 \kms), however, presumably correspond to the multiple components seen in CN in those sight lines (rather than the widths of single components).
Slight differences between the empirical and theoretical curves for several sight lines may be due to the more complex component structure (seen in the higher resolution optical spectra) in those cases.
Comparisons of the observed equivalent widths with the theoretical curves for $T_{\rm ex}$ $\sim$ 3 K (which fall slightly below the corresponding curves for $T_{\rm ex}$ $\sim$ 5 K) generally exhibited the best agreement for somewhat higher column densities at the same effective $b$ or else for similar column densities and slightly higher effective $b$.
While the column densities derived via the COG thus depend on knowledge of $T_{\rm ex}$ (obtained via detailed fits to the high-resolution line profiles), they do provide useful confirmation of the total $N$(CO) derived in the profile fits.

\subsection{Apparent Optical Depth}
\label{sec-aod}

The Apparent Optical Depth (AOD) analysis was applied to all detected $^{12}$CO and $^{13}$CO bands, over the same velocity ranges used for the equivalent width measurements.  
In the absence of unresolved saturated structure(s) within the line profile, the AOD method yields the ``apparent'' (instrumentally smeared) column density as a function of velocity, with some improvement over the simplest assumption of a linear relationship between equivalent width and column density.  
If the line is somewhat saturated, the AOD analysis yields a lower limit to the actual column density, especially if the spectral resolution is not very high.
In principle, the degree of saturation in the profiles of a given species can be assessed if multiple lines, with $f$-values differing by at least a factor of 2, are available (Jenkins 1996). 
For CO, the distribution of the total column density over multiple rotational levels (some blended) renders the specific $N$($v$) not very meaningful, but it does delay somewhat the onset of saturation.  
As for the COG analysis, the total column density (over the whole profile and for all rotational levels) can still be estimated.
 
For most of the sight lines, smaller total column densities are obtained from the stronger CO bands (relative to those obtained from the weakest observed bands) --- indicative of more severe saturation in the stronger bands.
Toward HD~207198, for example, the AOD analysis of the strongest CO band observed with STIS [A-X (7-0) at 1344 \AA] yields a total $N$(CO) smaller by a factor of about 1.5 than that found for the weaker CO bands. 
Toward HD~210839 and HD~206267, saturation effects are already apparent in the A-X (9-0) (1301 \AA) band, and the total column densities inferred from the A-X (7-0) band at 1344 \AA\ are factors of 3--4 lower than those obtained from the weakest bands. 
In such cases, however, accurate column densities still can be determined by using only the weaker A-X (12-0), (11-0), and (10-0) bands (at 1246, 1263, and 1281 \AA), which yield concordant results. 
The AOD column densities inferred from the weakest CO bands are listed in Columns (3) and (7) of Table~\ref{tab:nco}; the associated error bars include contributions from both photon noise and continuum-fitting uncertainty. 
In most cases, the total column densities derived from the AOD analysis are consistent with those obtained with the COG method. 

\subsection{Profile Fitting}
\label{sec-fits}

\begin{figure}
\epsscale{1.0}
\plotone{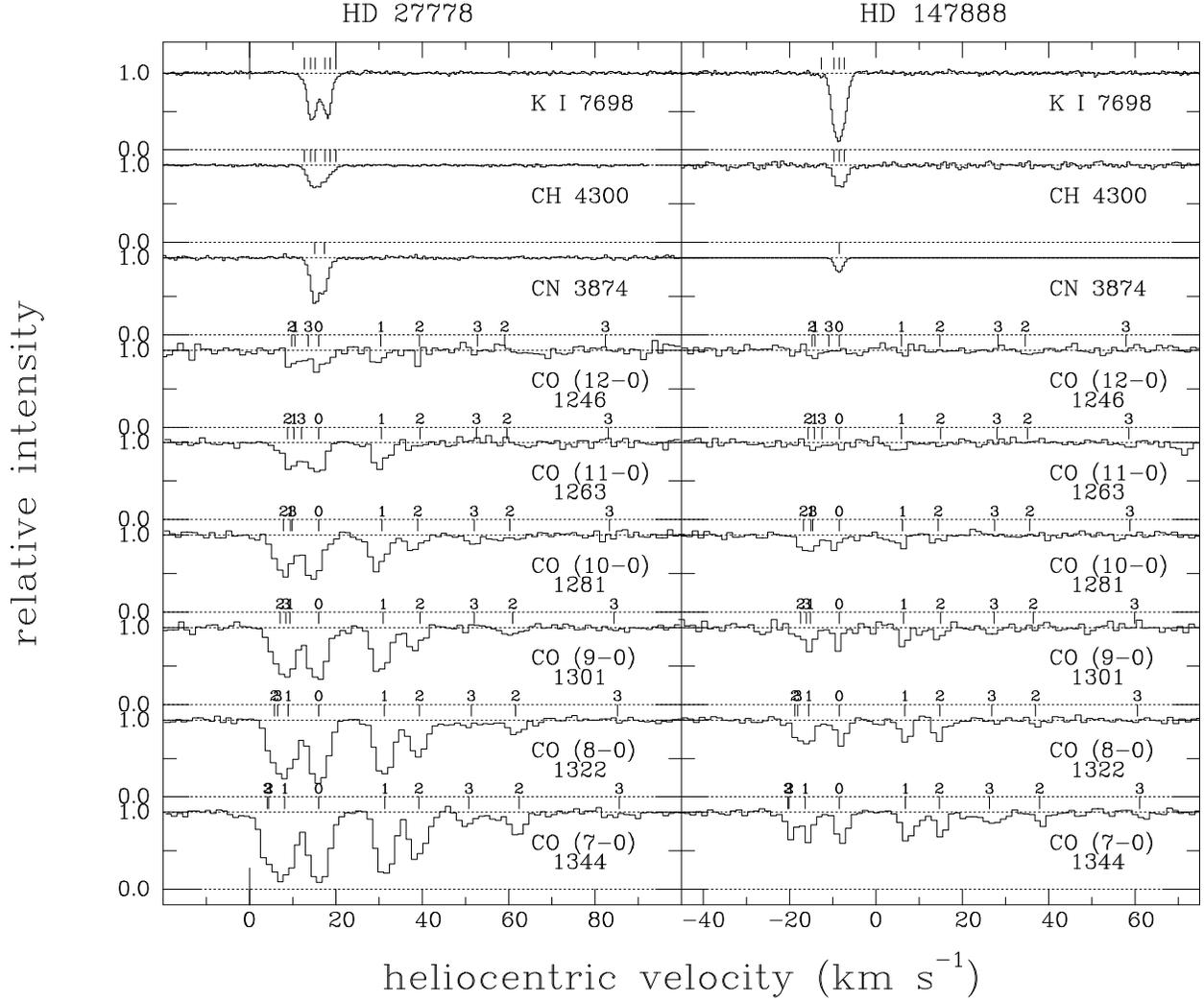}
\caption{STIS spectra of $^{12}$CO A-X bands toward HD~27778 and HD~147888.
At the top are higher resolution optical spectra of K~I, CH, and CN, which show more detailed component structure; the CN profile toward HD~147888 was calculated from the component structure in Pan et al. 2004.
CO rotational levels are marked above each spectrum (R, Q, and P branches are left-to-right).}
\label{fig:stis1}
\end{figure}

\begin{figure}
\epsscale{1.0}
\plotone{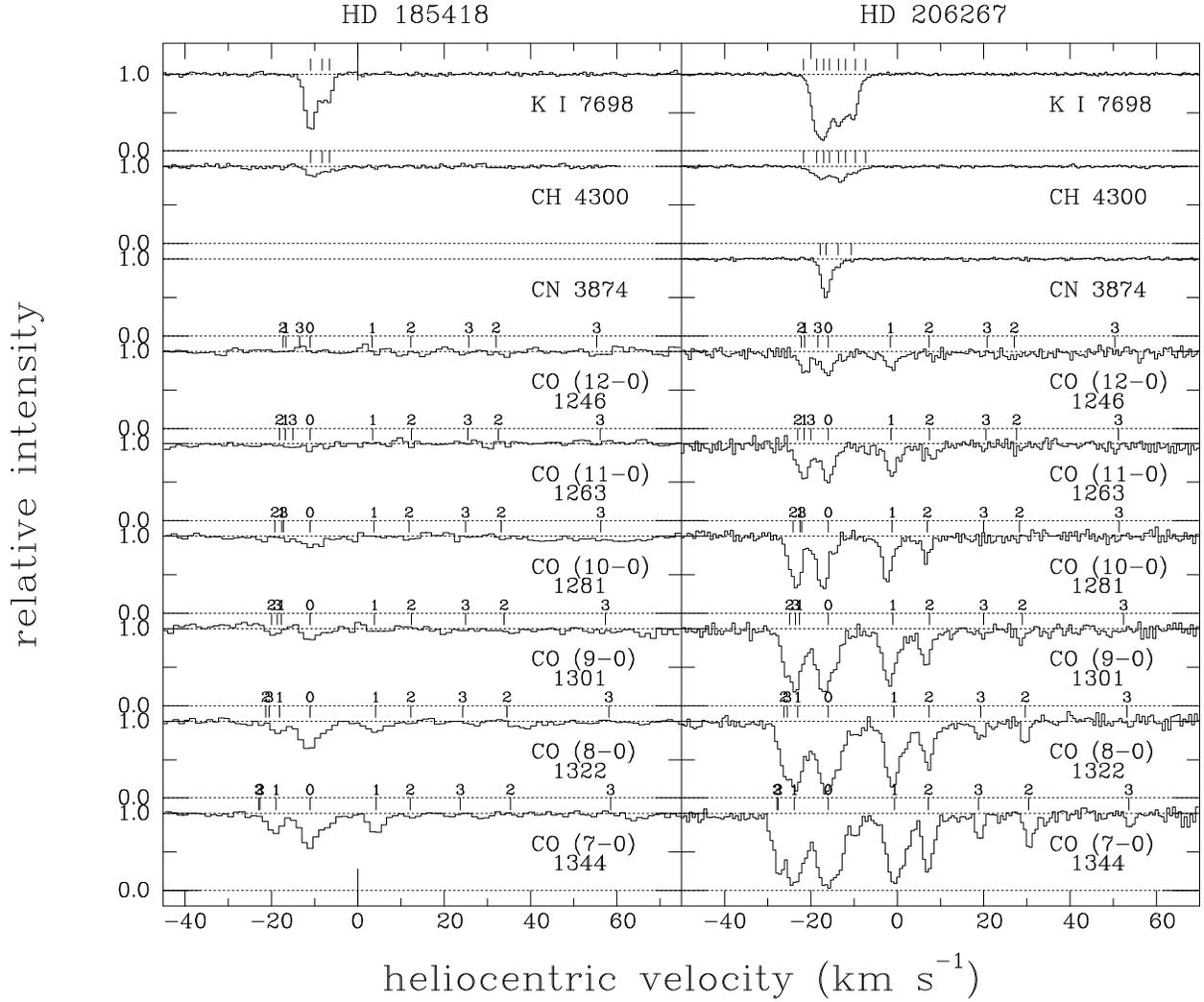}
\caption{STIS spectra of $^{12}$CO A-X bands toward HD~185418 and HD~206267. 
At the top are higher resolution optical spectra of K~I, CH, and CN, which show
more detailed component structure.
CO rotational levels are marked above each spectrum (R, Q, and P branches are left-to-right).}
\label{fig:stis2}
\end{figure}

\begin{figure}
\epsscale{1.0}
\plotone{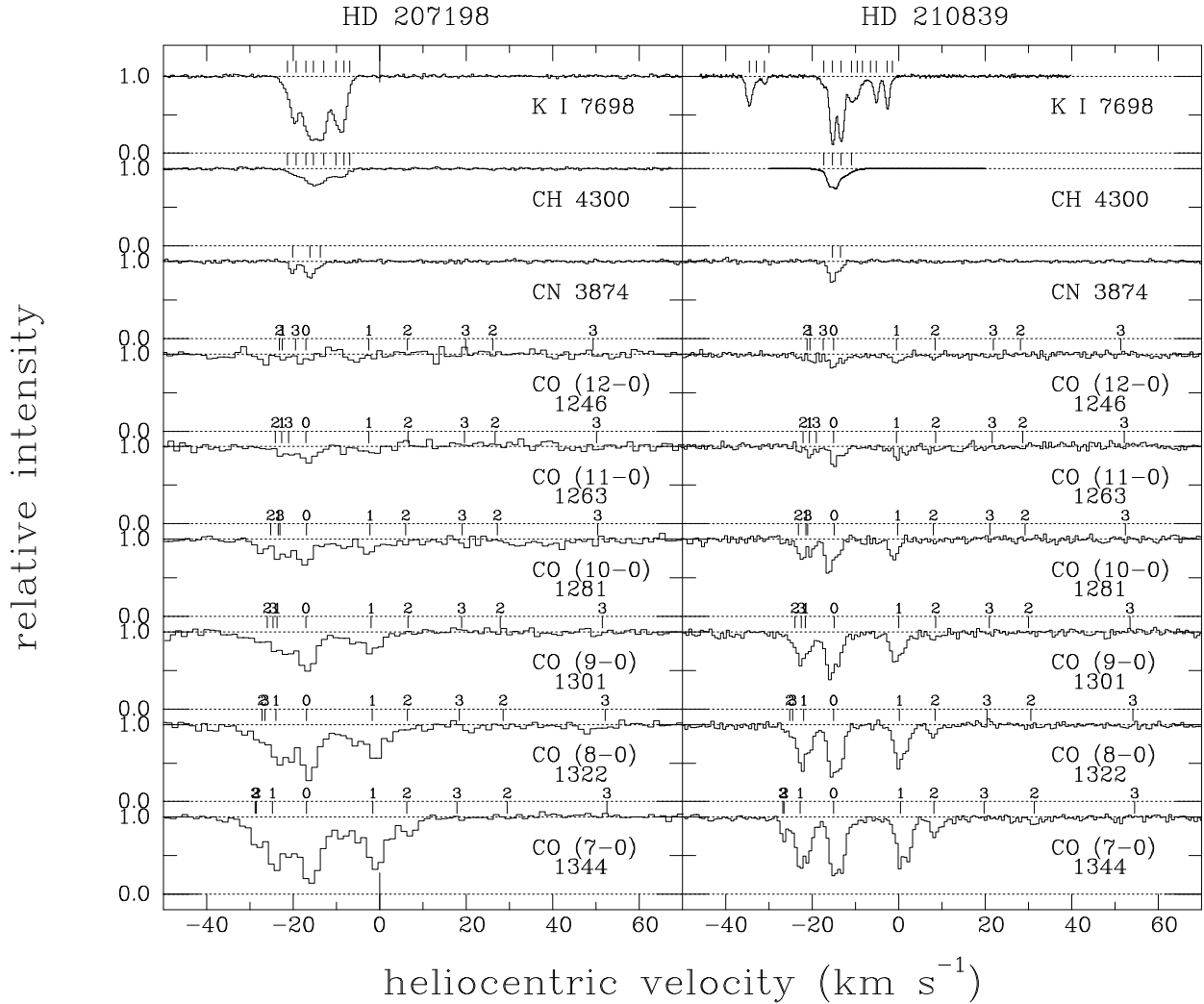}
\caption{STIS spectra of $^{12}$CO A-X bands toward HD~207198 and HD~210839. 
At the top are higher resolution optical spectra of K~I, CH, and CN, which show
more detailed component structure.
CO rotational levels are marked above each spectrum (R, Q, and P branches are left-to-right).}
\label{fig:stis3}
\end{figure}

\begin{figure}
\epsscale{1.0}
\plotone{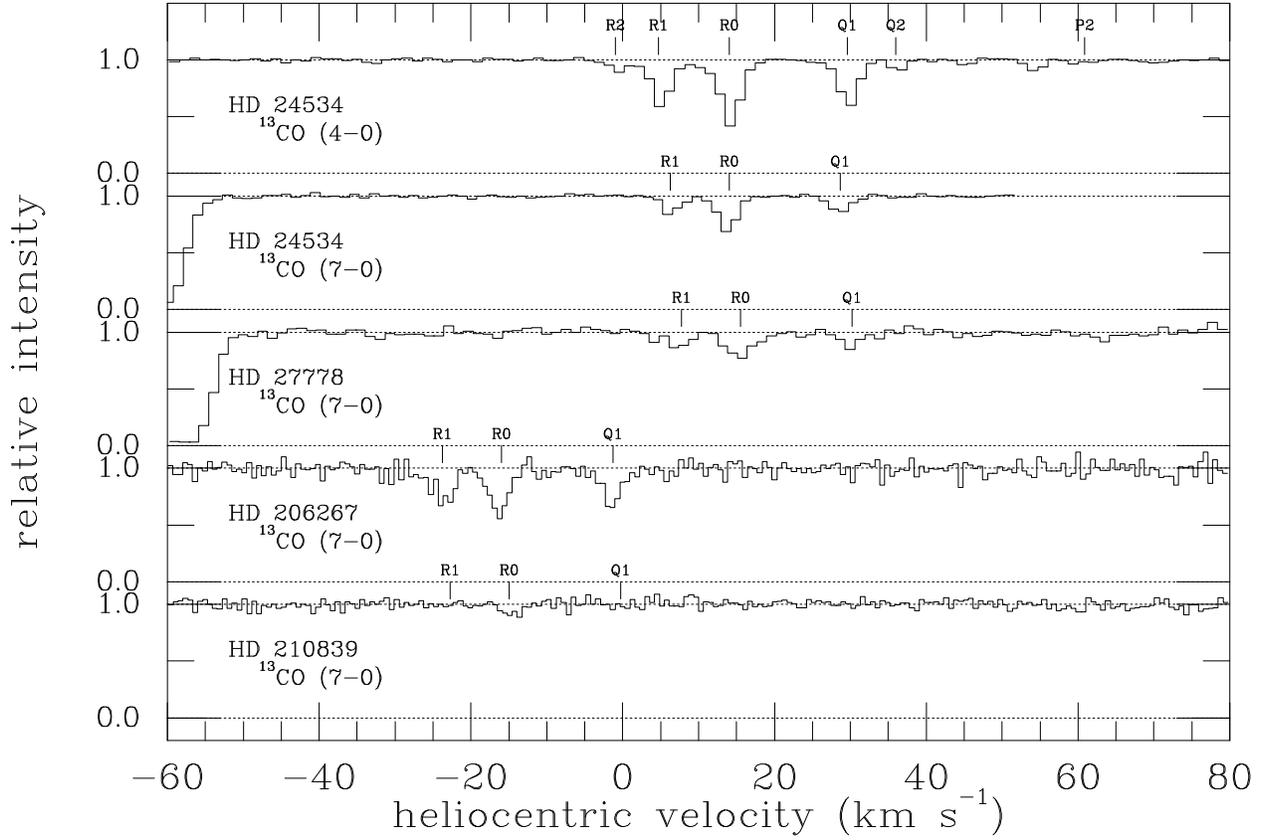}
\caption{STIS spectra of the $^{13}$CO A-X (7-0) (1347 \AA) band toward four stars and the $^{13}$CO A-X (4-0) (1421 \AA) band toward HD~24534.
Rotational levels are marked above each spectrum; lines from $J$ = 2 are evident in the A-X (4-0) band.
Additional weak absorption red-ward of the A-X (4-0) band is from $^{12}$C$^{18}$O (Sheffer et al. 2002b).}
\label{fig:13co}
\end{figure}

\begin{figure}
\epsscale{1.0}
\plotone{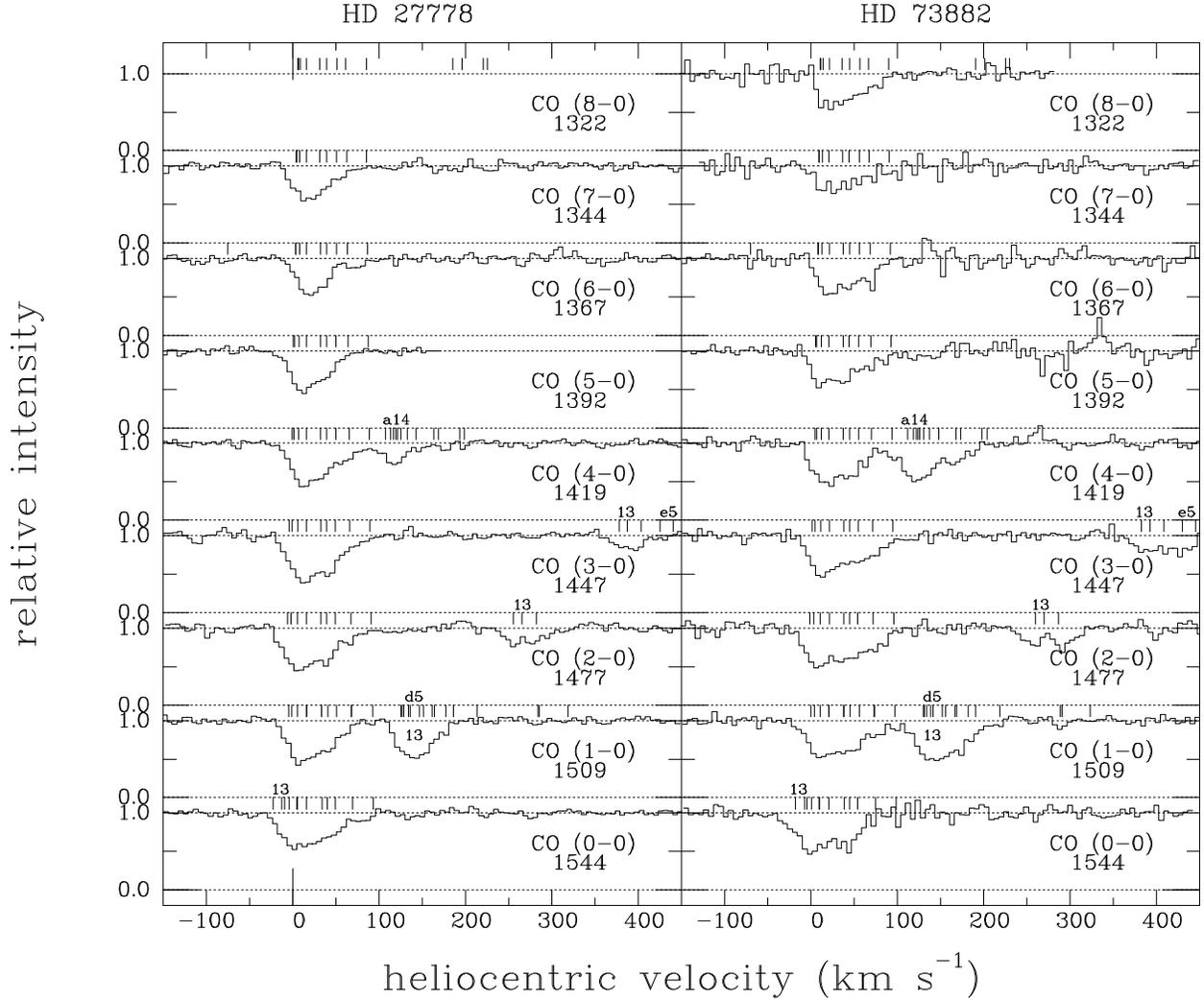}
\caption{{\it IUE} spectra of $^{12}$CO A-X bands toward HD~27778 and HD~73882.
CO rotational levels are marked above each spectrum (R, Q, and P branches are left-to-right).
Additional absorption from $^{13}$CO A-X bands and strongest $^{12}$CO intersystem bands is also marked.
Note the larger velocity range covered in this plot, compared to plots of STIS spectra.}
\label{fig:iue}
\end{figure}

\begin{figure}
\epsscale{1.0}
\plotone{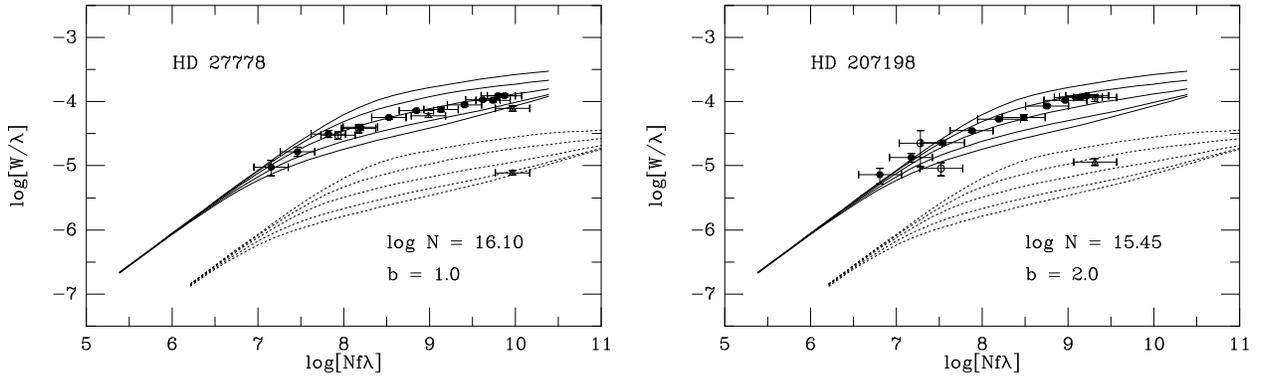}
\caption{Curves of growth for $^{12}$CO toward HD~27778 and HD~207198.
Solid lines are theoretical curves [single component; using structure for A-X (6-0) band] for $b$ = 0.3, 0.5, 1.0, 3.0, and 5.0 km~s$^{-1}$ (bottom to top).
Dotted lines are corresponding theoretical curves for the E-X (0-0) band, displaced downward by 1.0 for clarity.
Filled circles represent values for permitted A-X bands; open circles represent intersystem bands, triangles represent B-X, C-X, and E-X (0-0) bands; the point for the E-X (0-0) band appears on both sets of theoretical curves.
Column density and (approximate) effective $b$-value are given for each sight line.
}
\label{fig:cog}
\end{figure}

\begin{figure}
\epsscale{1.0}
\plotone{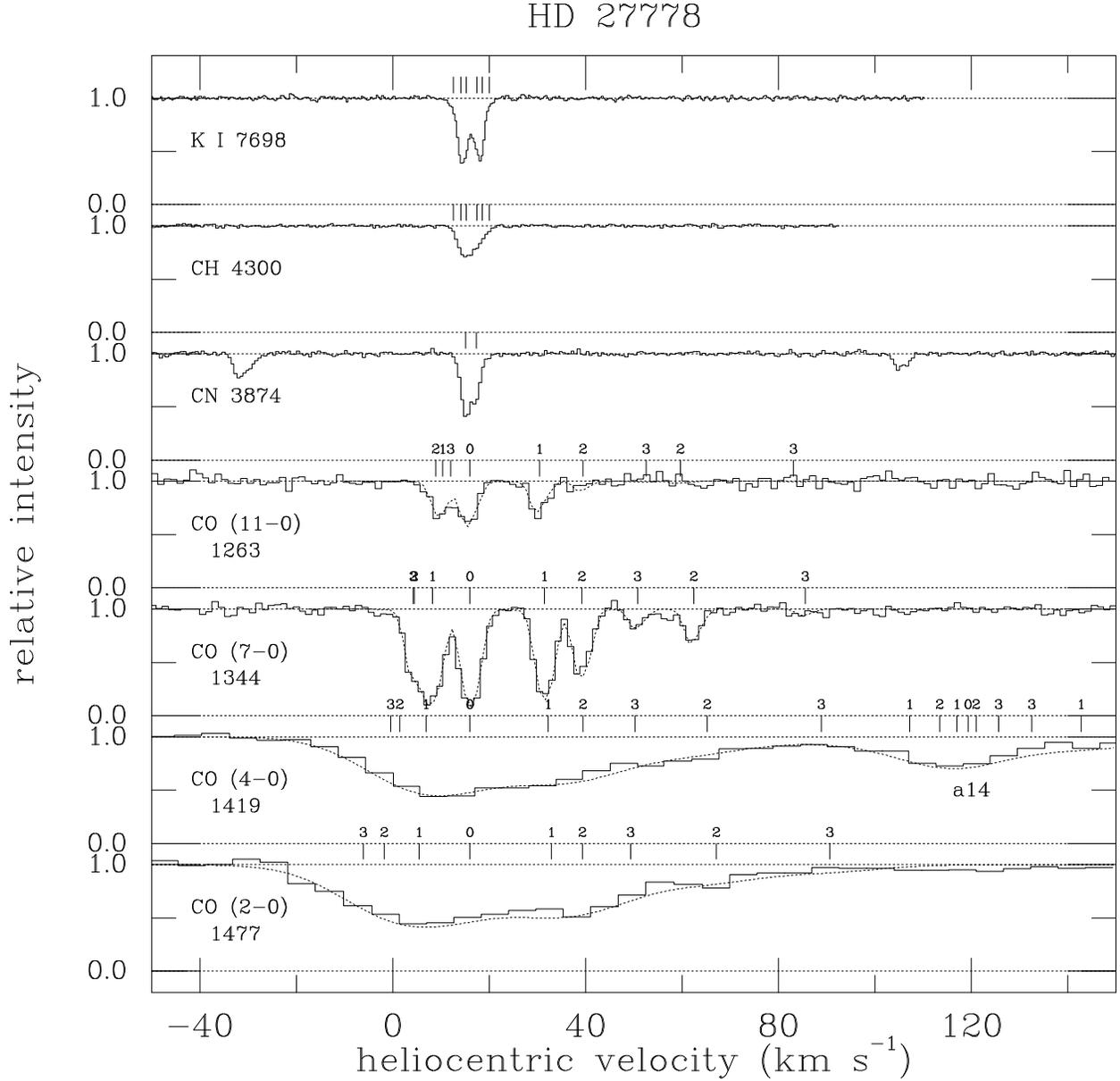}
\caption{Fits to UV line profiles toward HD~27778.
Observed spectra (from KPNO coude feed, STIS, and {\it IUE}) are given by solid histograms; fits to the UV profiles are given by dotted lines.
Note the $^{12}$CO a'14 intersystem band centered at $v$ $\sim$ 120 km~s$^{-1}$ near the permitted A-X (4-0) band.
Individual components are marked for higher resolution spectra of K~I, CH, and CN R(0) lines; note also the CN R(1) and P(1) lines.
CO rotational lines (R, Q, and P branches; $J$ = 0--3) are marked at the mean velocities of the unresolved component blends; two components of roughly equal column density, separated by 2.3 km~s$^{-1}$, were used in fitting the CO profiles.}
\label{fig:fit1}
\end{figure}

\begin{figure}
\epsscale{1.0}
\plotone{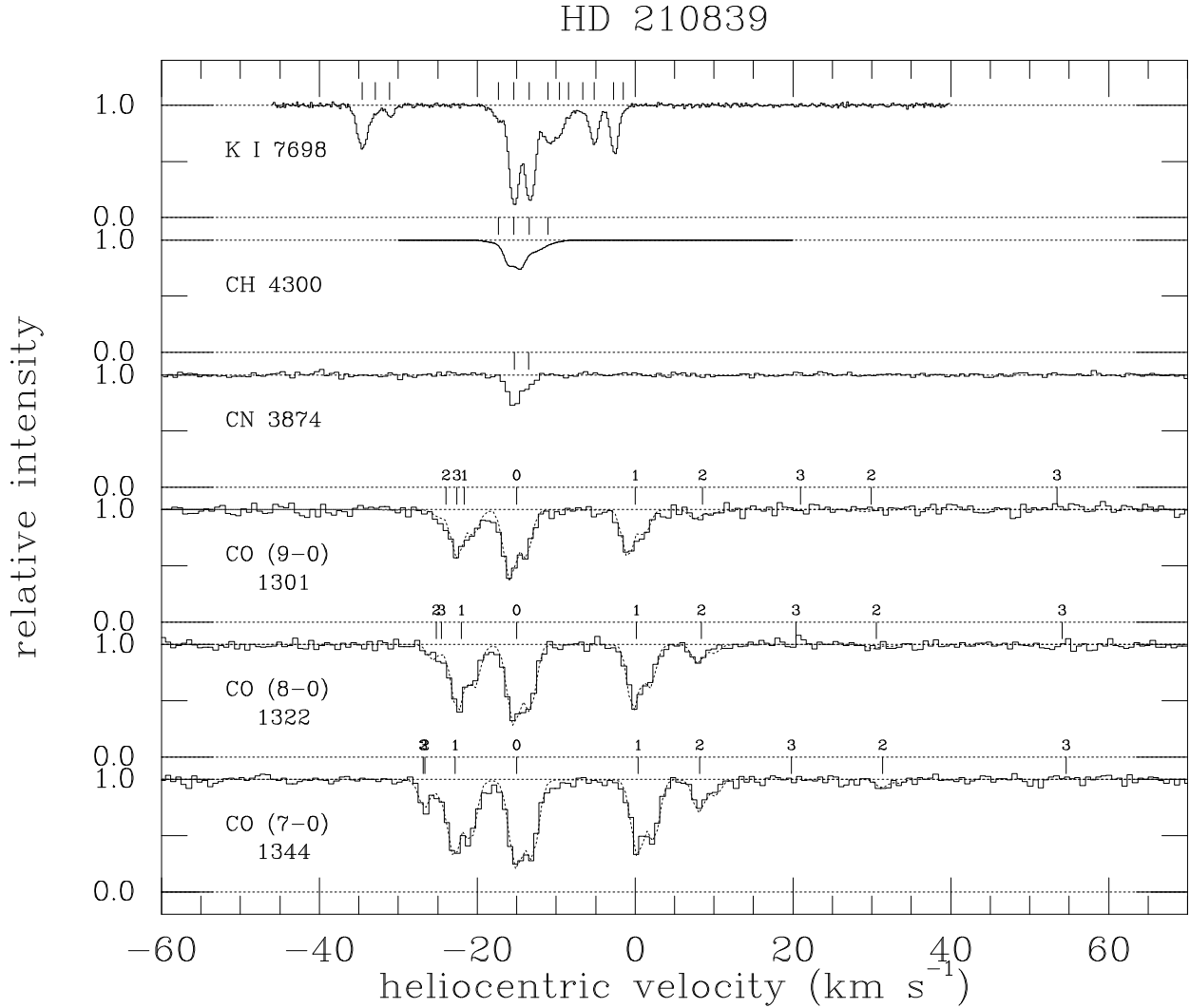}
\caption{Fits to UV line profiles toward HD~210839.
Observed spectra (from KPNO coude feed and STIS) are given by solid histograms; fits to the UV profiles are given by dotted lines.
Individual components are marked for higher resolution K~I, CH, and CN lines.
The CH profile is based on the component structure in Crane et al. (1995).
CO rotational lines (R, Q, and P branches; $J$ = 0--3) are marked at the mean velocities of the unresolved component blends; two components, separated by 2.1 km~s$^{-1}$, were used in fitting the CO profiles.}
\label{fig:fit2}
\end{figure}

While the STIS resolution is high enough to at least partially resolve the individual rotational levels in the CO and C$_2$ band structures (Figs.~\ref{fig:stis1}--\ref{fig:13co} and \ref{fig:fit1}--\ref{fig:c2uv}), it is not sufficient to separate or resolve all of the multiple, narrow gas components that produce the observed absorption lines. 
The blending between components --- both interstellar and rotational --- is particularly severe for the shorter wavelength R-branch lines of the CO A-X bands (Figs.~\ref{fig:stis1}--\ref{fig:stis3}) and the C$_2$ F-X bands (Fig.~\ref{fig:c2uv}). 
In order to determine accurate (total) column densities for the individual rotational levels contributing to the molecular bands, the absorption-line profiles were fitted with multi-component models based on fits to higher resolution optical spectra.   
The profile fitting program ``Owens'' (Lemoine et al. 2002) and a variant of the program FITS6P (e.g., Welty, Hobbs, \& Morton 2003) --- both of which perform iterative least-squares fits to the observed line profiles --- were used for the detailed component analyses of the optical and UV spectra. 

While lines of molecular species such as CH, C$_2$, and CN are the most direct optical tracers of diffuse molecular gas, the good correlations observed among the column densities of H$_2$, CH, and \ion{K}{1} --- and the often striking correspondence between the line profiles of CH and \ion{K}{1} --- suggest that component information from \ion{K}{1} may be relevant as well (Welty \& Hobbs 2001). 
More detailed structure is often discernible in the \ion{K}{1} line profiles because (1) the potassium atom is heavier than the CH molecule, (2) the hyperfine splitting of the \ion{K}{1} $\lambda$7698 line ($\sim$ 0.3 \kms) is smaller than the separation between the lambda-doubled components of the CH $\lambda$4300 line (1.43 \kms), and (3) the \ion{K}{1} line is typically stronger.
The more detailed \ion{K}{1} profile structures (i.e., the relative velocities and $b$-values for each discernible component) were therefore used to model the corresponding CH profiles.  
For the ten sight lines in this study, between three and thirteen components were needed to fit the \ion{K}{1} profiles; not all were detected in CH, however.  
Because CN is likely to be more concentrated in the colder, denser regions of each sight line, independent fits were performed to the CN profiles; one to four components (typically corresponding to the main components seen in \ion{K}{1} and CH) were required.
The structures found from the high-resolution optical spectra of \ion{K}{1}, CH, and/or CN (Appendix~\ref{sec-comp}; Table~\ref{tab:comp}) were then taken to represent the component distributions for the UV CO and C$_{2}$ bands in each sight line. 
Knowledge of the component structure can be crucial for determining column densities from the stronger CO lines, but is generally less important for the corresponding analyses of the much weaker C$_2$ lines.

\subsubsection{CO Column Densities}
\label{sec-nco}

The component structures found for \ion{K}{1} and CH were used for the initial fits to the CO band profiles observed with GHRS and STIS. 
In the Owens fits, the relative velocity and the $b$-value of each component were fixed, while the continuum placement and individual component column densities were allowed to vary. 
The FITS6P fits were performed on the normalized spectra, fixing the relative velocities (usually) and the $b$-values but allowing the component column densities to vary.
In both cases, all the CO A-X band profiles were fitted simultaneously.
For sight lines where both Owens and FITS6P were used to analyze the CO bands, the two methods generally yielded consistent results (within the respective uncertainties) both for the column densities of the individual rotational levels ($J$ = 0--3) and for the total $N$(CO).

Toward HD~192639, HD~185418, and HD~207198, all the CO bands were adequately fitted using the \ion{K}{1} component structures.
Toward HD 27778, HD~206267, and HD~210839, however, fits using the \ion{K}{1} structures tended to under-produce the weakest bands and overproduce the strongest bands. 
In an attempt to obtain mutually consistent column densities for all the bands in those sight lines, two additional series of fits were undertaken. 
In the first additional series, the relative velocities of the various gas components were again fixed to those of \ion{K}{1}, but both the column densities and the $b$-values were allowed to vary. 
For all three sight lines, the best fits to the CO bands were realized with $b$-values between 0.3 and 0.7 \kms\ for the main components --- i.e., somewhat smaller than the 0.6--1.1 \kms\ values found for \ion{K}{1}. 
The second additional series of fits, using the (simpler) CN velocity structures (but with slightly smaller $b$-values), also yielded more consistent fits to the ensemble of observed profiles. 
Inspection of the observed line profiles suggests that the profiles of the individual CO rotational lines seen in the higher resolution STIS spectra of HD~206267 and HD~210839 generally are more similar to the profiles of CN than to those of \ion{K}{1} and CH.
A close correspondence between the component structures for CO and CN has also been found by Pan et al. (2005) toward a number of stars in the Cep OB2 and Cep OB3 associations, including HD~206267, HD~207198, and HD~210839. 
For those three lines of sight, our total CO column densities are consistent with those determined by Pan et al.

Examples of the fits to some of the CO A-X system profiles toward HD~27778 and HD~210839 are shown in Figures~\ref{fig:fit1} and \ref{fig:fit2}, respectively, together with the corresponding higher resolution optical profiles of \ion{K}{1}, CH, and CN.
The fits to the profiles for HD~27778 utilize two components separated by 2.3 \kms, with $b$ $\sim$ 0.5--0.6 \kms\ and relative column densities $\sim$ 3:1 --- i.e., similar to the structure found for CN, but with slightly smaller $b$-values and somewhat different relative strengths (Table~\ref{tab:comp}).  
That structure (with the rotational level column densities $N_{J}$ obtained from fitting the ensemble of lines) provides acceptable fits to both weak and strong CO bands, including the stronger A-X bands observed at lower resolution with {\it IUE}.  
The best fits to the profiles for HD~210839 employ two components separated by 2.1 \kms, with $b$ $\sim$ 0.4 \kms\ (slightly smaller than the $b$-values adopted for CN) and relative strengths roughly 3:1.

Because the lower resolution {\it IUE} spectra of HD~27778, HD~73882, HD~207198, and HD~210121 do not resolve the CO rotational structure, the line profiles and total equivalent widths of the various permitted A-X and intersystem CO bands listed in Table~\ref{tab:ew2} were fitted by assuming rotational excitation temperatures of 3 and 5 K, using the component structures determined for \ion{K}{1} and CN.
The resulting total sight line CO column densities toward HD~27778 and HD~207198 are consistent with those derived from the higher resolution STIS spectra, but differ somewhat from the values determined by Joseph et al. (1986) (and Federman et al. 1994a) from curve of growth analyses of more limited {\it IUE} spectra of the permitted A-X bands (only).
While no high-resolution CN spectra are available for HD~73882, three components are seen in both \ion{K}{1} absorption and mm-wave CO emission (van Dishoeck et al. 1991).
Three-component fits to the intersystem and weakest permitted A-X bands of $^{12}$CO, assuming $T_{\rm ex}$ = 5 K and several choices for the relative column densities in each component, yield column densities consistent both with those determined from the COG analysis and with the value estimated from the CO emission lines (van Dishoeck et al. 1991).
Because the higher resolution GHRS spectra of HD~210121 allow better recognition of narrow stellar absorption features blended with several of the interstellar CO lines, the column density of $^{13}$CO is significantly smaller than that estimated by Welty \& Fowler (1992) from the {\it IUE} spectra alone.
Adoption of a smaller CO $b$-value (0.5 \kms vs. 1.0 \kms) for that sight line, however, yields a somewhat higher column density for $^{12}$CO.

The results of the CO profile fitting (FIT) analyses are summarized in Table~\ref{tab:nco}, which lists the total $^{12}$CO and $^{13}$CO column densities (for all gas components and all rotational levels), and in Table~\ref{tab:conj}, which lists the column densities $N_{J}$ for the individual rotational levels ($J$ = 0--3) for the eight sight lines with high-resolution UV spectra. 
Where only $N_0$ has been measured for $^{13}$CO, we have assumed that $N_0$/$N_1$ (or the excitation temperature $T_{01}$) is the same as for $^{12}$CO.
The uncertainties in the $N_{J}$ include contributions from photon noise, continuum uncertainty, and uncertainties in the component $b$-values ($\pm$ 0.05--0.10 \kms).
The continuum uncertainties are most significant for the weaker lines; the $b$-value uncertainties generally are most significant for the stronger lines (especially for $b$ $\la$ 0.5 \kms).
While the column densities derived from the AOD, COG, and FIT methods agree within the mutual uncertainties, we have adopted the $N$(CO) determined in the profile fits, as they account explicitly for saturation (using the detailed component structures), they use information from all the observed transitions, and they yield values for the individual rotational levels. 
The better fits to the CO bands obtained both with smaller $b$-values and with the CN component structure suggest that CO (like CN) may preferentially sample colder, denser gas than \ion{K}{1} and CH (see also Pan et al. 2005).  
Discussions of the physical conditions in the gas in the following sections seem consistent with that interpretation.

\subsubsection{C$_{2}$ Column Densities}
\label{sec-nc2}

Nearly all of the existing studies of interstellar C$_2$ have been based on optical and/or near-IR spectra of the A-X (Phillips) bands [(1-0) at 10144 \AA, (2-0) at 8757 \AA, and (3-0) at 7719 \AA] (e.g., Souza \& Lutz 1977; Hobbs \& Campbell 1982; van Dishoeck \& de Zeeuw 1984; van Dishoeck \& Black 1986b; Federman et al. 1994a; Gredel 1999).
In one exception, high-resolution GHRS ECH-B spectra of the D-X (Mulliken) (0-0) band at 2313 \AA\ and lower resolution G160M spectra (FWHM $\sim$ 16 \kms) of the F-X (0-0) band at 1341 \AA\ toward $\zeta$ Oph were used to obtain the C$_2$ rotational level populations up to $J$ = 24 and to estimate the relative $f$-values for the C$_2$ A-X, D-X, and F-X systems (Lambert, Sheffer, \& Federman 1995).  
Many of the lines from the individual rotational levels in the F-X (0-0) band are inextricably blended in the G160M spectrum, however.
Kaczmarczyk (2000) has discussed GHRS echelle spectra of the F-X (0-0) and D-X (0-0) bands toward HD~24534.

\begin{figure}
\epsscale{1.0}
\plotone{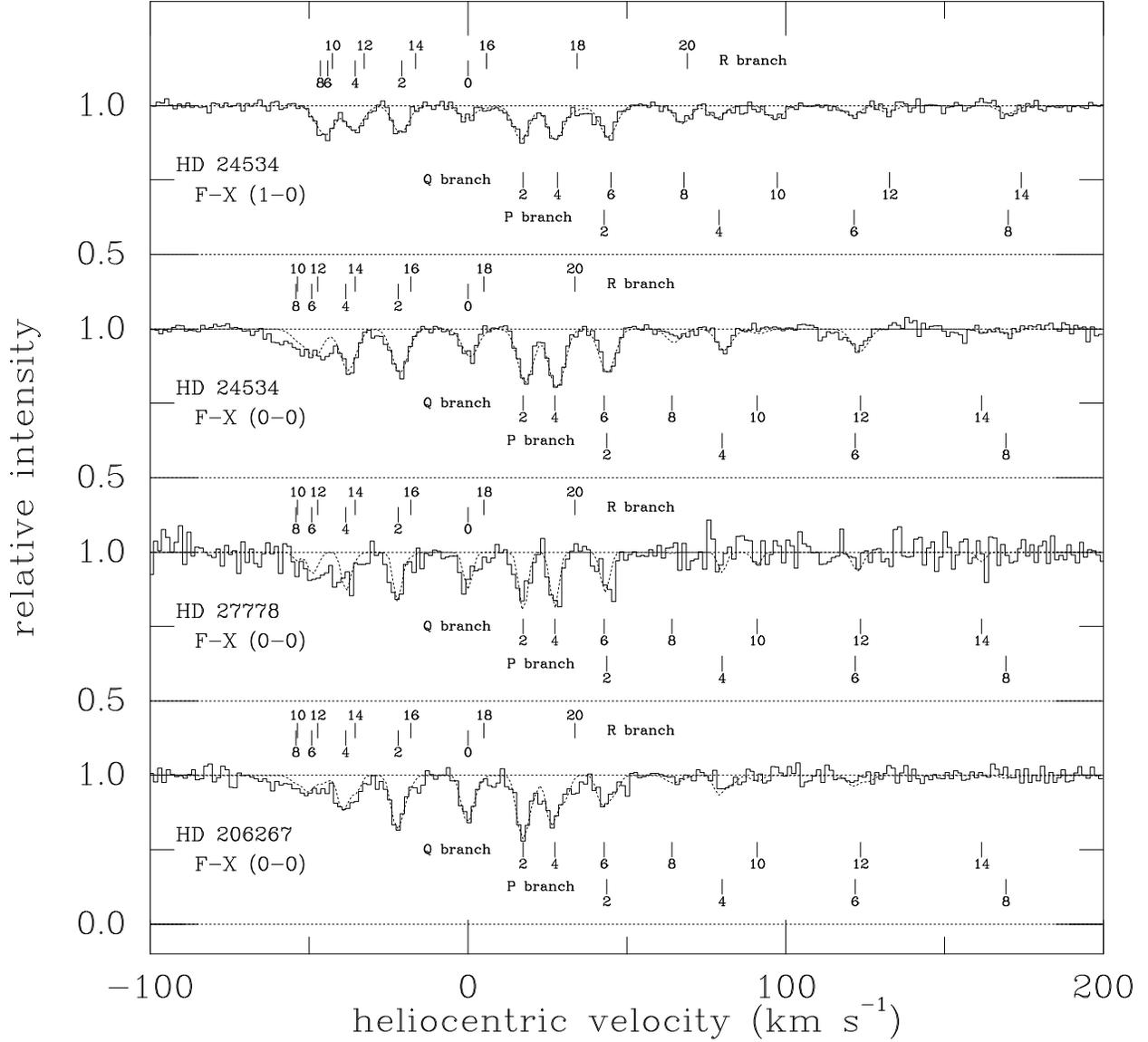}
\caption{UV C$_{2}$ F-X (1-0) and (0-0) (1314, 1341 \AA) bands toward HD~24534, HD~27778 and HD~206267.
Spectra have been shifted so that R(0) lines are at $v$ = 0 km~s$^{-1}$; note the expanded vertical scale for the top three profiles.
Observed spectra (from STIS) are given by solid histograms; fits to the profiles are given by dotted lines.
C$_2$ rotational lines (R, Q, and P branches; $J$ = 0, 2, 4, ...) are marked at the mean velocities of the unresolved component blends.
Note the poor fit to the F-X (0-0) R-branch ``pile-up'' for $J$ = 6--12 near $-$50 \kms (toward all three stars) and the relative weakness of the lines from $J$ = 8 and 10 in the F-X (0-0) band [relative to those in the F-X (1-0) band] toward HD~24534.} 
\label{fig:c2uv}
\end{figure}

\begin{figure}
\epsscale{1.0}
\plotone{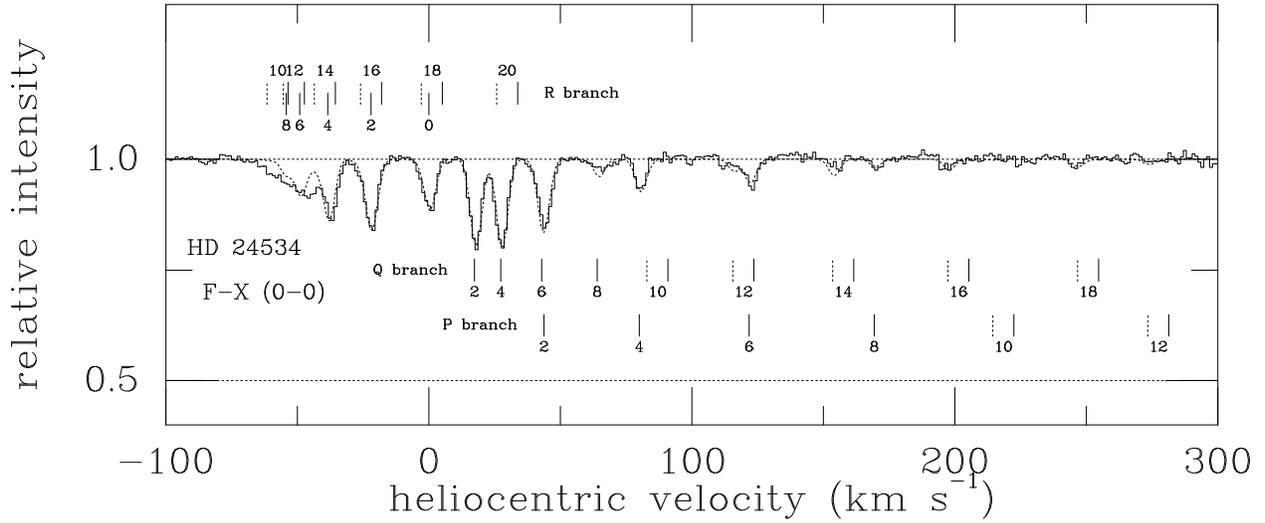}
\caption{C$_2$ F-X (0-0) (1341 \AA) band toward HD~24534 (combined spectra from three epochs).
The spectrum has been shifted so that the R(0) line is at $v$ = 0 km~s$^{-1}$; note the expanded vertical scale.
The observed spectrum (from STIS) is given by the solid histogram; a fit to the spectrum is given by the dotted line.
C$_2$ rotational lines (R, Q, and P branches; $J$ = 0, 2, 4, ...) are marked at the mean velocities of the unresolved component blends.
The solid tick marks have all been shifted by 8 km~s$^{-1}$, relative to the predicted wavelengths (see Sec.~3.3.2); the dotted tick marks for $J$ $\ge$ 10 show the unshifted locations.
In the fit, only the lower $J$ lines ($J$ $<$ 10) have been shifted.
Note the somewhat better fit to the R-branch ``pile-up'' and the weak absorption features at the unshifted positions of the higher $J$ Q-branch lines.}
\label{fig:fx00}
\end{figure}

\begin{figure}
\epsscale{1.0}
\plotone{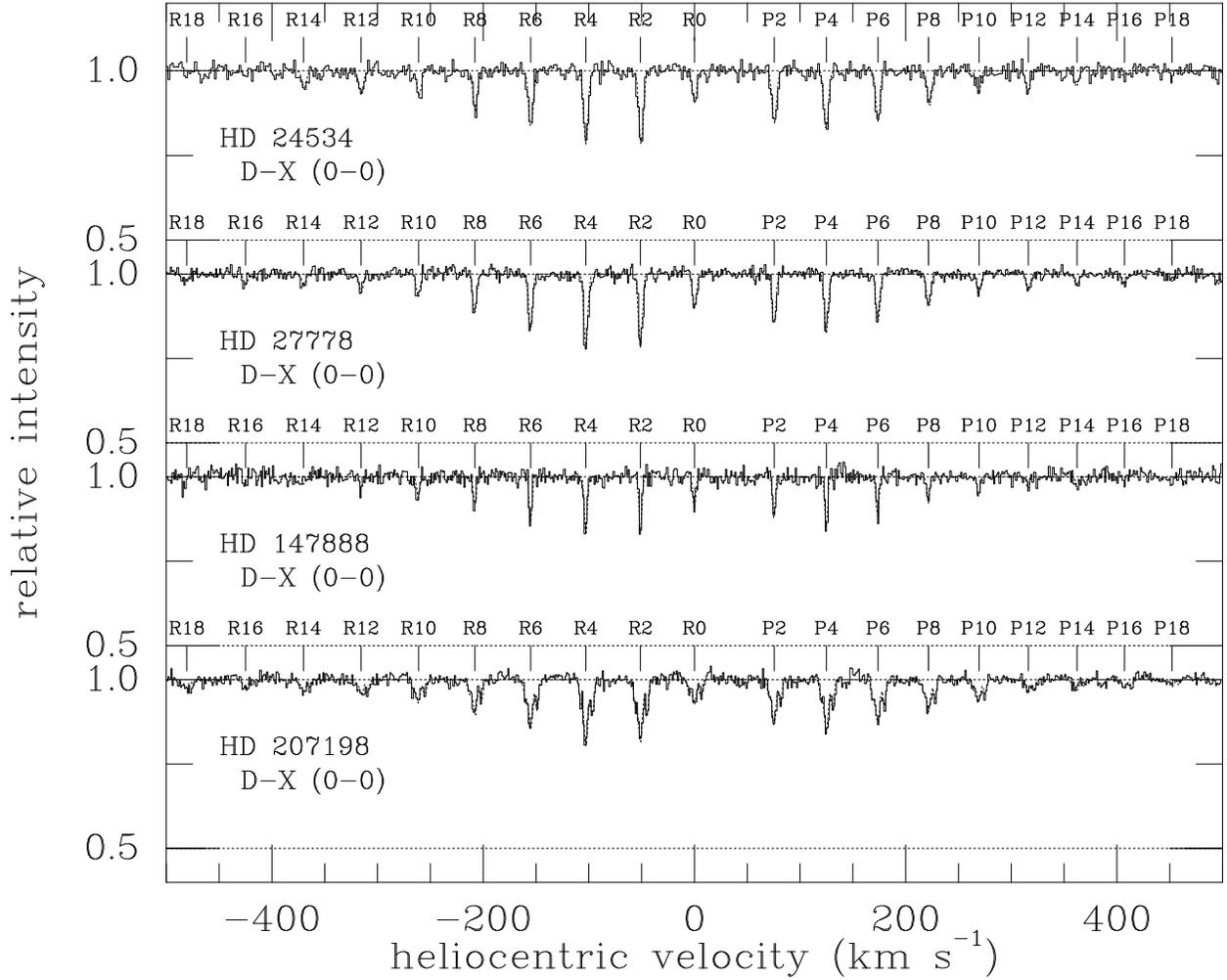}
\caption{UV C$_{2}$ D-X (0-0) (2313 \AA) band toward HD 24534, HD~27778, HD~147888, and HD~207198.
Spectra have been shifted so that R(0) lines are at $v$ = 0 km~s$^{-1}$; note the expanded vertical scale.
Observed spectra (from GHRS or STIS) are given by the solid histograms; fits to the profiles are given by the dotted lines.
C$_2$ rotational lines (R and P branches; $J$ = 0, 2, 4, ...) are marked at the mean velocities of the unresolved component blends; note the three components (or component groups) discernible toward HD~207198.}
\label{fig:c2dx}
\end{figure}

\begin{figure}
\epsscale{1.0}
\plotone{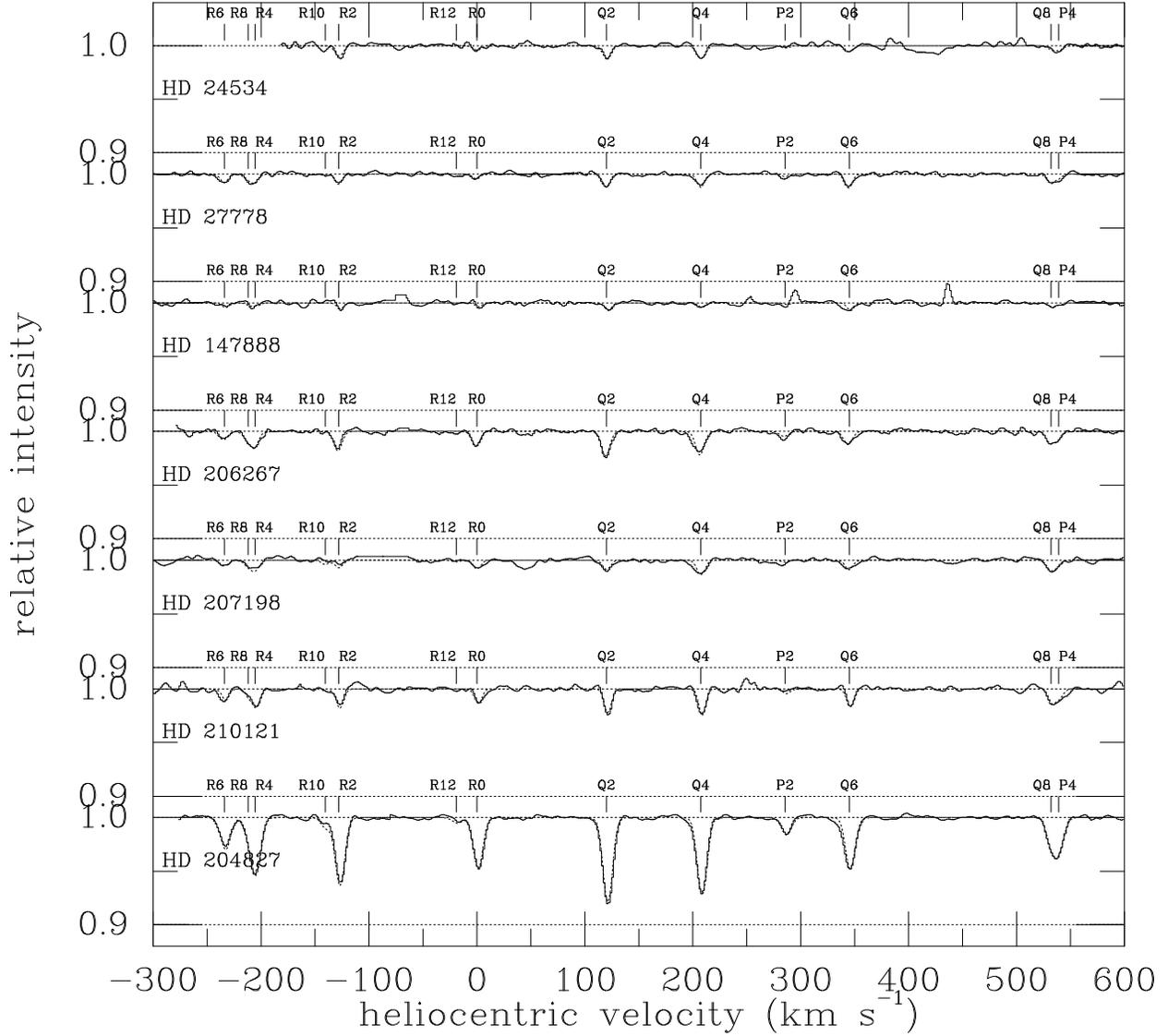}
\caption{Optical C$_{2}$ A-X (2-0) band at 8757 \AA\ toward seven stars.
Spectra have been shifted so that the R(0) line is at $v$ = 0 km~s$^{-1}$; note the expanded vertical scale for all profiles. 
Sections of the spectrum of HD~24534 within the stellar H Paschen 12 (emission) feature were not normalized (several R-branch lines are thus missing here).
C$_2$ rotational lines (R, Q, and P branches; $J$ = 0, 2, 4, ...) are marked at the mean velocities of the unresolved component blends.
The spectrum of HD~204827, with much stronger C$_2$ lines, is included to show the band structure more clearly.
Observed spectra (from ARCES) are given by solid histograms; fits to the profiles are given by dotted lines.} 
\label{fig:c2opt}
\end{figure}

\begin{figure}
\epsscale{1.0}
\plotone{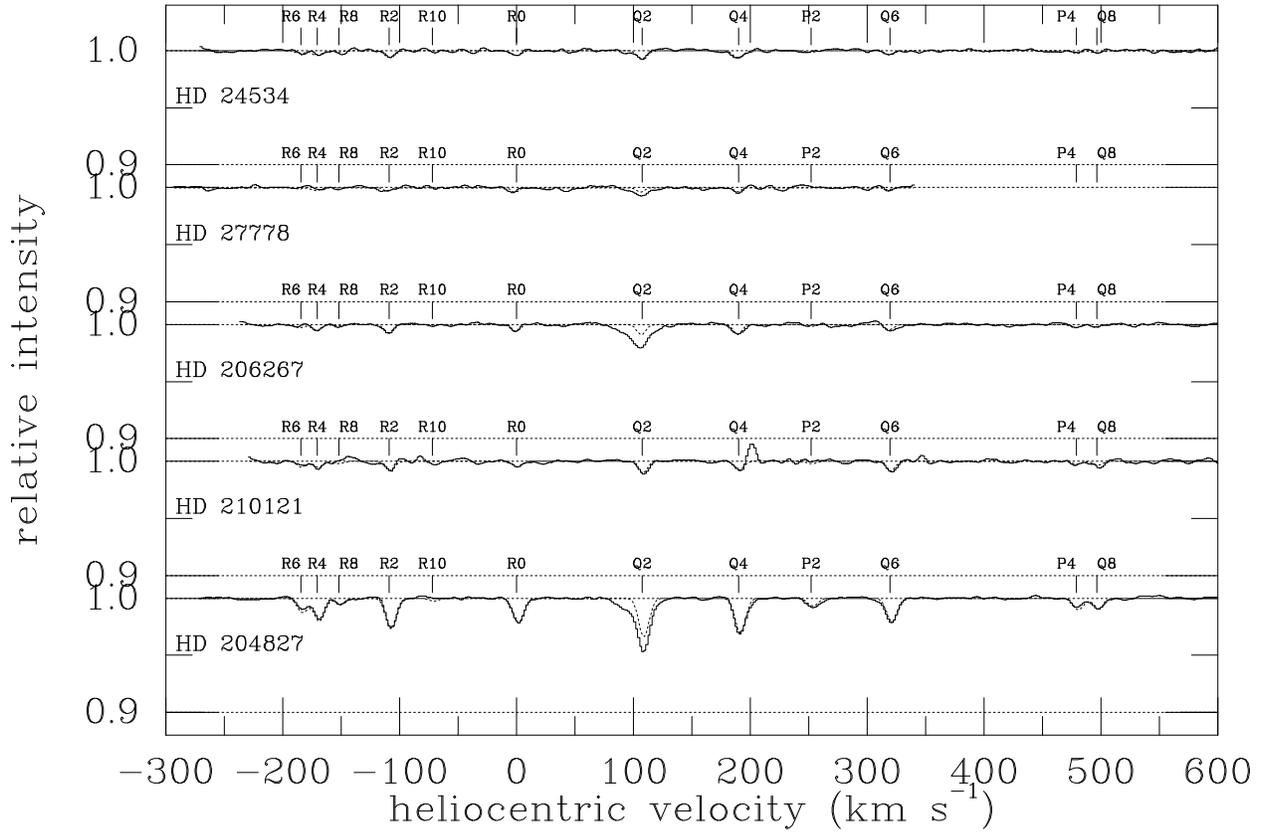}
\caption{Optical C$_{2}$ A-X (3-0) band at 7719 \AA\ toward five stars.
Spectra have been shifted so that the R(0) line is at $v$ = 0 km~s$^{-1}$; note the expanded vertical scale for all profiles. 
C$_2$ rotational lines (R, Q, and P branches; $J$ = 0, 2, 4, ...) are marked at the mean velocities of the unresolved component blends.
The spectrum of HD~204827, with much stronger C$_2$ lines, is included to show the band structure more clearly.
Observed spectra (from ARCES) are given by solid histograms; fits to the profiles are given by dotted lines; note the additional broad absorption feature (a weak diffuse interstellar band) nearly coincident with the Q(2) line in several cases.} 
\label{fig:c2opt3}
\end{figure}

In this paper, we present high-resolution profiles of the UV C$_{2}$ F-X (0-0) and (1-0) bands (as seen in the STIS E140H spectra of HD~24534, HD~27778, and HD~206267; Figures~\ref{fig:c2uv} and \ref{fig:fx00}), along with high-resolution GHRS ECH-B and STIS E230H spectra of the D-X (0-0) band toward HD~24534, HD~27778, HD~147888, and HD~207198 (Figure~\ref{fig:c2dx}) and high S/N optical spectra of the A-X (3-0) and (2-0) bands toward seven stars (Figures~\ref{fig:c2opt} and \ref{fig:c2opt3}).
All these UV and optical spectra at least partially resolve the rotational structure of the C$_2$ bands --- enabling detailed line profile analyses of the C$_2$ absorption. 
The D-X (0-0) band is particularly useful, as the individual rotational lines from both the R and P branches are well separated and relatively strong.
Equivalent widths for the individual D-X (0-0) lines are given in Table~\ref{tab:c2dx}; equivalent widths for lines in the A-X (2-0) and (3-0) bands are listed in Table~\ref{tab:c2optw}.
Wavelengths and relative strengths for the individual rotational lines in the various C$_2$ bands --- as well as the adopted total band $f$-values --- are listed in Table~\ref{tab:c2bands} (Appendix~\ref{sec-c2bands}).

By comparing column densities derived from the various UV and optical bands, a set of mutually consistent band oscillator strengths may be determined.
Lambert et al. (1995) adopted a band oscillator strength $f$ = 0.0545 for the D-X (0-0) band, based on two independent (and closely concordant) theoretical estimates.
Comparison of the overall strengths of the D-X (0-0) and F-X (0-0) bands toward $\zeta$ Oph then yielded $f$ = 0.10$\pm$0.01 for the latter band --- in excellent agreement with the theoretical value 0.098 determined by Bruna \& Grein (2001).
With those $f$-values, fits to the profiles of the D-X (0-0) and F-X (0-0) bands toward HD~24534 and HD~27778 yield reasonably consistent C$_2$ column densities for rotational levels $J$ = 0--6 (Table~\ref{tab:nc2}), though there appear to be some systematic discrepancies for $J$ = 8 and 10 (see below).
Similar fits to the F-X (1-0) band at 1314 \AA\ toward HD~24534 then suggest a total oscillator strength $f$ $\sim$ 0.06 for that band.

As for CO, the \ion{K}{1} velocity structure was initially adopted to determine the column densities of the individual C$_2$ rotational levels present in the spectra of the three UV bands. 
Additional fits were performed using either a single component (with an effective $b$) or the structure that best accounted for the saturation in the stronger CO bands; a separate three-component fit was made to the D-X (0-0) band toward HD~207198, where more complex structure (similar to that seen in \ion{K}{1} and CH) is evident (Figs.~\ref{fig:stis3}~and~\ref{fig:c2dx}). 
In each case, the various fits led to similar results (within the uncertainties), indicating that the total C$_2$ column densities derived from the C$_{2}$ D-X (0-0) and F-X (0-0) and (1-0) bands generally are not very sensitive to the details of the gas velocity distribution along these lines of sight. 
Column densities for individual C$_2$ rotational levels up to $J=$ 12 were thus obtained from the F-X band(s) toward HD~24534, HD~27778, and HD~206267 and up to $J=$ 18 from the D-X (0-0) band toward HD~24534, HD~27778, HD~147888, and HD~207198 (Table~\ref{tab:nc2}). 
The listed uncertainties include contributions from photon noise, continuum-fitting uncertainty, and uncertainty in the effective $b$; the uncertainties for the D-X (0-0) band are similar to those that would be inferred from the uncertainties in the equivalent widths of the individual rotational lines.
The relative rotational level populations $N_{J}$/$N_2$ appear to be fairly similar for these five lines of sight (but see the caveats below).
Toward HD~210839, the C$_2$ F-X (0-0) band absorption is significantly weaker, so that detailed rotational distributions could not be determined from the UV spectra. 

Detailed examination of the fits to the various C$_2$ band profiles and comparisons of the C$_2$ column densities obtained from those bands, however, indicate that the strength and structure of the UV C$_2$ bands --- particularly the F-X (0-0) band --- are not completely understood. 
On the one hand, the fits to the D-X (0-0) and F-X (1-0) bands toward HD~24534 are quite good (Figs.~\ref{fig:c2dx} and \ref{fig:c2uv}) and yield mutually consistent column densities for the individual rotational levels $J$ = 0--10.  
Moreover, while there are some ``discrepant'' $N_{J}$ for individual sight lines, there do not appear to be any systematic differences in the $N_{J}$ determined from the D-X (0-0), A-X (3-0), and A-X (2-0) bands (Table~\ref{tab:nc2}).
The fits to the F-X (0-0) band toward HD~24534, HD~27778, and HD~206267, however, all are rather poor near the ``pile-up'' of the R(6)--R(12) lines; some additional broad (?) absorption appears to be present there.  
In addition, the calculated wavelengths of the clearly detected F-X (0-0) lines ($J$ $<$ 10) seem to be too small by about 36 m\AA\ (8 km~s$^{-1}$).  
Finally, the lines from $J$ = 6--10 seem to be weaker in the F-X (0-0) band than in the intrinsically weaker F-X (1-0) band; compare, for example, the relatively unblended Q(8), Q(10), and P(8) lines of the two bands toward HD~24534 in the upper half of Figure~\ref{fig:c2uv}. 

The discrepancies between the C$_2$ D-X (0-0) and F-X (0-0) bands previously noted in GHRS spectra of $\zeta$ Oph (Lambert et al. 1995) and HD~24534 (Kaczmarczyk 2000) may reflect similar effects.  
Lambert et al. might not have recognized either the ``extra'' absorption near the F-X (0-0) band R-branch ``pile-up'' or the 8 km~s$^{-1}$ velocity offset in the lower resolution G160M spectrum.  
Their fits to the whole F-X (0-0) band, which uniformly scaled the rotational level column densities determined from the D-X (0-0) band at 2313 \AA, would thus have tended to predict too much absorption over the rest of the band --- as seen, for example, for the Q(10)-P(4), Q(12)-P(6), and Q(14)-P(8) blends in their Figure~3.
Kaczmarczyk found that the $N_J$ derived from the D-X (0-0) band did not yield a good fit to the Q(12)-P(6) and Q(14)-P(8) blends in the F-X (0-0) band and noted (unspecified) problems with the F-X (0-0) band wavelengths.
While the ``extra'' absorption near the F-X (0-0) band R-branch ``pile-up'' is not evident in the lower S/N GHRS spectrum analyzed by Kaczmarczyk, he did mention a ``lower than expected'' continuum near several of the stronger lines.

The causes of the discrepant behavior observed for the F-X (0-0) band are not understood.
In principle, the ``extra'' broad absorption near the R-branch pile-up could be a stellar feature, but the three stars observed (HD~24534, HD~27778, and HD~206267) have somewhat different spectral types, radial velocities, and projected rotational velocities.
Moreover, the ``extra'' absorption toward HD~24534 is very similar in three separate sets of STIS spectra taken at epochs separated by about 3 and 33 days, respectively.
The wavelengths adopted for both of the F-X bands (Table~\ref{tab:c2bands}) were calculated from molecular constants derived from flash discharge spectra analyzed by Herzberg, Lagerqvist, \& Malmberg (1969).
The 8 km~s$^{-1}$ velocity offset for the clearly detected (and unblended) lines in the F-X (0-0) band (all with $J$ $<$ 10) does not appear to be due to errors in the STIS wavelengths, as the (0-0) band appears in two adjacent orders, one of which also contains the $^{12}$CO A-X (7-0) band at 1344 \AA\ and the \ion{Cl}{1} line at 1347 \AA.
While the velocity offset for those lower $J$ F-X (0-0) band lines would be consistent with a slightly smaller value for the upper level $\nu_0$ (74530.9 cm$^{-1}$ instead of 74532.9 cm$^{-1}$), the wave numbers for the higher $J$ lines measured by Herzberg et al. (1969) generally are consistent (within $\pm$ 0.2 cm$^{-1}$) with the values predicted using their tabulated constants.
Unfortunately, Herzberg et al. apparently could not measure the positions of any of the lower $J$ lines in the F-X (0-0) band, and there are no clear detections of any of the unblended higher $J$ lines in the spectra shown in Figure~\ref{fig:c2uv}.
There are, however, tantalizing hints of weak absorption at the predicted (i.e., unshifted) positions of several of the higher $J$ Q-branch lines in the spectrum of HD~24534 [which was taken from data set o66p01020, the longest single STIS exposure of HD~24534 covering the F-X (0-0) band].
When STIS spectra of HD~24534 obtained at two other epochs are combined with the spectrum shown in Figure~\ref{fig:c2uv} (more than doubling the total exposure time), those weak features persist. 
Fits to that combined spectrum with only the lines from $J$ $<$ 10 shifted yield low values of $N_J$ for $J$ = 8--12, but quite reasonable values of $N_J$ for $J$ = 14--18 (Figure~\ref{fig:fx00}, Table~\ref{tab:nc2}).

Column densities for the individual C$_2$ rotational levels were also obtained from the A-X (2-0) and (3-0) bands at 8757 and 7719 \AA\ (Thorburn et al. 2003), using both the measured equivalent widths of the individual rotational lines in the medium-resolution ARCES spectra and fits to the ensemble of lines (including blended features).
Figure~\ref{fig:c2opt} shows normalized spectra of the optical C$_2$ A-X (2-0) band toward seven stars; Figure~\ref{fig:c2opt3} shows corresponding spectra of the A-X (3-0) band toward five stars.  
The profiles observed toward HD~204827, which has high column densities of both C$_2$ and C$_3$ (Thorburn et al. 2003; Oka et al. 2003; \'{A}d\'{a}mkovics et al. 2003) are included to show the band structures more clearly.
Table~\ref{tab:c2optw} lists the wavelengths, log($f\lambda$) values, and equivalent widths for the A-X (2-0) and (3-0) band transitions toward seven of our stars (plus HD~204827); C$_2$ was not detected toward HD~185418 or HD~192639 and was not observed toward HD~73882 (though see Gredel et al. 1993).
The equivalent widths of A-X (2-0) R-branch lines located in the steep wing or deep core of the stellar Paschen 12 line of \ion{H}{1} can be rather uncertain; in some cases (e.g., toward HD~24534), some of those R-branch lines could not be reliably measured.
There is also a weak diffuse interstellar band nearly coincident with the A-X (3-0) Q(2) line in several cases (Herbig \& Leka 1991).
The equivalent widths in Table~\ref{tab:c2optw} are consistent with most of the (generally lower precision) values previously reported for HD~27778 (Federman et al. 1994a); HD~24534, HD~206267, HD~207198, and HD~210839 (Federman \& Lambert 1988); and HD~210121 (Gredel et al. 1992).

Unfortunately, the band oscillator strengths for the C$_2$ A-X bands have been somewhat uncertain, with persistent differences between theoretical and experimental estimates (e.g., van Dishoeck \& Black 1989; Langhoff et al. 1990; Lambert et al. 1995; though see Erman \& Iwamae 1995).
Lambert et al. (1995) adopted $f$ = 0.00123$\pm$0.00016 for the A-X (2-0) band, based on comparisons with the UV D-X (0-0) and F-X (0-0) bands toward $\zeta$ Oph and the D-X (0-0) band $f$-value noted above.
That empirical value lies roughly halfway between the theoretical estimate $f$ = 0.00144$\pm$0.00007 of Langhoff et al. (1990) and the experimental value $f$ = 0.00100 adopted by van Dishoeck \& Black (1989).
Similar comparisons based on our fits to the profiles of the F-X, D-X, and A-X bands observed toward HD~24534, HD~27778, HD~147888, HD~206267, and HD~207198 suggest a somewhat higher $f$-value for the A-X (2-0) band, however.
We have therefore adopted the value $f$ = 0.00140 --- an average of the $f$ = 0.00136$\pm$0.00015 measured by Erman \& Iwamae (1995) and the value calculated by Langhoff et al. (1990).
On average (toward HD~24534, HD~27778, HD~204827, HD~206267, HD~210121, and several other stars with data for these two Phillips bands), the strength of the C$_2$ A-X (3-0) band then is consistent with $f$ $\sim$ 0.00065 --- i.e., weaker than the (2-0) band by the expected factor $\sim$ 2.2 (van Dishoeck \& Black 1982).
The C$_2$ column densities derived from fits to the A-X (2-0) and (3-0) bands (with the adopted $f$-values) are listed in Table~\ref{tab:nc2}.
The values for $\zeta$ Oph given in Table~\ref{tab:nc2} were derived from published equivalent widths (Hobbs \& Campbell 1982; Danks \& Lambert 1983; van Dishoeck \& Black 1986b; Lambert et al. 1995).
The adopted C$_2$ band $f$-values yield quite good agreement for the column densities derived from the D-X (0-0), A-X (3-0), and A-X (2-0) bands toward $\zeta$~Oph as well.

\begin{figure}
\epsscale{1.0}
\plotone{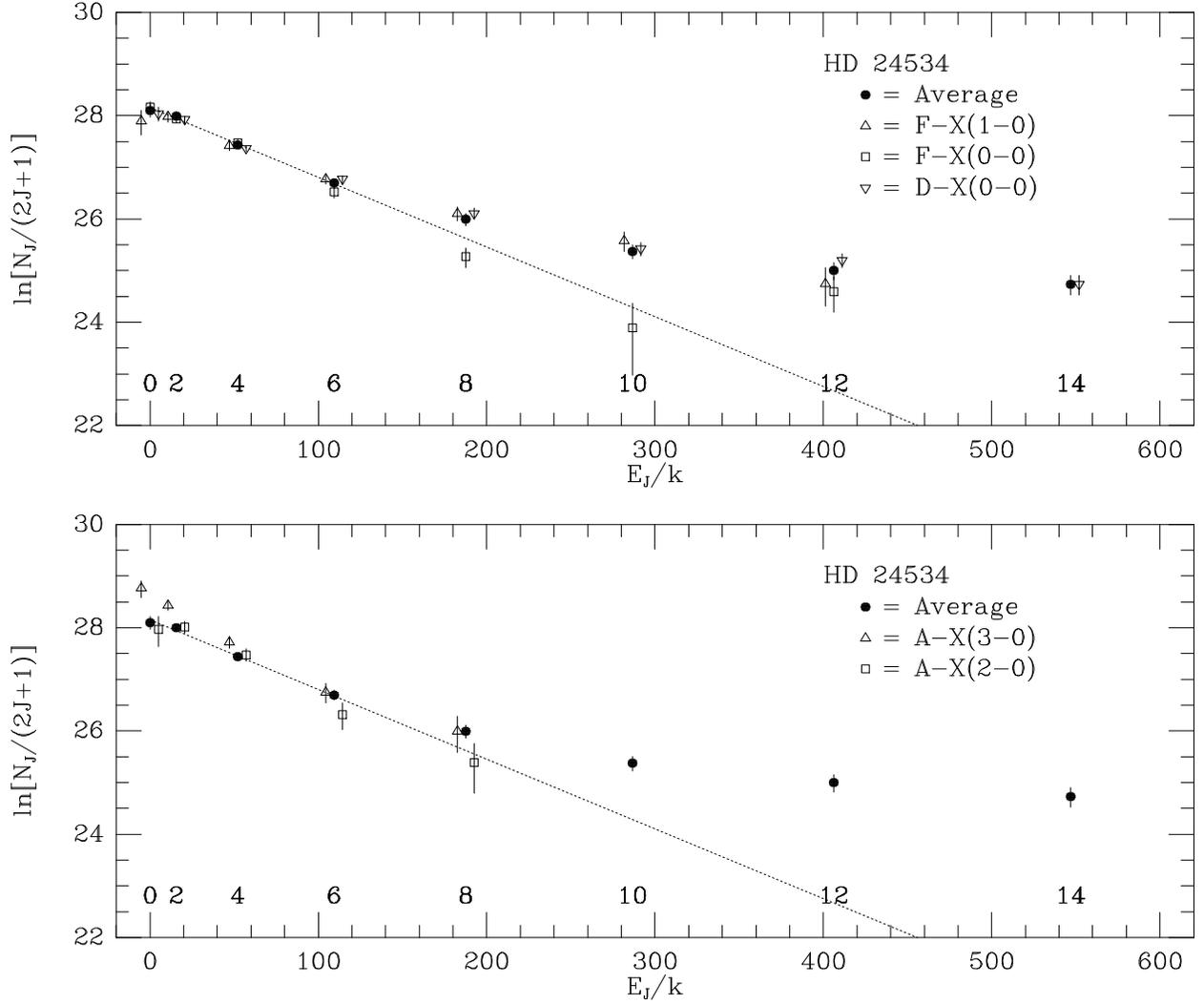}
\caption{Rotational populations of C$_2$ ($J$ = 0--14) toward HD~24534.
The upper panel shows the normalized $N_{J}$/(2$J$+1) obtained from the UV bands --- F-X (1-0) at 1314 \AA\ ({\it upright triangles}), F-X (0-0) at 1341 \AA\ ({\it squares}), and D-X (0-0) at 2313 \AA\ ({\it inverted triangles}) --- compared with the average values derived from all the UV and optical bands ({\it filled circles}).
Note the apparent deficits for $J$ = 8 and 10 in the F-X (0-0) band; similar differences are seen for those levels in the F-X (0-0) bands toward HD~27778 and HD~206267.
The lower panel shows the corresponding $N_{J}$/(2$J$+1) obtained from the optical bands --- A-X (3-0) at 7719 \AA\ ({\it triangles}) and A-X (2-0) at 8757 \AA\ ({\it squares}) --- again compared with the average values ({\it filled circles}).
In both panels, the rotational levels $J$ are noted at the bottom.
The dotted lines show the extrapolated straight line fit to the average $N_{J}$ for the lowest three levels; the corresponding excitation temperature, $T_{04}$ $\sim$ 74 K, is the negative of the inverse of the slope of that line.
The higher rotational levels are characterized by progressively higher excitation temperatures.}
\label{fig:c2ex}
\end{figure}

For most of the sight lines in this study, observations of the UV and/or optical C$_2$ bands, together with the corresponding adopted $f$-values, appear to yield fairly reliable and mutually consistent column densities $N_{J}$ for the various rotational levels.
We have therefore adopted the weighted average $N_{J}$ listed in Table~\ref{tab:nc2} for sight lines with data for multiple bands.
Figure~\ref{fig:c2ex} compares the average C$_2$ rotational populations toward HD~24534 with those derived from the individual UV ({\it upper panel}) and optical ({\it lower panel}) bands.
In general, there is good agreement; the main exceptions are the low values found for $J$ = 8 and 10 from the F-X (0-0) band.
[In estimating the average C$_2$ in levels $J$ = 8 and $J$ = 10 toward HD~24534, HD~27778, and HD~206267, we have doubled the contributions from the F-X (0-0) band, in view of that apparent (and unexplained) deficiency, compared to the values obtained for those levels in the other C$_2$ bands.]

While we have C$_2$ data for $J$ = 0--8 for most of the sight lines in our sample, examination of the C$_2$ rotational population distributions toward HD~24534, HD~27778, and HD~207198 (for $J$ = 0--18) and $\zeta$ Oph (for $J$ = 0--24; Lambert et al. 1995) --- and extrapolation of those distributions to even higher $J$ --- suggests that a non-negligible fraction of the total C$_2$ is in rotational levels higher than $J$ = 8.
Toward HD~24534 and $\zeta$ Oph, for example, roughly 20\% and 40\% of the total C$_2$, respectively, appear to be in rotational levels above $J$ = 8.
In order to estimate the total C$_2$ column density in all rotational levels for each line of sight, we have multiplied the observed total for $J$ = 0--6 ($N_{0-6}$, which we have for all sight lines) by a factor $N_{0-20}$/$N_{0-6}$ derived from the theoretical rotational distribution that best fits the observed $N_{J}$ for that line of sight (see below, \S~\ref{sec-c2tex}). 
Although the theoretical distributions are calculated only to $J$ = 20, $N_{0-20}$ is within 10\% of $N_{\rm tot}$ for all the sight lines in our sample; the factor $N_{0-20}$/$N_{0-6}$ ranges between about 1.1 and 2.3 for those sight lines.
The estimated total $N$(C$_2$) (listed in the last columns of Tables~\ref{tab:c2dx} and \ref{tab:nc2}) are generally somewhat smaller than the corresponding values given in Thorburn et al. (2003) --- due mostly to differences in the adopted A-X band $f$-values, but also to differences in the extrapolation to all $J$, in the estimation of saturation effects, and (in some cases) in the measured equivalent widths.

\subsubsection{Adopted Column Densities}
\label{sec-adopt}

Table~\ref{tab:ntot} summarizes the total column densities of H$_2$, $^{12}$CO, $^{13}$CO, and C$_2$ found for the ten lines of sight in this study, together with the resulting $^{12}$CO/H$_{2}$, $^{12}$CO/$^{13}$CO, and C$_2$/H$_2$ ratios. 
The H$_2$ column densities are from Rachford et al. (2002), except for the value toward HD~147888 (Cartledge et al. 2004). 
The column densities for CO and C$_2$ are from this paper, except for the $^{12}$CO and $^{13}$CO values toward HD~24534 (Sheffer et al. 2002a).
Because the STIS and {\it FUSE} data do not resolve the individual gas components seen in the higher resolution optical spectra, a component-by-component analysis is rendered uncertain (especially for H$_2$). 
The discussions below therefore consider only the total column densities --- which should in any case represent the primarily molecular material in each line of sight.  

\section{CO, C$_2$, and Physical Conditions in Translucent Sight lines}
\label{sec-phys}

The so-called ``translucent'' clouds are thought to be intermediate or transitional objects between diffuse, primarily atomic clouds and dense molecular clouds (e.g., van Dishoeck \& Black 1988, 1989).  
It has been difficult, however, to isolate and characterize specific examples of ``true'' translucent clouds using optical/UV absorption lines, even with the more sensitive spectroscopic capabilities provided by {\it FUSE} and {\it HST}.  
In a recent {\it FUSE} survey of interstellar H$_2$ in translucent sight lines, Rachford et al. (2002) obtained accurate measurements of the H$_{2}$ column densities in the $J=$ 0-1 rotational levels of the ground electronic and vibrational state for 23 lines of sight with $A_{\rm V}$ between 0.6 and 3.4 mag.  
While all the sight lines exhibit strong H$_2$ absorption, with $N$(H$_2$) $>$ 3 $\times$ 10$^{20}$ cm$^{-2}$, none is characterized by the very high molecular fractions $f$(H$_2$) $\ga$ 0.8 expected for ``true'' translucent clouds.
Rachford et al. concluded that most of the observed sight lines contained multiple components consisting of mixtures of diffuse and denser gas, with ``little evidence for the presence of individual translucent clouds''.
That assessment has been confirmed toward HD~185418 and HD~192639 via detailed studies of those sight lines (Sonnentrucker et al. 2002, 2003). 
Similarly, Snow, Rachford, \& Figoski (2002) did not find significantly enhanced depletion of iron (as would have been expected for translucent clouds), in essentially the same sample of sight lines (though it is possible that individual components with very severe depletions could be masked by other components with less severe depletions in the low-resolution {\it FUSE} spectra).
Nine of the ten stars considered here were included in the {\it FUSE} H$_2$ and \ion{Fe}{2} surveys. 

\subsection{Behavior of CO}
\label{sec-co}

Because of the abundance of CO and its importance as a tracer and diagnostic of molecular gas, a number of studies have been undertaken to understand and predict its behavior in diffuse, translucent, and denser clouds.  
Semi-analytic models (e.g., Federman \& Huntress 1989) identify the most significant channels for the formation and destruction of each molecular species and their dependences on the local physical conditions. 
Comparisons of the predictions of such simplified chemical models with observed column densities, for an ensemble of sight lines, can be used to constrain (sometimes poorly known) rate coefficients and other parameters (e.g., the cosmic ray ionization rate) of the models.  
More detailed numerical models (e.g., van Dishoeck \& Black 1988, 1989; Viala et al. 1988; Warin, Benayoun, \& Viala 1996; Le Petit et al. 2006) typically include more extensive reaction networks and radiative transfer effects, and attempt to determine the abundances of various species (including the individual rotational levels for H$_2$ and CO) as functions of depth within the model clouds.
All those detailed models, however, are for single, isolated plane-parallel clouds (often of uniform density).

In diffuse and translucent clouds with $A_{\rm V}$ $\la$ 3 mag, the primary route for the production of CO begins with the reaction C$^+$~+~OH~$\rightarrow$~CO$^+$~+~H.  
The CO$^+$ then reacts with H$_2$ to form HCO$^+$, which subsequently forms CO via dissociative recombination.  
In thicker, denser clouds, reactions involving O, CH, C$_2$, and other species may also contribute.  
The destruction of CO is via line absorption into predissociated bound states; the relevant transitions are at far-UV wavelengths between 912 and 1100 \AA\ (e.g., van Dishoeck \& Black 1988; Viala et al. 1988; Warin et al. 1996).  
At high enough column densities [$N$($^{12}$CO) $\ga$ 10$^{14}$ cm$^{-2}$], those transitions become saturated, self-shielding begins to reduce the photodissociation rate, and the column density of $^{12}$CO increases rapidly.  
Recent observations of some of the far-UV CO bands suggest that the self-shielding by those bands may be more effective than previously thought (Sheffer, Federman, \& Andersson 2003).
In addition, some of the $^{12}$CO transitions coincide with strong absorption lines of \ion{H}{1} and/or H$_2$, which can dominate the shielding of $^{12}$CO at lower column densities.  
The other isotopomers (e.g., $^{13}$CO, C$^{18}$O) are much less abundant, and so do not self shield (nor are they effectively shielded by either H$_2$ or $^{12}$CO).
At much higher column densities, reactions between CO and other species become the dominant destruction mechanism for CO.

In the sections below, the observed column densities of CO, H$_2$, and several other species will be compared with predictions from both the models of van Dishoeck \& Black (1988, 1989), Warin et al. (1996), and Le Petit et al. (2006).
The T, H, and I model sequences of van Dishoeck \& Black cover the range of extinctions and H$_2$ column densities relevant to most current observations of CO absorption ($A_{\rm V}$ $\la$ 2.5 mag). 
The T models are characterized by the average interstellar radiation field, total hydrogen densities $n_{\rm H}$ = $n$(\ion{H}{1}) + 2$n$(H$_2$) of 300--1000 cm$^{-3}$, temperatures of 15--60 K, H$_2$ column densities of 0.5--4.0 $\times$10$^{21}$ cm$^{-2}$, and visual extinctions of 0.7--5.2 mag.  
Several of the models allow for some variation in the temperature and density with depth.
The H models have half the average interstellar field, $n_{\rm H}$ = 500 cm$^{-3}$, $T$ = 40 K, $N$(H$_2$) = 0.3--3.0 $\times$ 10$^{21}$ cm$^{-2}$, and $A_{\rm V}$ = 0.4--3.8 mag.  
The I models have ten times the average interstellar field, $n_{\rm H}$ = 2000 cm$^{-3}$, $T$ = 30 K, $N$(H$_2$) = 0.5--10.0 $\times$ 10$^{21}$ cm$^{-2}$, and $A_{\rm V}$ = 0.8--12.8 mag.
All three series assume depletions of carbon, nitrogen, and oxygen of 0.1--0.4, 0.5, and 0.6, respectively (with respect to the solar or ``cosmic'' abundances adopted at that time); all three use the average Galactic extinction curve [though van Dishoeck \& Black (1989) also considered the effects of curves that are shallower (HD 147889) and steeper (HD 204827) in the UV].

Warin et al. (1996) computed models for diffuse clouds [$A_{\rm V}$ = 1.09 mag, $N$(H) = 2 $\times$ 10$^{21}$ cm$^{-2}$, $n_{\rm H}$ = 500 cm$^{-3}$], translucent clouds [$A_{\rm V}$ = 4.34 mag, $N$(H) = 8.1 $\times$ 10$^{21}$ cm$^{-2}$, $n_{\rm H}$ = 1000 cm$^{-3}$], and dark clouds [$A_{\rm V}$ = 10.86 mag, $N$(H) = 20 $\times$ 10$^{21}$ cm$^{-2}$, $n_{\rm H}$ = 10$^4$ cm$^{-3}$] --- both at constant $T$ and in thermal equilibrium.  
All the Warin et al. models are for uniform density slabs; all assume depletions of 0.1 for both carbon and oxygen.  
Unlike the van Dishoeck \& Black models, the Warin et al. models allow for differences in rotational excitation temperature for different $J$ --- resulting from the coupling between level populations, self-shielding, and photodissociation rates.
The total unshielded photodissociation rate adopted by Warin et al. (1996), based on more recent CO spectroscopic data, is of order 20\% smaller than that used by van Dishoeck \& Black (1988).

Le Petit et al. (2006) give an overview of the current Meudon photon-dominated region (PDR) code (e.g., Le Bourlot et al. 1993), which computes the radiative transfer, thermal balance, and chemistry (including the excitation of H$_2$ and CO) for a plane-parallel cloud in steady state.
In principle, both temperature and density can vary with depth in the cloud.
While Le Petit et al. do not give extensive numerical results, they do provide a coarsely sampled table of column densities of H$_2$, CO, and several other species (for $n_{\rm H}$ = 100 cm$^{-3}$, the standard interstellar radiation field, and total $A_{\rm V}$ ranging from 0.2 to 7.0) and several figures showing the column densities of CH, CN, and CO, as functions of $n_{\rm H}$/$\chi$, for a set of diffuse cloud models [$A_{\rm V}$ = 1, $n_{\rm H}$ ranging from 20 to 2000 cm$^{-3}$, and $\chi$ (the scaling factor for the radiation field) from 0.1 to 10].
All the models use the current observed gas-phase interstellar abundances of carbon, nitrogen, and oxygen.

The various models of cloud chemistry suggest that the column density of CO generally will increase with increasing local density, H$_2$ column density, and optical depth.  
For diffuse clouds, Federman et al. (1980) concluded that the local density of CO would be proportional to [$n$(H$_2$)]$^2$ (from formation) and to exp[$\tau$(OH)+$\tau$(CO)] (from destruction; the optical depths $\tau$ are for the wavelength ranges appropriate for photodissociation of OH and CO), but that the density dependence would be only linear with $n$(H$_2$) once reactions with other species dominate the destruction of CO.  
Isobaric models with mean densities of 150 and 2500 cm$^{-3}$ appeared able to account for most of the range in $N$(CO) observed at any given $N$(H$_2$).
(The role of line absorption and self-shielding in the dissociation of CO was not yet appreciated, however.)
For clouds with $A_{\rm V}$ $\la$ 2.5, the more detailed models of van Dishoeck \& Black (1988), which include self-shielding, predict that:
(1) increasing the gas-phase carbon abundance by a factor of four [for constant density and $N$(H$_2$)] increases the total CO column density by a factor of 1.5--3.0 (models T1--T4); 
(2) decreasing the radiation field by a factor of 2 [for constant depletion, density, and $N$(H$_2$)] increases $N$(CO) by a factor of 3--4 (models H2--H4 vs. models T1--T3); and 
(3) increasing $N$(H$_2$) by a factor of 2 (for constant depletion and radiation field) increases $N$(CO) by more than an order of magnitude (models T1--T4, H2--H5, and I1--I4). 
The Warin et al. (1996) translucent cloud model has twice the density, four times the H$_2$ column density, and roughly 3000 times the CO column density of the diffuse cloud model.  
The earlier models suggest that CO becomes the dominant carbon-containing species for clouds with $A_{\rm V}$ $\ga$ 3, with the transition between C$^+$ (and C$^0$) and CO typically occurring at optical depths $\tau_{\rm V}$ between 0.5 and 1.0 (van Dishoeck \& Black 1988; Warin et al. 1996) --- where the hydrogen is already almost completely in molecular form.  
The very high predicted molecular fractions $f$(H$_2$) $\ga$ 0.9 --- even for the total (integrated) column densities in the diffuse cloud models --- have not yet been observed, however (Rachford et al 2002). 
In the Meudon PDR models with $n_{\rm H}$ =100 cm$^{-3}$ and $\chi$ = 1, C$^+$ remains the dominant form of carbon even for $A_{\rm V}$ = 7; in the diffuse cloud models, $N$(CO) does not show a very strong dependence on density, for $n_{\rm H}$/$\chi$ between 10 and 1000 cm$^{-3}$.
Additional aspects of the behavior of CO (e.g., rotational excitation, isotopic abundances), as predicted by the various models, will be discussed further below (\S~\ref{sec-cotex} and \S~\ref{sec-1213}, respectively).

\subsection{Behavior of C$_2$}
\label{sec-c2}

While C$_2$ is not as abundant as CO, it does play a role in the networks of chemical reactions occurring in diffuse and translucent clouds, and analyses of its absorption lines can provide constraints on the density, temperature, and radiation field in those clouds.  
Federman \& Huntress (1989) have described the key reactions controlling the abundance of C$_2$; van Dishoeck \& Black (1982) have provided a detailed discussion of its excitation in diffuse and translucent clouds.  
A number of papers (e.g., van Dishoeck \& de Zeeuw 1984; Federman \& Lambert 1988; van Dishoeck \& Black 1989; Federman et al. 1994a) have discussed observations of C$_2$ absorption in the context of models for its abundance and/or excitation.  

In diffuse and translucent clouds, the formation of C$_2$ begins with the reaction C$^+$~+~CH~$\rightarrow$~C$_2^+$~+~H; subsequent hydrogen abstraction reactions and dissociative recombinations yield C$_2$ via several channels.  
Destruction of C$_2$ can occur by photodissociation (more important at relatively low optical depths) or by reactions with atomic oxygen (yielding CO) or nitrogen (yielding CN), which become dominant at higher $\tau$.  
The detailed models T1--T6 of van Dishoeck \& Black (1989), with total $A_{\rm V}$ = 0.7--5.2 mag, predict a roughly constant C$_2$/H$_2$ ratio $\sim$ 3.3--5.7 $\times$ 10$^{-8}$ over that range.  
The C$_2$ abundance is somewhat enhanced if the oxygen depletion is more severe than the value of 0.5 usually assumed in the models, is essentially unaffected by the assumed nitrogen depletion, and is somewhat affected by the shape of the far-UV extinction.  
As a ``second generation'' molecule dependent on the prior existence of H$_2$ and CH, C$_2$ is likely to be more concentrated in somewhat colder, denser regions.  
The models of van Dishoeck et al. (1991) suggest that the average density traced by C$_2$ may be either higher or lower than that traced by CO, depending on the structure of the cloud and how it is illuminated by the ambient radiation field.  
The rotational excitation of C$_2$ will be discussed below (\S~\ref{sec-c2tex}).

\subsection{Column Density Comparisons}
\label{sec-coldens}

In order to compare the column densities of $^{12}$CO and $^{13}$CO obtained in this paper with those found for other Galactic sight lines, we have collected CO column densities derived from UV absorption-line data from the literature.
Table~\ref{tab:coldens} (Appendix~\ref{sec-app}) lists total column densities for H$_2$, CH, C$_2$, CN, and CO for 74 sight lines for which CO has been measured (53 detections, 21 upper limits) --- a reasonably large sample which includes recently improved $N$(CO) values for both diffuse and more heavily reddened sight lines.  
For a number of those sight lines, column densities also have been recently reported for HD (Lacour et al. 2004) and for C$_3$ (Maier et al. 2001; Roueff et al. 2002; Galazutdinov et al. 2002; Oka et al. 2003; \'{A}d\'{a}mkovics, Blake, \& McCall 2003). 
Table~\ref{tab:dq} (Appendix) lists various derived quantities (e.g., excitation temperatures, column density ratios) for the sight lines with data for CO, H$_2$, CN, and CH.
There are now 19 sight lines with relatively well determined $N$(CO) $\ga$ 10$^{15}$ cm$^{-2}$ derived from UV absorption lines [plus a few others with values from mm-wave CO absorption (Liszt \& Lucas 1998, 2000)], compared to one such sight line ($\zeta$ Oph) in the early survey of Federman et al. (1980).  
This enlarged higher column density sample allows more meaningful comparisons with the predictions of theoretical models of translucent clouds.  
Comparisons among these molecular species (and of various ratios of these species) may provide some clues as to their behavior in different environments and to the physical conditions characterizing the clouds in which they reside (e.g., Federman et al. 1994a).  
While there are likely to be some differences in the spatial distribution of these species, they all trace the main components in each sight line, where the bulk of the molecular material is located.

\subsubsection{$N$(CO) and $N$(C$_2$) vs. $N$(H$_2$)}
\label{sec-coh2}

\begin{figure}
\epsscale{0.9}
\plotone{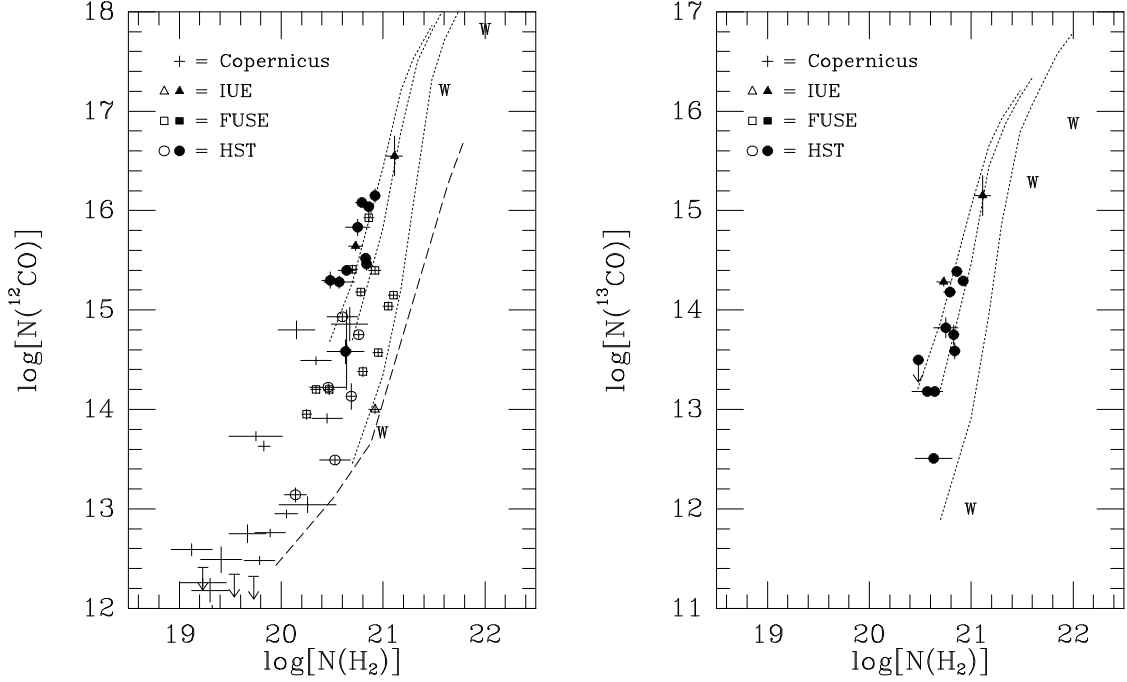}
\caption{$N$($^{12}$CO) ({\it left}) and $N$($^{13}$CO) ({\it right}) vs. $N$(H$_2$).
Circles denote CO data from {\it HST}, squares from {\it FUSE}, triangles from {\it IUE}, and crosses from {\it Copernicus}.
Filled symbols are for sight lines with measured $^{13}$CO.
The dotted lines show the predictions from models H1--H6, T1--T6, and I1--I8 of van Dishoeck \& Black (1988), for 0.5, 1.0, and 10 times the average Galactic radiation field (left to right).
Letters ``W'' show the results the for ``standard'' diffuse, translucent, and dark cloud models of Warin et al. (1996).
The dashed line (for $^{12}$CO only) shows the predictions from the Meudon PDR code (Le Petit et al. 2006) for $n_{\rm H}$ = 100~cm$^{-3}$ and the average Galactic field.}
\label{fig:comod}
\end{figure}

Figure~\ref{fig:comod} shows plots of the column densities of $^{12}$CO and $^{13}$CO versus $N$(H$_2$).
In the plots, the different symbols refer to the sources of the CO data ({\it Copernicus}, {\it IUE}, {\it FUSE}, or {\it HST}); the filled points represent sight lines for which $^{13}$CO has been detected.
The dotted lines in both figures indicate the relationships predicted by the models of van Dishoeck \& Black (1988); series H, T, and I (increasing radiation field; all with carbon depletion $\delta_{\rm C}$ = 0.4) are left-to-right in each case.
The letters ``W'' indicate the values predicted by the diffuse, translucent, and dense cloud models of Warin et al. (1996). 
The dashed lines show the predictions of the Meudon PDR models with $n_{\rm H}$ = 100 cm$^{-3}$, $\chi$ = 1, and $A_{\rm V}$ ranging from 0.2 to 7 (Le Petit et al. 2006). 
The observed column densities confirm the steep increase of $N$(CO) with $N$(H$_2$), for $N$(H$_2$) $\ga$ 3 $\times$ 10$^{20}$ cm$^{-2}$, as predicted by the various models and as noted by Federman et al. (1980).  
The much more extensive data for $N$($^{12}$CO) $\ga$ 10$^{15}$ cm$^{-2}$, however, appear to indicate a somewhat steeper increase for the higher column density sight lines than that predicted by Federman et al. --- presumably due to the effects of self-shielding.  
The models of van Dishoeck \& Black (1988) appear to reproduce the trends in the observed data reasonably well, but predict too little $^{12}$CO by a factor of order 2--3 [for a given $N$(H$_2$)].
For the sight lines with detected $^{13}$CO, the observed $N$($^{13}$CO) appear to agree somewhat better with the model predictions than do the observed $N$($^{12}$CO).  
The models of Warin et al. (1996) and Le Petit et al. (2006) predict much smaller CO column densities than the observed values.  
While the density in these models from Le Petit et al. is relatively low, the Meudon diffuse cloud models ($A_{\rm V}$ = 1) do not show an increase in $N$(CO) until $n_{\rm H}$ is greater than about 1500 cm$^{-3}$ (for $\chi$ = 1).
More extensive exploration of the parameter space available in the Meudon code (though beyond the scope of this paper) might prove instructive. 

As suggested by the three series of models (H, T, I), some of the scatter in the relationship between $N$(CO) and $N$(H$_2$) may be due to differences in the strength of the ambient radiation field. 
For sight lines with column density data for the lowest few rotational levels of H$_2$, the relative amount in $J$ = 4 (which is assumed to be populated primarily via photon pumping) may be used to estimate the strength of the radiation field (Jura 1975; Lee et al. 2002; Welty et al. 2006).
The four open squares just to the left of the high radiation field (I) model sequence correspond to four sight lines in Cep OB3 (Pan et al. 2005).
For three of those sight lines, the H$_2$ rotational populations determined by Pan et al. do suggest fairly strong ambient fields. 
The molecular material seen toward HD~37903 [open triangle near $N$(H$_2$) = 8 $\times$ 10$^{20}$ cm$^{-2}$ and $N$(CO) = 10$^{14}$ cm$^{-2}$)] also is likely subject to an enhanced field (Lee et al. 2002).
However, there are other sight lines with similarly high inferred fields whose CO column densities lie closer to the model sequences characterized by average or low fields.
In addition, Lacour et al. (2005b) obtained progressively higher $b$-values for the higher rotational levels of H$_2$ in several moderately reddened lines of sight (including HD~73882, HD~192639, and HD~206267) and argued that the column densities of those higher $J$ levels could be significantly overestimated under the usual assumption of constant $b$ for all $J$. 
If $N_4$(H$_2$) is too high, then the radiation field will be similarly overestimated.
While differences in the radiation field probably do affect the abundances of CO observed in different sight lines, other factors must also play a role. 

Increases in the gas-phase carbon and oxygen abundances and/or in the self-shielding of $^{12}$CO (Sheffer et al. 2003) in the models could yield somewhat better agreement between the observed $^{12}$CO column densities and those predicted by the models.  
While interstellar \ion{C}{2} column density data are still rather limited, there does appear to be a fairly well-defined average value C/H = 161$\pm$17 ppm for both diffuse and translucent sight lines (Sofia et al. 2004).
For a total (gas + dust) carbon abundance of 245 ppm (Lodders 2003), the carbon depletion $\delta_{\rm C}$ is thus about 0.66 [or a logarithmic depletion D(C) $\sim$ $-$0.18 dex].
Only the sight line toward HD~27778 shows evidence for a gas-phase carbon abundance significantly below that average value [$\le$ 108 ppm (3-$\sigma$)]\footnotemark; the carbon abundance toward X~Per (155$\pm$45 ppm), with higher $A_{\rm V}$ and $f$(H$_2$) and similar molecular abundances, is quite consistent with the average (Sofia, Fitzpatrick, \& Meyer 1998).
\footnotetext{While the CO column density toward HD~27778 estimated by Federman et al. (1994a) seemed suggestive of an unusually high CO abundance in that line of sight (Sofia et al. 2004), our value for $N$(CO) is a factor of 2 smaller.}
The contribution of CO to the overall gas-phase carbon abundance in those two sight lines is only $\sim$ 6 ppm.  
The Warin et al. (1996) models assumed a total carbon abundance of 360 ppm and $\delta_{\rm C}$ = 0.1.
The resulting gas-phase carbon abundance ($\sim$ 36 ppm) is a factor of $\sim$ 4.5 lower than the observed average interstellar value. 
Increasing the gas-phase carbon abundance would move the predicted $N$(CO) closer to the observed values, but would generally not be enough to yield good agreement.  
The van Dishoeck \& Black (1988, 1989) models assumed a total carbon abundance of 467 ppm and $\delta_{\rm C}$ = 0.1--0.4. 
For $\delta_{\rm C}$ = 0.4, the resulting gas-phase carbon abundance ($\sim$ 190 ppm) is slightly higher than the average interstellar value, so that other adjustments (e.g., increasing the $^{12}$CO self-shielding) would be required to obtain better agreement with the observed $N$(CO).

Several other suggested modifications to the (relatively) simple early models may be able to increase the CO abundances predicted for diffuse and translucent clouds.
(1) Both van Dishoeck \& Black (1989) and Warin et al. (1996) noted that one way would be to reduce the (relatively poorly known) intensity of the radiation field below 1200 \AA.  
The field would need to be weaker by an order of magnitude (or more), however, for the Warin et al. diffuse cloud model to reproduce the observed abundances of $^{12}$CO and $^{13}$CO (even with an increased carbon abundance).  
(2) Liszt \& Lucas (1994) concluded that the low predicted values of $N$(CO) toward $\zeta$ Oph should be ascribed to insufficient production of CO (rather than excessive photodissociation), in view of the similar underprediction of its immediate precursor HCO$^+$.
An increase in the rate of the initiating reaction C$^+$~+~OH~$\rightarrow$~CO$^+$~+~H (e.g., Dubernet, Gargaud, \& McCarroll 1992; Federman et al. 2003) would yield corresponding increases in both HCO$^+$ and CO.
(3) In view of observational evidence for complex, inhomogeneous structure in molecular clouds, Spaans (1996) explored the consequences of having higher density clumps embedded in a lower density interclump medium.
On the one hand, dissociating radiation may penetrate further into the interior of such a clumpy structure than is the case for the previously modeled homogeneous slabs.
Both theoretical considerations (\S~\ref{sec-co}) and empirical evidence (\S~\ref{sec-var}), however, suggest that CO will be enhanced at higher densities.
Spaans found that CO (which is located mostly in the denser clumps) could actually be enhanced by factors of 5--6 in the clumpy models, compared to the abundances found for the corresponding homogeneous models.
(4) Because the photodissociation of CO is dominated by line absorption, it can depend on details of the velocity distribution in the molecular gas.
Consideration of the effects of a turbulent velocity field has suggested that resulting changes to the CO absorption-line profiles could significantly increase the self-shielding of CO --- thus increasing the CO abundance in moderately dense molecular clouds (e.g., R\"{o}llig, Hegmann, \& Kegel 2002).
Much smaller effects are seen, however, when the full CO rotational structure is incorporated (instead of the single-line approximation used in the initial calculations).
Finally, however, if many sight lines contain a significant amount of H$_2$ without much associated CO (e.g., in more diffuse molecular gas), then the agreement with the models would not be as good as Figure~\ref{fig:comod} would otherwise seem to suggest.
Detailed studies of individual lines of sight (to separate the contributions from different molecular components) and comparisons with more sophisticated models (incorporating these additional physical effects) may yield better agreement between observed and predicted CO abundances.

\begin{figure}
\epsscale{0.45}
\plotone{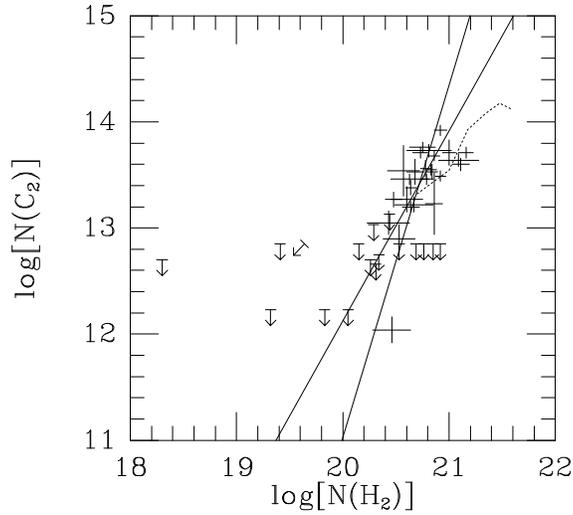}
\caption{$N$(C$_2$) vs. $N$(H$_2$).
The solid lines are linear least-squares fits (weighted and/or unweighted) to the data.
The dotted line shows the predictions from models T1--T6 of van Dishoeck \& Black (1988).}
\label{fig:c2mod}
\end{figure}

Figure~\ref{fig:c2mod} shows the relationship between the column densities of C$_2$ and H$_2$.
The solid lines are linear least-squares fits (weighted and unweighted) to the points representing detections of both quantities, taking into account that both quantities have associated uncertainties.\footnotemark\   
\footnotetext{A slightly modified version of the subroutine regrwt.f, obtained from the statistical software archive maintained at Penn State (http://www.astro.psu.edu/statcodes), was used to perform the fits; see Welty \& Hobbs (2001).}
Table~\ref{tab:corr} lists the linear correlation coefficients ($r$), the slopes of the best-fit lines, and the residual scatter about the best-fit lines found for this comparison (and for others discussed below).
The dotted line, which shows the roughly linear relationship predicted by models T1--T6 of van Dishoeck \& Black (1988, 1989), falls slightly below many of the observed data points.  
Much of the systematic difference between the models and the observed data may be ascribed to the depletion of carbon ($\delta_{\rm C}$ = 0.1--0.15) assumed in the models --- as the corresponding gas-phase carbon abundances are factors of 2--3 lower than the values observed for nearly all diffuse and translucent clouds.  
Apart from that systematic offset, the observed data seem to follow a somewhat steeper relationship (slope $\sim$ 1.8) than the roughly linear predicted trend (though the ranges in both column densities are rather limited); the much steeper slope found for the unweighted fit is due largely to the point for 40 Per.
While Federman \& Huntress (1989) suggested progressively more severe depletion of carbon (to a factor of 15 by $A_{\rm V}$ = 4 mag.) was required to explain the C$_2$ abundances in translucent sight lines, none of the sight lines in Fig.~\ref{fig:c2mod} has $A_{\rm V}$ higher than 2.7 mag.

\subsubsection{$N$(CO) vs. Column Densities of CN, CH, C$_2$, C$_3$, HD, and \ion{K}{1}}
\label{sec-molec}

\begin{figure}
\epsscale{0.9}
\plotone{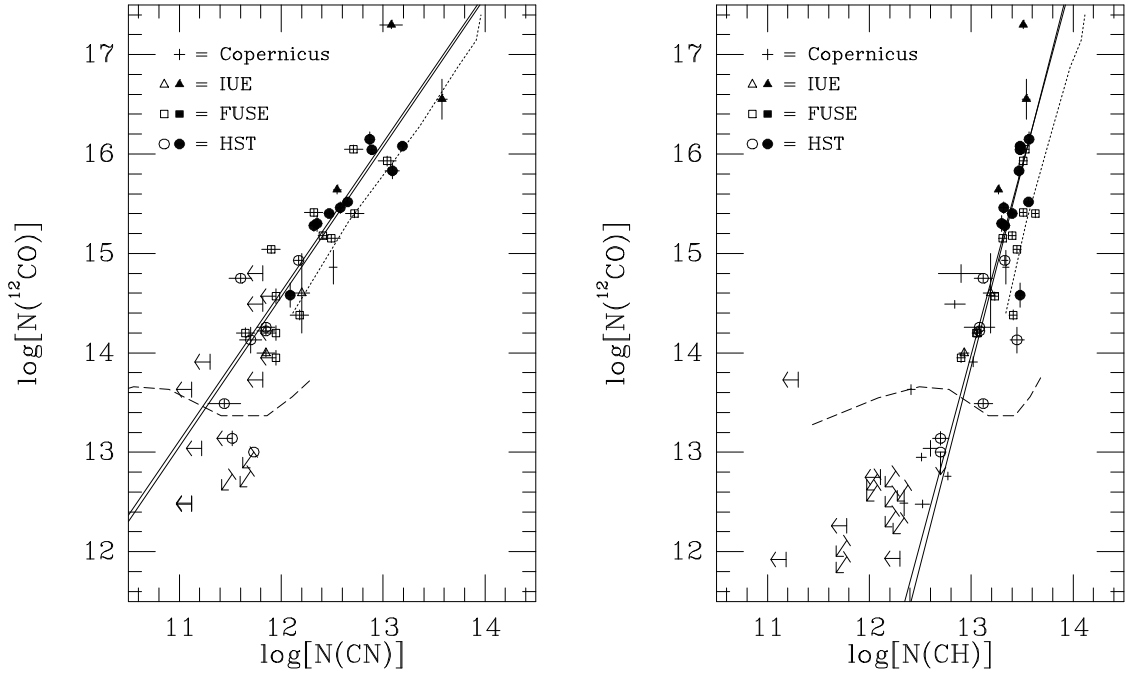}
\caption{$N$($^{12}$CO) vs. $N$(CN) ({\it left}) and vs. $N$(CH) ({\it right}).
Circles denote CO data from {\it HST}, squares from {\it FUSE}, triangles from {\it IUE}, and crosses from {\it Copernicus}.
Filled symbols are for sight lines with measured $^{13}$CO.
The solid lines are linear least-squares fits (weighted and/or unweighted) to the data.
The dotted lines are the predictions from models T1--T6 of van Dishoeck \& Black (1989).
The dashed lines show the predictions from the Meudon PDR code (Le Petit et al. 2006) for diffuse clouds [$A_{\rm V}$ = 1.0, $n_{\rm H}$ = 20--2000~cm$^{-3}$ (left to right), and the average Galactic field].}
\label{fig:cocnch}
\end{figure}

\begin{figure}
\epsscale{0.9}
\plotone{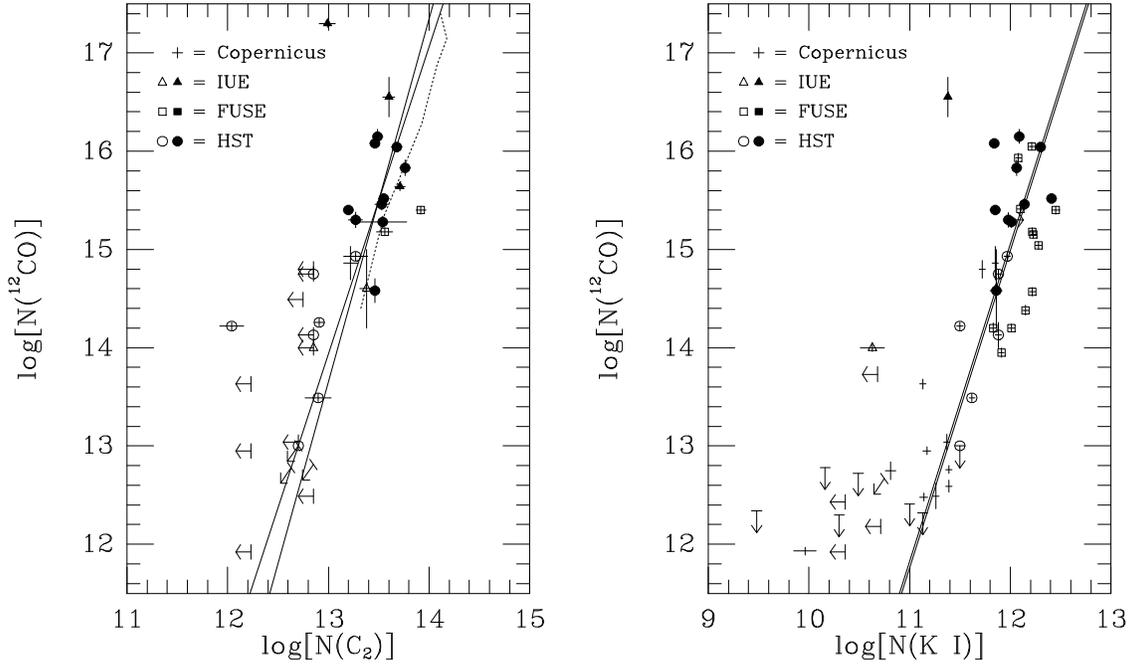}
\caption{$N$($^{12}$CO) vs. $N$(C$_2$) ({\it left}) and vs. $N$(K~I) ({\it right}).
Circles denote CO data from {\it HST}, squares from {\it FUSE}, triangles from {\it IUE}, and crosses from {\it Copernicus}.
Filled symbols are for sight lines with measured $^{13}$CO.
The solid lines are linear least-squares fits (weighted and/or unweighted) to the data.
The dotted line for C$_2$ shows the predictions of models T1--T6 of van Dishoeck \& Black (1989).}
\label{fig:coc2k1}
\end{figure}

Figures~\ref{fig:cocnch} and \ref{fig:coc2k1} show plots of the column density of $^{12}$CO versus the column densities of CN, CH, C$_2$, and \ion{K}{1}; similar comparisons were also made with the column densities of C$_3$ and HD.
Again, the different symbols denote the sources of the CO data, with the filled points representing sight lines for which $^{13}$CO has been detected, and the solid lines are the linear least-squares fits to the data.
For the full sample, there are clear, significant correlations ($r$ from 0.733 to 0.905) with CN, CH, HD, and \ion{K}{1} (Table~\ref{tab:corr}). 
The smaller correlation coefficients for C$_2$ and C$_3$ appear to be due to several sight lines (40 Per, HD~200775) with very high CO/C$_2$ ratios and to the small range in column density for the few sight lines with detections of C$_3$, respectively.  
The strongest correlations are with CN ($r$ = 0.900; slope $\sim$ 1.50) and with HD ($r$ = 0.905; slope $\sim$ 1.75); the relationship with CN shows significantly smaller scatter, however.  
The slopes of the relationships with CH, C$_2$, and \ion{K}{1} are much steeper (between about 3.1 and 4.0), though consideration of the upper limits for some sight lines suggests that the relationships may be somewhat shallower at lower column densities.  
In view of the scatter in those relationships, it would be difficult to accurately predict $N$(CO) from the column densities of CH, C$_2$, or \ion{K}{1} for any individual sight line.

The dotted lines in Figures~\ref{fig:cocnch} and \ref{fig:coc2k1} give the relationships between the column density of $^{12}$CO and those of CN, CH, and C$_2$ predicted by models T1--T6 of van Dishoeck \& Black (1988, 1989).  
In all three cases, the average slopes of the predicted relationships are very similar to those determined from the fits to the observed data, but the predicted $N$(CO) are lower than the observed (fitted) values by factors of 2--4.  
To first order, the low values assumed for the carbon abundance in the models presumably should not significantly affect the predicted ratios of CO to CN or CH.  
The gas-phase abundances assumed in the van Dishoeck \& Black (1988, 1989) models for oxygen and nitrogen are closer to the current observed values (Meyer, Cardelli, \& Sofia 1997; Cartledge et al. 2004).
The dashed lines in Figure~\ref{fig:cocnch} show the predictions of the Meudon diffuse cloud models (Le Petit et al. 2006), for $\chi$ = 1 and $n_{\rm H}$ ranging from 20 to 2000 cm$^{-3}$.
The predicted $N$(CO) would be consistent with the observed trends for densities of about 200--500 cm$^{-3}$ --- though there are sight lines with $A_{\rm V}$ $\la$ 1 and $N$(CO) much higher than those predicted values.

As was found from fitting the observed CO absorption-line profiles, these column density comparisons suggest that the distribution of CO is most similar to that of CN.  
The correlation coefficient with CN is larger than those for CH, C$_2$, C$_3$, and \ion{K}{1}, and the slope of the relationship with CN is closer to unity.
Weak components in \ion{K}{1} and CH, for example, are probably much weaker in CO, though they may still be detectable in the strongest CO bands (e.g., Sheffer et al. 2003; see also the profiles for HD~206267 in Fig.~\ref{fig:stis2}).
While the abundances of both CO and CN appear to increase strongly with overall column density, however, the reasons for the increase may be different for the two species.  
Because the production of CN depends on the prior existence of C$_2$, CH, and/or NH, the ``third-generation'' molecule CN is expected to trace colder, denser, more shielded regions where those precursor molecules have already become abundant (Federman, Danks, \& Lambert 1984; Federman et al. 1994).
On the other hand, the strong increase in $N$($^{12}$CO) is likely to be due both to increased formation at higher densities and to reduced photodissociation resulting from the onset of self-shielding, once sufficient $^{12}$CO is present [i.e., $N$($^{12}$CO) $\ga$ 10$^{14}$ cm$^{-2}$].

\subsection{Variation of the CO/H$_{2}$ Ratio}
\label{sec-var}

Both theoretical modeling and observations of the CO abundance indicate that the CO/H$_{2}$ column density ratio can vary by three orders of magnitude, from $\la$ 10$^{-7}$ in diffuse clouds to about 10$^{-4}$ in dense molecular clouds (van Dishoeck \& Black 1988; van Dishoeck \& Black 1990; Warin et al. 1996); the transition between those values occurs in the translucent regime. 
The analogous ratio of $N$(H$_2$) to CO emission strength [the so-called ``conversion factor'' X$_{\rm CO}$ = $N$(H$_2$)/I($^{12}$CO), of order 2 $\times$ 10$^{20}$ cm$^{-2}$(K~\kms)$^{-1}$ in our Galaxy] is often used to derive the H$_2$ column density in dense molecular clouds, since H$_2$ is often difficult to measure directly in such regions (Lacy et al. 1994).
That conversion factor reflects the relationship between the width of the (saturated) $^{12}$CO $J$ = 1--0 emission line and the mass of a gravitationally bound dense cloud.
A unique conversion factor cannot be defined, however, in diffuse and translucent clouds (e.g., Magnani \& Onello 1995) --- which are not gravitationally bound and where $N$(CO) is not high enough for the emission lines to be saturated. 

\begin{figure}
\epsscale{0.9}
\plotone{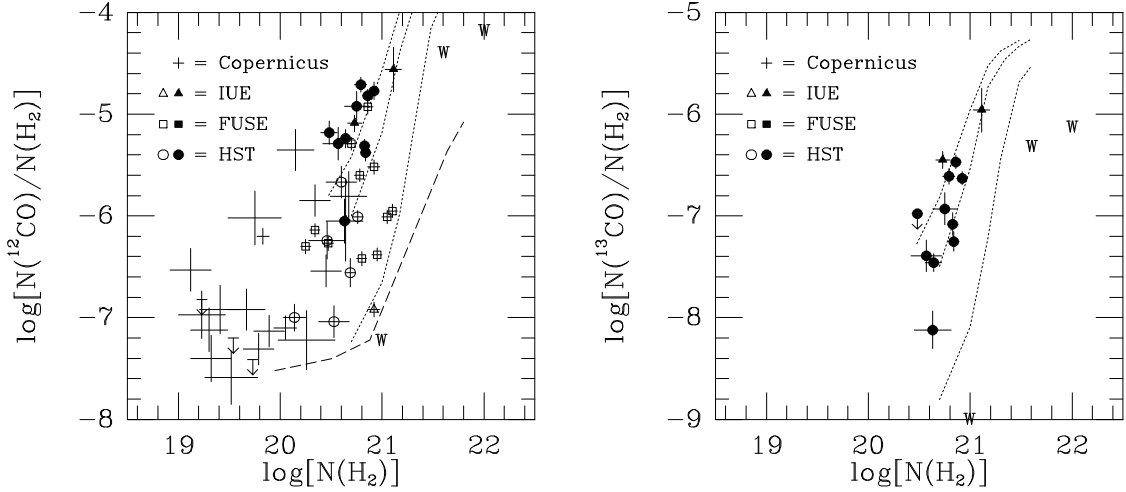}
\caption{$N$($^{12}$CO)/$N$(H$_2$) ({\it left}) and $N$($^{13}$CO)/$N$(H$_2$) ({\it right}) vs. $N$(H$_2$).
Circles denote CO data from {\it HST}, squares from {\it FUSE}, triangles from {\it IUE}, and crosses from {\it Copernicus}.
Filled symbols are for sight lines with measured $^{13}$CO.
Dotted lines show predictions from models H1--H6, T1--T6, and I1--I8 of van Dishoeck \& Black (1988), for 0.5, 1.0, and 10 times the average Galactic radiation field (left to right).
Letters ``W'' show the results for the ``standard'' diffuse, translucent, and dark cloud models of Warin et al. (1996).
The dashed line (for $^{12}$CO only) shows the predictions from the Meudon PDR code (Le Petit et al. 2006) for $n_{\rm H}$ = 100~cm$^{-3}$ and the average Galactic field.}
\label{fig:coh2mod}
\end{figure}

Figure~\ref{fig:coh2mod} shows the same data as in Figure~\ref{fig:comod}, but plotted as $N$(CO)/$N$(H$_2$) versus $N$(H$_2$).  
Again, the dotted lines show the relationships predicted by models H, T, and I of van Dishoeck \& Black (1988), the letters ``W'' show the values predicted by the models of Warin et al. (1996), and the dashed line (for $^{12}$CO) shows the predictions of the Meudon code (Le Petit et al. 2006).
In our sample of ten sight lines, the $^{12}$CO/H$_{2}$ ratio varies by a factor of about 100 --- from 2.8 $\times$ 10$^{-7}$ toward HD~192639 to 2.8 $\times$ 10$^{-5}$ toward HD~73882 (Table~\ref{tab:ntot}); values as low as 2.5 $\times$ 10$^{-8}$ are seen in the larger sample. 
For our sight lines, the variations in the overall column densities of both H$_2$ and total hydrogen (\ion{H}{1} + 2H$_2$) are much smaller; for example, seven of the ten have $N$(H$_2$) in the range 4.9--7.2 $\times$ 10$^{20}$ cm$^{-2}$ (Table~\ref{tab:ntot}). 
The differences in the CO/H$_2$ ratios in our sight lines thus are due primarily to variations in the CO column densities, which in turn depend on the physical conditions in the molecular gas in those particular sight lines. 
Since the total hydrogen content is similar for those sight lines, variations in the shielding of the CO bands by H and H$_2$, lines cannot explain the differences in the CO column density. 
It seems most likely that the range in $N$(CO) observed for our ten sight lines is due to differences in the CO photodissociation rate (ambient radiation field, self-shielding) and/or in the CO formation rate (local hydrogen density). 

In all ten of our sight lines, CO contains only a minor fraction of the total carbon abundance.
For the current solar carbon abundance C/H $\sim$ 2.5 $\times$ 10$^{-4}$ (e.g., Asplund, Grevesse, \& Sauval 2005) and typical interstellar depletions ($\delta_{\rm C}$ $\sim$ 0.66), the total gas-phase ratio C/H$_2$ would be about 3.2 $\times$ 10$^{-4}$ (if hydrogen were fully molecular).
The highest values of the CO/H$_2$ ratio in our sample --- $\sim$ 2--3 $\times$ 10$^{-5}$ toward HD 27778 and HD 73882 --- are more than an order of magnitude smaller.  
Most of the gas-phase carbon is still in the form of C$^+$ (Sofia et al. 2004; see also Liszt \& Lucas 1998).

In order to explore observationally the relationship between the CO/H$_2$ ratios and local physical conditions (e.g., $T$, $n_{\rm H}$, radiation field), empirical diagnostics of those parameters are needed.
The simplified chemical model of Federman \& Huntress (1989) implies, for example, that the CN/CH ratio in diffuse molecular gas should be roughly proportional to the square of the local hydrogen density ($n_{H}^2$) (Cardelli, Federman, \& Smith 1991) --- suggesting that the CN/CH column density ratio may be used as a density indicator.  
The more detailed models of van Dishoeck \& Black (1986a, 1989), however, predict a somewhat weaker increase of $N$(CN)/$N$(CH) with $n_{\rm H}$.  
While the densities estimated by Pan et al. (2005) for individual components toward stars in Cep OB2 suggest that the CN/CH ratio may be roughly proportional to $n^{1/2}_{\rm H}$ (with much scatter), the densities inferred below from C$_2$ (\S~\ref{sec-c2tex}) appear to indicate a somewhat steeper relationship.
The CN/CH ratio thus does appear to be correlated with density, but other factors (e.g., $A_{\rm V}$ and the steepness of the extinction curve in the far-UV) may contribute to the scatter (see also Cardelli 1988; Welty \& Fowler 1992).
Some CH may also be formed together with CH$^+$ in lower density gas (Federman, Welty, \& Cardelli 1997a; Zsarg\'{o} \& Federman 2003), but such contributions generally appear to be fairly minor for sight lines in which CN has been detected.

\begin{figure}
\epsscale{0.9}
\plotone{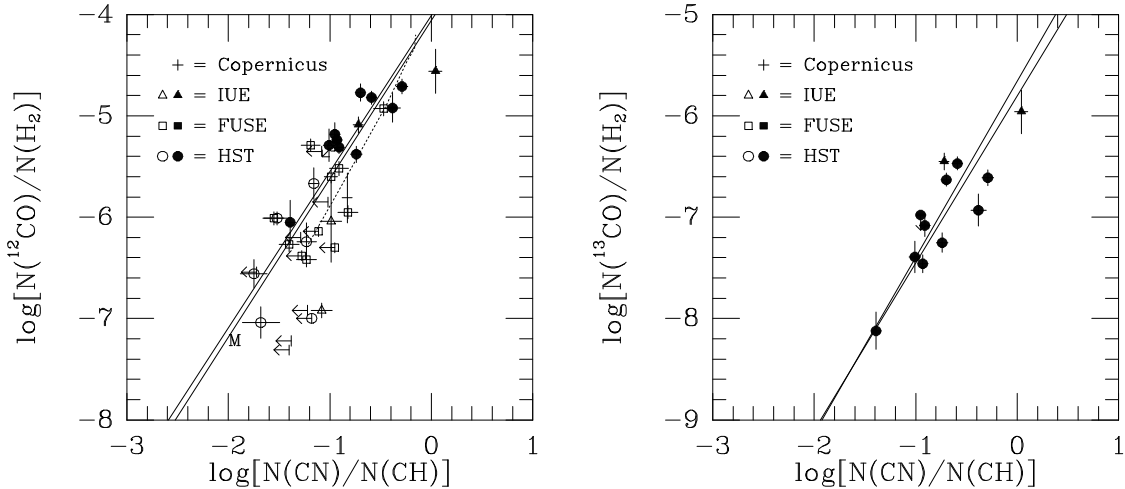}
\caption{Column density ratios $^{12}$CO/H$_2$ ({\it left}) and $^{13}$CO/H$_2$ ({\it right}) vs. density indicator CN/CH. 
The CO data are from {\it HST} (circles), {\it FUSE} (squares), {\it IUE} (triangles), and {\it Copernicus} (crosses); filled symbols are for sight lines with measured $^{13}$CO.
Solid lines are linear least-squares fits (weighted and/or unweighted) to the observed data.
Dotted line for $^{12}$CO/H$_2$ shows the predictions from models T1--T6 of van Dishoeck \& Black (1989).
The letter ``M'' (for $^{12}$CO only) shows the prediction from the Meudon PDR code (Le Petit et al. 2006) for a diffuse cloud with $A_{\rm V}$ = 1.0, $n_{\rm H}$ = 100~cm$^{-3}$, and the average Galactic field.}
\label{fig:coh2}
\end{figure}

Figure~\ref{fig:coh2} shows the observed ratios $N$($^{12}$CO)/$N$(H$_2$) and $N$($^{13}$CO)/$N$(H$_2$) versus $N$(CN)/$N$(CH) for sight lines with data for those species (Appendix~\ref{sec-app}; Table~\ref{tab:dq}).  
There are clear correlations in both cases ($r$ = 0.815 and 0.868); the solid lines show the linear least-squares fits to the data (with slopes of order 1.55--1.75; Table~\ref{tab:corr}).  
The dotted line for $^{12}$CO/H$_2$, showing the relationship predicted by models T1--T6 of van Dishoeck \& Black (1989) [for $\delta_{\rm C}$ = 0.15, $\delta_{\rm N}$ = 0.6, $\delta_{\rm O}$ = 0.5, and grain model 2], is very close to the best fit to the observed data.  
The letter ``M'', corresponding to the values predicted by the Meudon code for $A_{\rm V}$ = 1, $\chi$ = 1, and $n_{\rm H}$ = 100 cm$^{-3}$, also lies very close to the observed trend.
These correlations are perhaps not surprising --- in view of the good correlation between $N$($^{12}$CO) and $N$(CN) and the known roughly linear relationship between $N$(H$_2$) and $N$(CH) (e.g., Danks, Federman, \& Lambert 1984).  
Nonetheless, the correlations do suggest that differences in the local density $n_{\rm H}$ [as indicated by $N$(CN)/$N$(CH)] may play a significant role in explaining the variations in the CO/H$_2$ ratios observed in diffuse and translucent molecular gas.  
As the dispersions relative to the best fits to the data ($\sim$ 0.2 dex) are larger than the typical measurement uncertainties, it is likely that other factors (e.g., radiation field, UV extinction, cloud structure; see \S~\ref{sec-coh2}) also affect the CO abundances, however.
In contrast, the ratio $N$(C$_2$)/$N$(H$_2$) does not show any significant dependence on $N$(CN)/$N$(CH).

The column density ratios observed for our ten sight lines are quite consistent with the trends observed for the larger sample.  
The largest ratios of both CO/H$_2$ and CN/CH are observed toward HD~73882 and HD~27778 (Table~\ref{tab:dq}); the smallest ratios are found toward HD~185418 and HD~192639 (where CN is weak).  
For the three targets in the Cep OB2 association, the CN/CH ratio toward HD~206267 is higher than those toward HD~207198 and HD~210839 by factors of 1.4--2.1, while the CO/H$_2$ ratio is about 2.6--3.0 times larger.  
Simple chemical modeling of the CH and CN absorption by Pan et al. (2005) yields densities of 525--1150 cm$^{-3}$ toward HD~206267, 225--425 cm$^{-3}$ toward HD~210839, and 150 cm$^{-3}$ toward HD~207198 for the main components seen in CN.
Analysis of the C$_2$ rotational excitation (see below) implies slightly higher densities toward HD~206267 than toward HD~207198.
For those three Cep OB2 sight lines, the CN/CH and CO/H$_2$ ratios thus both appear to increase with density.

\subsection{Rotational Excitation of CO and C$_2$}
\label{sec-tex}

Because the populations of the ground and excited rotational levels of molecules such as H$_2$, CO, and C$_2$ can depend on both collisional and radiative processes, comparison of the observed rotational excitation of those molecules with predictions from theoretical models can provide information on the density, temperature, and radiation environment characterizing the molecular gas.
In general, measurements of both lower and higher rotational states are needed to disentangle the collisional and radiative effects.  
The relative rotational level populations are often expressed in terms of rotational excitation temperatures 
$T_{ij}$ = [($E_j$/$k$)~$-$~($E_i$/$k$)]~/~[ln($N_i$/$g_i$)~-~ln($N_j$/$g_j$)] (usually for adjacent levels $i$ $<$ $j$), where $E_i$, $N_i$, and $g_i$ are the energy, column density, and statistical weight of level $i$. 
Because H$_2$, CO, and C$_2$ each respond somewhat differently to the local physical conditions and may well trace somewhat different volumes of gas, determinations of their respective rotational excitations can provide both complementary constraints on those conditions and some information on the structure of the clouds.  
Tables~\ref{tab:texco} and \ref{tab:texc2} list various excitation temperatures derived from the rotational level populations of CO, H$_2$, and C$_2$, which will be used to constrain the physical conditions in the molecular gas. 
  
\subsubsection{CO Excitation}
\label{sec-cotex}

Because of the abundance and ubiquity of CO in interstellar space, observations of its emission lines have often been used to estimate the physical conditions in dense molecular gas ($n_{\rm H}$ $\ga$ 10$^{4}$ cm$^{-3}$, $T$ $\sim$ 10 K). 
Although CO is a fairly good density and temperature diagnostic in such dense clouds, where collisional processes dominate, the use of CO as a diagnostic of the physical conditions is more problematic in diffuse and translucent clouds still permeated by UV photons (Wannier et al. 1997).
While the detailed models of van Dishoeck \& Black (1988) incorporated both chemical reactions and various radiative and collisional processes in order to predict both the total CO abundance and its rotational excitation, the excitation temperature was assumed to be the same for all rotational levels. 
Warin et al. (1996) removed that constraint --- and were thus able to study the variations in the excitation of the various CO rotational levels for ''typical'' diffuse, translucent, and dense clouds. 

For diffuse clouds, the Warin et al. (1996) models predict that the photodissociation rate remains both large throughout the cloud and globally independent of the $J$ level. 
While CO in the excited rotational levels is easily photodissociated by UV photons, the relatively low gas densities do not allow efficient collisional population of those excited levels (achieved primarily via collisions with H$_2$ but also with H). 
The rotational population distribution of $^{12}$CO (and its isotopes) is therefore sub-thermal --- i.e., the excitation temperatures $T_{ij}$ of the various excited rotational levels ($\la$ 10 K for $J$ = 1--4) are well below the assumed temperature of the gas (50 K) --- even though the energies of those levels are less than or comparable to the thermal energy ($kT$) at that temperature (Table~\ref{tab:conj}). 
The models predict that the relative population of the $J=$ 1 level is $\sim$ 10--50\% higher than that of the $J=$ 0 level and that the populations of the higher levels decrease by nearly a factor 10 for each level above $J=$ 1. 
For the typical column densities measured in diffuse clouds (and the sensitivity of the instrumentation available to this point), the $J=$ 0, 1, and (perhaps) 2 rotational levels of $^{12}$CO should be detectable. 
Corresponding lines from the much less abundant isotopomers are expected to be weak in diffuse clouds.

For translucent clouds, the Warin et al. (1996) models predict a strong dependence of the photodissociation rate on the rotational level, as self-shielding becomes effective for the lowest-lying levels of $^{12}$CO. 
Because of this differential photodissociation effect, the excitation temperatures are predicted to vary both with rotational level and with depth in the cloud. 
Thermalization is still not achieved, however, and all excitation temperatures remain sub-thermal throughout the cloud. 
For example, the isothermal model predicts that the relative populations in $J$ = 0/1/2/3 at optical depth (due to dust) $\tau_{\rm V}$ = 0.5 are $\sim$ 1.0/2.0/0.5/0.07, while the relative populations at the cloud core ($\tau_{\rm V}$ = 2.0) are $\sim$ 1.0/2.3/2.3/1.5. 
The corresponding predicted excitation temperatures $T_{ij}$ for $J$ = 1--3 are 6--12 K at $\tau_{\rm V}$ = 0.5 and 19--21 K at the cloud core (compared to the assumed constant temperature $T$ = 25 K).
The $J$ = 0 and 1 levels of the isotopomers $^{13}$CO and C$^{18}$O also should be detectable, with the $^{13}$CO excitation temperature $T_{01}$ $\sim$ 9 K throughout the cloud.

\begin{figure}
\epsscale{0.45}
\plotone{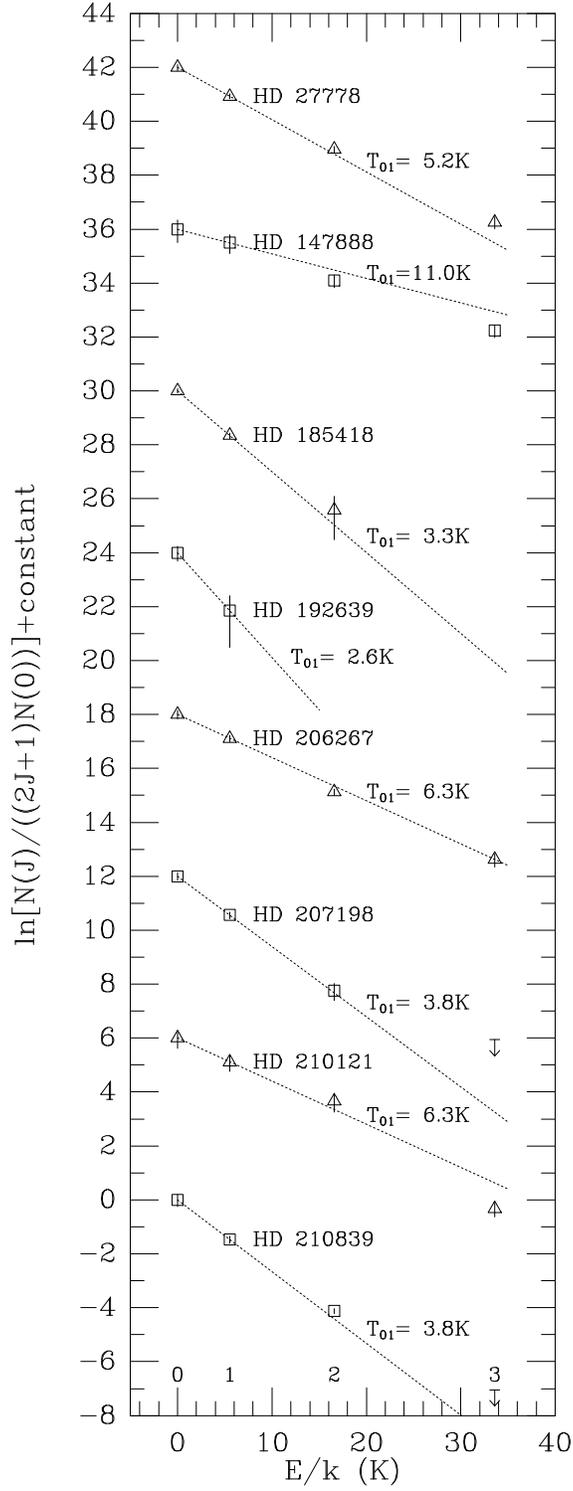}
\caption{Rotational excitation of $^{12}$CO toward eight stars.
Open squares show the normalized $^{12}$CO rotational level column densities, for $J$ = 0--3.
Dotted lines show the extrapolated straight line fits to $J$ = 0--1.
In each case, the excitation temperature $T_{01}$ is the negative of the inverse of the slope of that line.}
\label{fig:corot}
\end{figure}

Figure~\ref{fig:corot} shows the $^{12}$CO normalized excitation diagrams (ln[$N_{J}$/((2$J$+1)~$N_0$)] vs. $E_{J}$/k) for the eight sight lines whose rotational level column densities were derived from our fits to high-resolution UV spectra obtained with GHRS or STIS (Table~\ref{tab:conj}). 
The dotted lines show the extrapolated straight-line fit to the points for $J$ = 0 and 1, the slope of which is the negative of the inverse of the excitation temperature $T_{01}$.
Table~\ref{tab:texco} lists the corresponding $^{12}$CO (and $^{13}$CO) excitation temperatures for those eight sight lines, together with the CO excitation temperatures reported in the literature for eight other sight lines and the average values derived from the CO rotational populations in the diffuse and translucent cloud models listed in Table~2 of Warin et al. (1996).  
For $^{12}$CO, the previously reported $T_{01}$ are all between 2.7 and 3.6 K.  
The $T_{01}$ derived from our profile fits range from $\sim$ 2.6 K toward HD~192639 and 3.3 K toward HD~185418 to $\sim$ 11 K toward HD~147888, with the values for the other five sight lines all between 3.8 and 6.3 K; similar values (generally between 3 and 8 K) have been derived from mm-wave CO absorption observations toward extragalactic radio continuum sources (Liszt \& Lucas 1998).  
The $^{12}$CO $T_{01}$ values for all 16 observed sight lines are significantly lower than those predicted by the Warin et al. (1996) translucent cloud models --- which is perhaps not surprising, since none of the sight lines has $A_{\rm V}$ or $N$(H$_2$) as high as the values assumed for the models.  
A number of the values determined in this paper are similar to those predicted for the thermal equilibrium diffuse cloud models.
Even for those models, however, the predicted dependence of excitation temperature on rotational level ($T_{12}$ $<$ $T_{01}$ $\la$ $T_{23}$) is generally not seen for the eleven sight lines with data for $J$ $>$ 1, many of which have $T_{01}$ $\la$ $T_{12}$ $\la$ $T_{23}$.  
For most of the sight lines in our sample, those three observed excitation temperatures are equal, within the uncertainties.

Wannier et al. (1997) pointed out that CO line emission from dense molecular gas adjacent or near to that seen in absorption could in principle increase the populations of levels $J$ $>$ 0 above the values due to collisional and radiative processes in the observed gas alone.  
Analyses of the CO rotational excitation which neglect that emission would thus overestimate the density in the gas seen in absorption.  
For most of the sight lines in the upper part of Table~\ref{tab:texco}, Wannier et al. concluded that the proposed radiative excitation could account entirely for the observed rotational excitation --- which thus could not be used to infer $n_{\rm H}$ in the gas.  
One consequence (and test) of the proposed radiative excitation is that the excitation temperatures for $^{13}$CO should be lower than the corresponding values for $^{12}$CO, as the $^{13}$CO emission lines should be much weaker.  
That does not appear to be the case, however, for the seven sight lines in Table~\ref{tab:texco} with values for $T_{01}$($^{13}$CO). 
In six of the sight lines, $T_{01}$($^{13}$CO) is nominally greater than $T_{01}$($^{12}$CO); in all cases, the two temperatures are equal, within the mutual uncertainties.   
And while two of the six components detected in both $^{12}$CO and $^{13}$CO absorption by Liszt \& Lucas (1998) have $T_{01}$($^{12}$CO) $>$ $T_{01}$($^{13}$CO), the other four components have $T_{01}$($^{12}$CO) $\sim$ $T_{01}$($^{13}$CO).
The radiative excitation mechanism proposed by Wannier et al. thus may not be significant, in many cases.

It is not clear whether the higher values of $T_{ij}$ ($\sim$ 8--11 K) for $^{12}$CO observed toward HD~147888 can be associated with a higher density in the molecular gas in that sight line.
While a relatively high density could be suggested by the observed relatively strong absorption from the excited fine-structure states of \ion{C}{1} and \ion{O}{1}, the density inferred from the C$_2$ rotational excitation (\S~\ref{sec-c2tex}) is similar to the values found for many other sight lines.   
Unfortunately, $^{13}$CO was not detected toward HD~147888.

\subsubsection{C$_2$ Excitation}
\label{sec-c2tex}

In diffuse and translucent clouds, most C$_2$ is in various rotational levels of the ground vibrational state of the ground $X ^1\Sigma^+_g$ electronic state.  
Levels in upper electronic states are populated by absorption of either near-IR photons [to $A~^1\Pi_u$ (Phillips system)] or (less frequently) UV photons [to $D~^1\Sigma^+_u$ (Mulliken system) or $F~^1\Pi_u$].
Quadrupole and intersystem (singlet-triplet) transitions then return the molecule to rotational levels in the ground vibrational state, whose populations are also affected by collisions with both H and H$_2$ (van Dishoeck \& Black 1982; van Dishoeck \& de Zeeuw 1984).   

Because both collisional and radiative processes contribute to the population of the C$_2$ rotational levels in diffuse and translucent clouds, care must be taken when deriving the gas physical conditions from observations of C$_2$.  
The effect of the radiative excitation (and subsequent cascades back to the ground electronic state) is to increase the populations of the excited rotational levels above their thermal equilibrium values --- thus yielding excitation temperatures for those levels that are greater than the actual kinetic temperature. 
(While the energies of the excited C$_2$ rotational levels are similar to those of CO {\it for the same $J$}, the relative populations in those excited levels are much higher for C$_2$, primarily because C$_2$ has no dipole moment.) 
Because the radiation-induced departures from thermal equilibrium increase with $J$, it is important (1) to get accurate measures of the lowest rotational levels (which yield the best estimates for the kinetic temperature) and (2) to measure as many of the higher $J$ levels (typically $J$ $\ga$ 8) as possible (to determine the extent to which radiative processes affect the rotational population distribution and to get tighter constraints on the density). 

There are several ways of using the observed lower $J$ populations of C$_2$ to characterize the kinetic temperature $T_{\rm k}$.
In principle, the best estimate for $T_{\rm k}$ comes from $T_{02}$.
Unfortunately, however, $T_{02}$ can be somewhat uncertain, as $N_0$ is determined from the single, generally relatively weak R(0) line in each of the observed C$_2$ bands.
For the sight lines analyzed in this study, the relatively well determined $T_{02}$ range from about 20 K (HD~207198) to about 40 K (HD~206267, HD~210121, $\zeta$~Oph); the nominal values for HD~24534, HD~27778, and HD~147888 are all somewhat larger (but are also very uncertain).
Within the uncertainties in the $N_{J}$, it is often possible to fit a single excitation temperature to the lowest 3--4 levels (i.e., to $J$ = 0--4 or $J$ = 0--6), generally at a value somewhat larger than $T_{02}$ (e.g., for HD~24534 in Fig.~\ref{fig:c2ex}).  
Toward HD~206267, for example, the excitation temperatures for the lowest two, three, and four levels are $T_{02}$ = 36$\pm$9 K, $T_{04}$ = 51$\pm$5 K, and $T_{06}$ = 53$\pm$3 K, respectively.  
The populations of the higher $J$ levels, however, generally are increasingly higher than would be expected from thermal equilibrium at a local kinetic temperature of $T_{02}$ --- especially for $J$ $\ga$ 8.
As discussed above, such elevated populations of the higher $J$ levels can be produced by radiative excitation.  

\begin{figure}
\epsscale{0.9}
\plotone{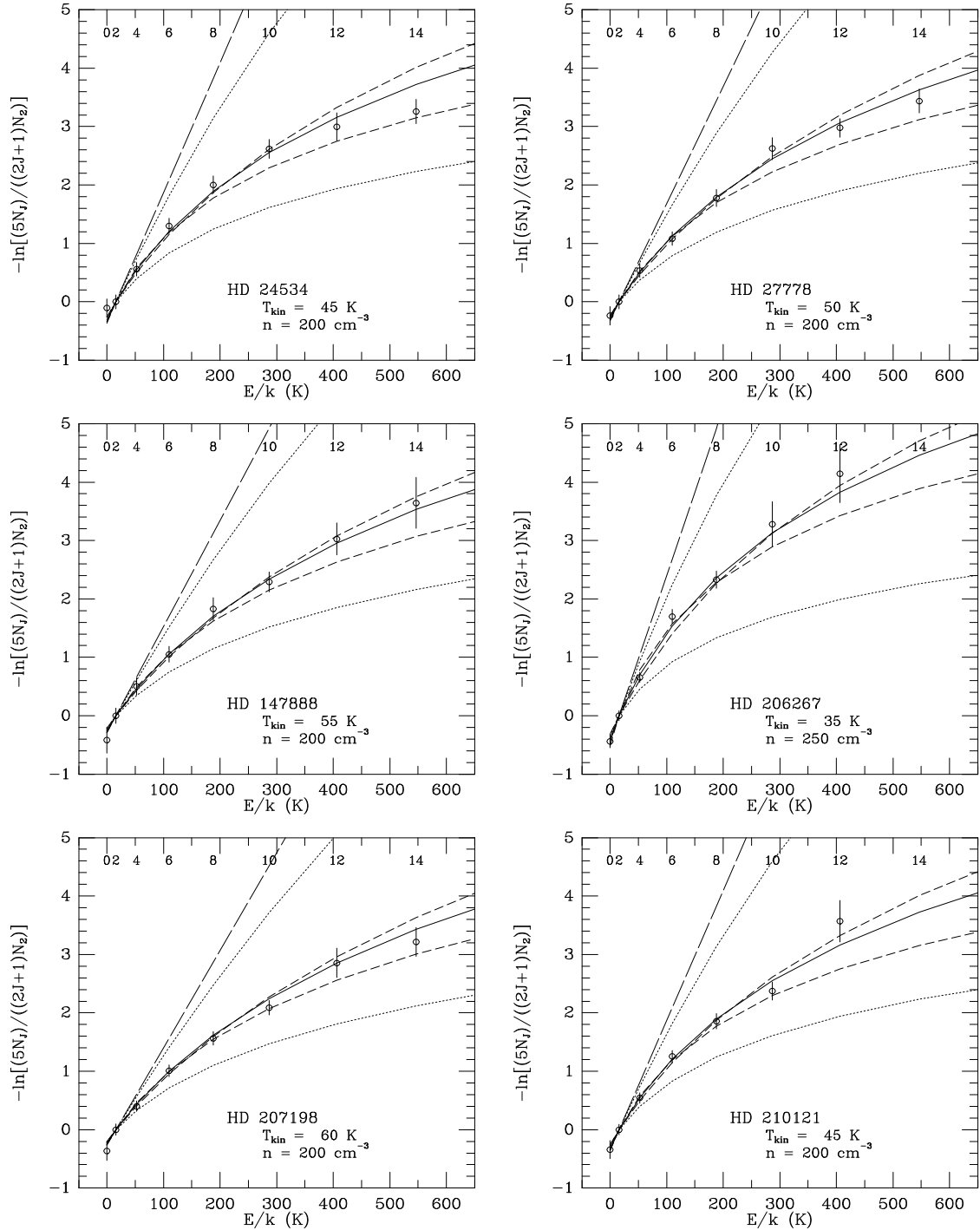}
\caption{Relative C$_2$ rotational population distributions toward six stars, as functions of excitation energy, compared to predicted values. 
In each case, the solid line represents the best-fit theoretical model (to the nearest 5 K in $T_{\rm k}$ and 50 cm$^{-3}$ in $n$), based on the analysis of van Dishoeck \& Black (1982) but adjusted for our adopted C$_2$ $f$-values.
The straight long-dashed line shows the thermal equilibrium distribution for the best-fit $T_{\rm k}$; the two dotted curves show the distributions for that $T_{\rm k}$ and $n$ = 1000 cm$^{-3}$ (nearer to the equilibrium line) and 100 cm$^{-3}$.
The two short-dashed curves show the distributions for ($T_{\rm k}$ $-$ 10, $n$ $-$ 50) and ($T_{\rm k}$ $+$ 10, $n$ $+$ 50).
The C$_2$ rotational distributions yield well determined $T_{\rm k}$ and $n$ in all six cases.}
\label{fig:c2rot}
\end{figure}

The set of observed C$_2$ rotational level populations may also be compared with models for the excitation in order to estimate both the actual kinetic temperature and the density (for an assumed ambient radiation field).
Figure~\ref{fig:c2rot} shows the normalized excitation diagrams (ln[(5~$N_{J}$)/((2$J$+1)~$N_2$)] vs. $E_{J}$/k) for six sight lines whose rotational level column densities were derived from our fits to the C$_2$ F-X, D-X, and/or A-X bands; the corresponding column densities were given in Table~\ref{tab:nc2}.
For each sight line, the observed relative C$_2$ rotational populations have been compared with those predicted by the models of van Dishoeck \& Black (1982; see also van Dishoeck \& de Zeeuw 1984), for $T$ ranging from 10 to 150 K (in steps of 5 K) and the density $n$ ranging from 50 to 1000 cm$^{-3}$ (in steps of 50 cm$^{-3}$)\footnotemark. 
\footnotetext{The comparisons were performed using C$_2$ rotational populations calculated via a web-based tool (available at http://dibdata.org) developed by B. McCall (see \'{A}d\'{a}mkovics et al. 2003), which reproduces the relative $N_{J}$ in Table~7 of van Dishoeck \& Black (1982) for densities smaller by a factor of 1.35 (to account for differences in the adopted C$_2$ band $f$-values; see van Dishoeck \& de Zeeuw 1984). 
Our derived densities were reduced by an additional factor of 1.2 to account for the C$_2$ $f$-values used in this study.}
The models may be characterized by the parameter $n\sigma$/$I$, which is a measure of the relative importance of collisional de-excitation and radiative excitation.
Here $n$ is the number density of collisional partners (H and H$_2$); $\sigma$ is the (assumed constant) collisional de-excitation cross-section for transitions from $J+2$ to $J$; and $I$ is a scaling factor for the strength of the interstellar radiation field in the near-IR (i.e., near the A-X bands), which is most important for the radiative excitation of C$_2$. 
The cross-section $\sigma$ has not been measured for conditions appropriate for diffuse or translucent clouds, but has been estimated to be about 2$\times$10$^{-16}$ cm$^{2}$ from a detailed analysis of the $\zeta$ Per sight line (van Dishoeck \& Black 1982).
Calculations by Lavendy et al. (1991), however, suggest a somewhat higher value for the de-excitation cross-section for collisions with H$_2$ (Federman et al. 1994).
Higher values of $n\sigma/I$ --- i.e., higher $n$ and/or lower $I$ --- yield predicted rotational populations closer to the thermal equilibrium values.
The solid curves in Figure~\ref{fig:c2rot} show the populations predicted for the best-fit $T_{\rm k}$ and $n$ (assuming $I$ = 1).
For that best-fit temperature, the long-dashed straight lines show the rotational populations predicted for thermal equilibrium.
For the adopted cross-section $\sigma$, the A-X band $f$-values noted above, and $I$ = 1, the two dotted curves in each panel of Figure~\ref{fig:c2rot} correspond to $n$ = 1000 (near the thermal equilibrium line) and 100 cm$^{-3}$.
The two short-dashed curves correspond to the rotational populations predicted for slight differences from the best-fit $T_{\rm k}$ and $n$ --- for ($T_{\rm k}$ $+$ 10 , $n$ $+$ 50) and ($T_{\rm k}$ $-$ 10 , $n$ $-$ 50).

The values for $T_{\rm k}$ obtained in this way usually are lower than those for $T_{04}$, and are more comparable to those for $T_{02}$.
Estimates for $T_{02}$, $T_{04}$, and $T_{\rm k}$ (including values derived from equivalent widths or column densities reported in the literature for some additional sight lines) are listed in Table~\ref{tab:texc2}.
It can sometimes be difficult, however, to find a unique, tightly constrained combination of $T$ and $n\sigma/I$ for which the observed and predicted rotational populations agree within the (often considerable) observational uncertainties --- especially if data are not available for rotational levels above $J$ = 8.
In some cases, ``acceptable'' fits are possible for a fairly wide range of values, while in others, none of the models seem able to reproduce all the observed $N_{J}$.
Toward HD~206267, for example, the best fit to most of the $N_{J}$ (for these particular models) is for $T =$ 35 K (very similar to the derived $T_{02}$) and $n$ = $n$(H) + $n$(H$_2$) $\sim$ 250 cm$^{-3}$, assuming the average interstellar radiation field.
Within the uncertainties, however, acceptable agreement between the observed and predicted populations is also possible for slightly lower temperatures and densities and also for somewhat higher temperatures and densities. 
[Note that $n$ (which counts collision partners) is smaller than $n_{\rm H}$ (which counts hydrogen nuclei) by a factor of (1 $-$ $f$(H$_2$)/2)$^{-1}$; estimates for both are given in columns (7) and (8) of Table~\ref{tab:texc2}.]
It seems somewhat unlikely, however, that $n_{\rm H}$ toward HD~206267 could be as high as 600--1200 cm$^{-3}$, as estimated from the CN/CH ratio (see \S~\ref{sec-var} above); CN may well trace denser gas, on average, than C$_2$.  
If the (near-IR) radiation field is stronger than average, however, then the estimated densities would be proportionally higher.  
The gas toward HD~206267 traced by H$_2$ probably has a somewhat higher temperature $\sim$ 65 K [$T_{01}$(H$_2$)] and somewhat lower average density $n_{\rm H}=$ 10-100 cm$^{-3}$ (obtained from an analysis of \ion{O}{1} fine-structure excitation, to be presented elsewhere). 
Much of the H$_2$ may thus reside in more diffuse gas, with the contributions from the denser clumps masked by blending in the lower resolution {\it FUSE} spectra.

Of the ten sight lines with detected C$_2$ (mostly from this paper) listed in the bottom section of Table~\ref{tab:texc2}, seven (HD~24534, HD~27778, HD~147888, $\zeta$~Oph, HD~204827, HD~206267, and HD~210121) have relative rotational populations roughly consistent with $T$ $\sim$ 35--55 K and $n$ $\sim$ 150--300 cm$^{-3}$ (or $n_{\rm H}$ $\sim$ 200--400 cm$^{-3}$). 
On average, the C$_2$-containing gas toward HD~207198 appears to be slightly warmer ($T_{\rm k}$ $\sim$ 60 K) but of comparable density;
the gas toward HD~73882 appears to be somewhat colder ($T_{\rm k}$ $\sim$ 20 K) and denser.
In many of the cases, somewhat lower $T_{\rm k}$ and slightly lower $n$ and/or somewhat higher $T_{\rm k}$ and higher $n$ would also be possible, within the uncertainties in the $N_{J}$.
(And recall that since the theoretical curves are for values of $n\sigma$/$I$, changes to the adopted $\sigma$ or $I$ will imply corresponding changes in $n$.)
The values for $T_{\rm k}$ and density in Table~\ref{tab:texc2} are consistent with those estimated by van Dishoeck et al. (1991), for nine sight lines in common, except that $T_{\rm k}$ is higher toward $\zeta$~Oph and HD~210121 (based on more recent C$_2$ data) and lower toward HD~62542.
We note that relatively few sight lines reported thus far in the literature have sufficiently accurate $N_{J}$ for high enough $J$ for very strong constraints to be placed on $T_{\rm k}$ and $n$; a number of these are included in the top part of Table~\ref{tab:texc2}.
More accurate column densities --- especially for $J$ = 0 and for the higher $J$ levels ($\ge$ 8) --- would be needed to better constrain the temperature and density in those sight lines.

Comparing the excitation temperatures derived for C$_2$ with the corresponding values for H$_2$ (Tables~\ref{tab:texc2} and \ref{tab:dq}) indicates that $T_{02}$(C$_2$) and $T_{\rm k}$(C$_2$) are almost invariably less than $T_{01}$(H$_2$) --- sometimes significantly so.
For the H$_2$ column density regime sampled by the sight lines with detected C$_2$, $T_{01}$(H$_2$) is thought to be an accurate measure of the local kinetic temperature, while (as noted above) $T_{02}$(C$_2$) may represent only an upper limit to $T_{\rm k}$ if radiative excitation is significant.
The systematically lower excitation and kinetic temperatures estimated for C$_2$ suggest that the C$_2$ is preferentially concentrated in somewhat colder, denser gas than that traced (on average) by H$_2$ --- consistent with predictions of the various models of cloud chemistry (e.g., van Dishoeck \& Black 1986a).

Given the indications that many (or most) sight lines are comprised of mixtures of gas with different characteristics --- and given the differences in physical conditions (e.g., density) often inferred from different tracers --- we may ask whether such mixtures can be detected and/or characterized via analyses of C$_2$ rotational excitation.
Cecchi-Pestellini \& Dalgarno (2002) have examined the behavior of the integrated sight line C$_2$ rotational populations when some fraction of the C$_2$ in the sight line is in dense clumps characterized by thermal equilibrium rotational populations.
Not surprisingly, the presence of such dense clumps is most noticeable when both the temperature and density of the more abundant diffuse gas are low --- e.g., for $T_{\rm k}$ $\la$ 20 K and $n$ $\la$ 100 cm$^{-3}$.
In such cases, the low-$J$ populations ($J$ $\la$ 8) for the sight line as a whole remain closer to the thermal equilibrium values, and the higher $J$ levels are somewhat enhanced as well.
Cecchi-Pestellini \& Dalgarno conclude, however, that such dense clumps will be very difficult to discern via analysis of the C$_2$ excitation (given the typical observational uncertainties) when the bulk of the more diffuse gas is warmer than $T_{\rm k}$ $\sim$ 20 K and/or denser than $n$ $\sim$ 100 cm$^{-3}$.
All of the sight lines in our primary sample (with the possible exception of that toward HD~210839) appear to fall in that latter category.
Several of the sight lines listed in the upper part of Table~\ref{tab:texc2} have $T_{\rm k}$ $\la$ 20 K, but $n$ $\ga$ 150 cm$^{-3}$; the sight line toward HD~110432 has $n$ $\sim$ 100 cm$^{-3}$, but uncertain temperature.

\subsection{Isotopic Ratio: $^{12}$C/$^{13}$C}
\label{sec-1213}

The carbon isotopic ratio $^{12}$C/$^{13}$C in the interstellar medium is often used as an indicator of the degree of stellar processing and, hence, of chemical evolution in the Galaxy. 
It is thought that as the Galaxy evolves more $^{13}$C should be produced relative to the dominant isotope $^{12}$C. 
While the current average local Galactic interstellar value of $^{12}$C/$^{13}$C is about 70 $\pm$ 3 (Stahl \& Wilson 1992), variations in that ratio of factors of 2--3 have been observed (Wilson 1999; Casassus, Stahl, \& Wilson 2005). 
Estimates of this ratio have been based on observations of the isotopomers of species such as CH$^+$ (e.g. Hawkins et al. 1993), CO (Lambert et al. 1994), H$_2$CO (Henkel et al. 1985), and CN (Savage et al. 2002) --- each of which presents its own difficulties. 
For example, while CH$^+$ is thought to be unaffected by chemical fractionation, measurement of $^{13}$CH$^+$ is often difficult because the (optical) lines can be quite weak. 
CO, on the other hand, is subject to chemical fractionation --- either from isotope exchange reactions or from selective photodissociation. 
Observations of mm-wave absorption spectra of the isotopomers of HCO$^+$, HCN, and HNC (which should not be significantly fractionated) have yielded $^{12}$C/$^{13}$C = 59 $\pm$ 2 (Lucas \& Liszt 1998).

The models of Warin et al. (1996) predict a somewhat complicated dependence of the $^{12}$CO/$^{13}$CO ratio on density, temperature, and $A_{\rm V}$ for clouds with $n_{\rm H}$ between 10$^2$ and 10$^4$ cm$^{-3}$ and $A_{\rm V}$ $\la$ 5 (see also van Dishoeck \& Black 1988).
The isotope exchange reaction $^{13}$C$^+$~+~$^{12}$CO $\rightarrow$ $^{13}$CO~+~$^{12}$C$^+$ [which is exothermic by only E/k $\sim$ 35 K (Watson, Huntress, \& Anicich 1979)] can enhance $^{13}$CO in low temperature gas at intermediate densities and $A_{\rm V}$ $\la$ 3 by as much as a factor of 5.  
At higher temperatures, lower densities, and low $A_{\rm V}$, however, $^{13}$CO is more readily photodissociated than $^{12}$CO because of its much lower abundance, which does not allow effective self-shielding. 
At higher densities ($\ga$ 10$^5$ cm$^{-3}$) and at higher $A_{\rm V}$, where chemical reactions dominate the destruction of CO and there is little $^{13}$C$^+$ present, the $^{12}$CO/$^{13}$CO ratio is restored to the average value corresponding to the carbon isotopic abundances.
Comparison of the $^{12}$CO/$^{13}$CO ratio with the corresponding average $^{12}$C/$^{13}$C ratio therefore yields indirect information concerning the physical conditions in the gas containing CO. 

Reasonably reliable column densities of $^{13}$CO from UV absorption are now available for twelve sight lines (six from the literature and six from this paper; Table~\ref{tab:coldens}).  
Of those twelve sight lines, three ($\rho$ Oph, $\chi$ Oph, $\zeta$ Oph) have $^{12}$CO/$^{13}$CO ratios significantly above the average value of 70, five (X Per, HD~27778, HD~207198, HD~210121, HD~210839) have $^{12}$CO/$^{13}$CO ratios consistent with the average value, and four (HD~73882, 20 Aql, HD~200775, HD~206267) have $^{12}$CO/$^{13}$CO ratios significantly below the average value.  
Federman et al. (2003) ascribed the higher than average $^{12}$CO/$^{13}$CO ratios toward $\rho$ Oph, $\chi$ Oph, and $\zeta$ Oph to selective photodissociation and proposed that the ratio $N$($^{12}$CO)/$I_{\rm UV}$, where $I_{\rm UV}$ gives the relative strength of the UV radiation field, could be viewed as a measure of the degree of fractionation.  
Liszt \& Lucas (1998) obtained $^{12}$CO/$^{13}$CO ratios between 15 and 54 (smaller than the average $^{12}$C/$^{13}$C ratio derived from other species) for a small set of mm-wave absorption components, with smaller ratios for higher $N$(CO).

\begin{figure}
\epsscale{0.9}
\plotone{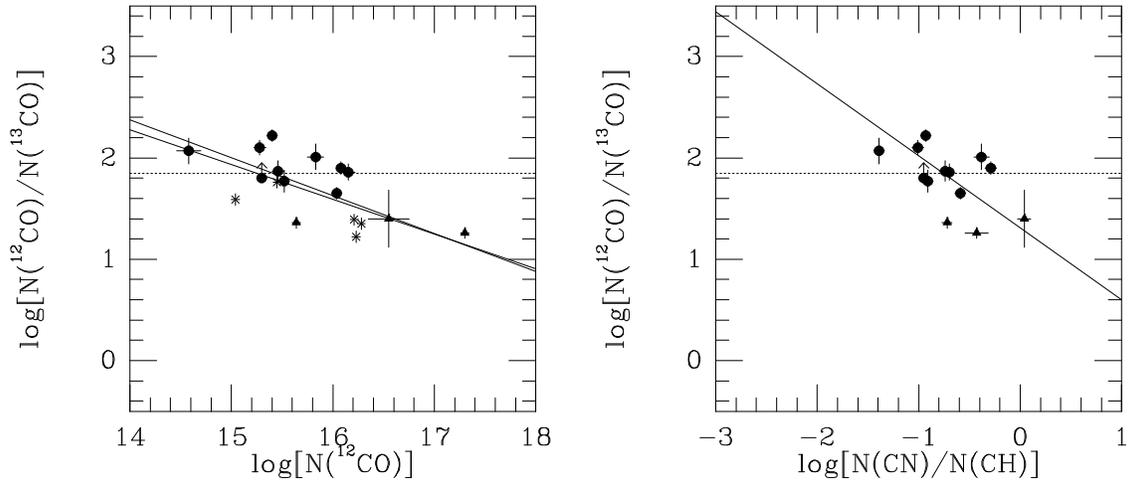}
\caption{Column density ratio $^{12}$CO/$^{13}$CO vs. $N$(CO) ({\it left}) and vs. CN/CH ({\it right}). 
The UV absorption-line data are from {\it HST} (circles), {\it FUSE} (squares), and {\it IUE} (triangles); mm-wave absorption-line data (asterisks) are from Liszt \& Lucas (1998).
The solid lines are linear least-squares fits (weighted and unweighted) to the observed data.
The dotted horizontal line indicates the average Galactic value (70) for the $^{12}$C/$^{13}$C ratio.}
\label{fig:1213}
\end{figure}

Figure~\ref{fig:1213} plots the $^{12}$CO/$^{13}$CO ratios versus the corresponding $^{12}$CO column densities ({\it left}) and CN/CH ratios ({\it right}) for the twelve sight lines in Table~\ref{tab:coldens} with measured $N$($^{13}$CO). 
For comparison, the asterisks in the left-hand panel show the total sight line values derived from mm-wave absorption by Liszt \& Lucas (1998).  
For this relatively small sample of sight lines, the $^{12}$CO/$^{13}$CO ratio appears to be anti-correlated with both $N$($^{12}$CO) (as noted by Liszt \& Lucas 1998) and the CN/CH ratio, with correlation coefficients $r$ = $-$0.647 and $-$0.549, respectively (Table~\ref{tab:corr}).
If the CN/CH ratio is a density indicator, that anti-correlation would suggest that (in general) the $^{12}$CO/$^{13}$CO ratio decreases with increasing density --- presumably because the isotope exchange reaction has enhanced $^{13}$CO in denser, colder gas --- consistent with the behavior predicted by Warin et al. (1996) for densities in the range 100--1000 cm$^{-3}$.
The excitation and kinetic temperatures derived from observations of C$_2$ (Table~\ref{tab:dq}) suggest that $T$ might in fact be low enough in the clouds with low values of the $^{12}$CO/$^{13}$CO ratio for the isotope exchange process to be effective --- particularly if the effective temperatures are lower for the more centrally concentrated, density-sensitive CN and CO than for C$_2$.
Three of the four sight lines with lower than average $^{12}$CO/$^{13}$CO ratios (toward HD~73882, HD~200775, and HD~206267) have $T_{02}$(C$_2$) and/or $T_{\rm k}$(C$_2$) $\la$ 35 K; the four sight lines with average or above average $^{12}$CO/$^{13}$CO ratios and well-determined $T_{\rm k}$ all have $T_{\rm k}$ $\ga$ 45 K.
We note, however, that three of the four low $^{12}$CO/$^{13}$CO ratios (toward HD~73882, 20 Aql, and HD~200775) were derived from {\it IUE} spectra alone, and may be less reliable than the values obtained from the higher resolution, higher S/N {\it HST} spectra.
As there are also indications that the interstellar $^{12}$C/$^{13}$C ratio may be somewhat variable (e.g., Casassus et al. 2005), additional data on the $^{12}$CO/$^{13}$CO ratio and corresponding physical conditions are needed for a more secure understanding of the fractionation of CO in diffuse and translucent clouds.
Several of the other sight lines listed in the upper part of Table~\ref{tab:texc2} with $T_{\rm k}$(C$_2$) $\la$ 30 K might be interesting candidates for further study.

\section{Summary}
\label{sec-summ}

Archival UV spectra from {\it HST} (GHRS and STIS), {\it FUSE}, and/or {\it IUE}, together with high and medium resolution optical spectra, were analyzed to study the molecular abundances and physical conditions in the molecular gas along ten translucent sight lines. 
Absorption from $^{12}$CO was detected in the far-UV E-X (0-0) (1076 \AA), C-X (0-0) (1087 \AA), and B-X (0-0) (1150 \AA) bands, in a number of bands in the A-X system (between 1240 and 1550 \AA), and/or in several of the stronger intersystem bands (between 1360 and 1550 \AA); corresponding A-X bands of $^{13}$CO were detected in six of the sight lines.
Accurate column densities were determined for $^{12}$CO and $^{13}$CO using a combination of curve of growth, apparent optical depth, and profile fitting methods. 
High-resolution optical spectra of \ion{K}{1}, CH, and CN were used to determine the velocity structure of the gas along the sight lines in order to account for potential saturation effects in the UV spectra. 

The UV C$_2$ F-X (0-0) absorption band at 1341 \AA\ was detected in high-resolution STIS echelle spectra of HD~24534, HD~27778, and HD~206267; the weaker F-X (1-0) band at 1314 \AA\ was detected toward HD~24534; and the stronger D-X (0-0) band at 2313 \AA\ was detected toward HD~24534, HD~27778, HD~147888, and HD~207198. 
These constitute nearly all of the reported high-resolution spectra of the UV bands of C$_2$. 
Absorption from the optical C$_2$ A-X (3-0) and/or (2-0) bands at 7719 and 8757 \AA\ was detected toward seven stars. 
Comparisons among the various C$_2$ bands have yielded a (reasonably) mutually consistent set of band $f$-values; the value adopted for the A-X (2-0) band is consistent with several recent theoretical and experimental determinations.
There appear to be some discrepancies between the observed and predicted wavelengths and/or $f$-values for (at least) some individual lines in the F-X (0-0) band.

Reliable column densities (or limits) are now available for $^{12}$CO, CN, CH, C$_2$, and H$_2$ for at least 30 diffuse or translucent sight lines (11 with detections of $^{13}$CO; 13 with detections of HD; 7 with detections of C$_3$; 15 with $A_{\rm V}$ between 1.0 and 2.7 mag) --- enabling investigations of correlations between the various species and providing more stringent tests for models of the structure and chemistry of translucent clouds.
Of those molecular species, CN shows the best correlation with CO --- both in overall column density and in component structure.  
Strong correlations between the ratios $^{12}$CO/H$_2$ and $^{13}$CO/H$_2$ and the density indicator CN/CH suggest that the abundances of both CO isotopomers increase with increasing local density. 
Differences in the strength of the ambient radiation field (as inferred from H$_2$ rotational excitation) also are likely to affect the CO abundances.  
For our sample of ten translucent sight lines, the $N$(CO)/$N$(H$_{2}$) ratio varies over a factor of 100, due primarily to differences in $N$(CO). 
For the total sample, the $^{12}$CO/H$_2$ ratio ranges from 2.5 $\times$ 10$^{-8}$ to 2.8 $\times$ 10$^{-5}$; CO is not yet the dominant carbon-containing species in any of the sight lines.
The models of van Dishoeck \& Black (1988, 1989) predict relationships with H$_2$, CN, CH, and C$_2$ that have slopes similar to those obtained in fits to the observed data, but the predicted $N$(CO) generally are lower than the observed values by factors of 2--4.
The models of Warin et al. (1996) and Le Petit et al. (2006) predict much less CO than is observed, for a given $N$(H$_2$) --- though more of the parameter space available in the Meudon PDR code should be explored.
Inclusion of the effects of inhomogeneous/clumpy structure (e.g., Spaans 1996) and increasing the CO self-shielding and (for the Warin et al. models) the gas-phase carbon abundance should yield better agreement with the observed CO abundances.

Rotational level populations and corresponding excitation temperatures were determined up to $J$ = 3 for $^{12}$CO (nine sight lines) and up to $J$ = 2 for $^{13}$CO (four sight lines).  
The $^{12}$CO excitation temperatures $T_{01}$ typically are between 3 and 7 K (well below the likely kinetic temperature of the gas); the corresponding values for the higher rotational levels are generally equal to $T_{01}$, within the uncertainties.  
Some of these excitation temperatures are slightly higher than the values previously reported for eight other comparably reddened sight lines, but they are similar to the values found via mm-wave CO absorption by Liszt \& Lucas (1998) and to the values predicted for diffuse clouds with $A_{\rm V}$ $\sim$ 1.1 mag by Warin et al. (1996).
The excitation temperatures $T_{01}$ for $^{13}$CO are equal (within the mutual uncertainties) to the corresponding values for $^{12}$CO --- suggesting that the radiative excitation mechanism proposed by Wannier et al. (1997) may not be significant, in many cases.

Rotational level populations and excitation temperatures were also determined for C$_2$ --- up to $J$ = 12 from the UV F-X (0-0) and (1-0) bands (three sight lines), up to $J$ = 18 from the near-UV D-X (0-0) band (four sight lines), and up to $J$ = 12 from the optical A-X (2-0) and (3-0) bands (seven sight lines).  
The excitation temperature for the lowest rotational level ($T_{02}$) ranges from about 20 to 45 K for those sight lines where it is reasonably well determined; corresponding values for the higher rotational levels increase with $J$ (especially for $J$ $\ga$ 8), due to radiative excitation of the upper states.
In most cases, the C$_2$ rotational excitation seems consistent with model predictions for kinetic temperatures of 35--55 K and local total hydrogen densities of 200--400 cm$^{-3}$, assuming the typical near-IR interstellar radiation field.

Of the twelve sight lines with measured $N$($^{13}$CO), three have $^{12}$CO/$^{13}$CO ratios significantly above the average Galactic $^{12}$C/$^{13}$C value of 70, five have $^{12}$CO/$^{13}$CO ratios consistent with the average value, and four have $^{12}$CO/$^{13}$CO ratios significantly below the average value.  
There appears to be an anti-correlation between the $^{12}$CO/$^{13}$CO ratio and the CN/CH ratio --- suggestive of a decrease in $^{12}$CO/$^{13}$CO with increasing local density, as predicted by the models of Warin et al. (1996) for densities in the range 100-1000 cm$^{-3}$.
The $^{12}$CO/$^{13}$CO ratio is also anti-correlated with the column density of CO, as found (for a smaller sample) by Liszt \& Lucas (1998).
Sight lines with below average $^{12}$CO/$^{13}$CO ratios also appear to have kinetic temperatures (from analysis of the C$_2$ rotational excitation) less than about 35 K; sight lines with average or above average $^{12}$CO/$^{13}$CO ratios appear to be somewhat warmer.
While $^{12}$CO/$^{13}$CO ratios above the average value likely result from selective photodissociation (Federman et al. 2003), which favors the self-shielded $^{12}$CO, the below-average ratios probably are due to isotope exchange reactions, which favor $^{13}$CO in cold gas ($T$ $\la$ 35 K).

While the H$_2$ excitation temperature $T_{01}$ ranges from about 50 K to 110 K for our ten sight lines, there are several indications that C$_2$ and (especially) CO trace colder, denser gas in many cases.
The excitation temperature $T_{02}$ of C$_2$ (which provides an upper limit to the local kinetic temperature for the gas containing the C$_2$) and the $T_{\rm k}$ inferred from comparisons with models of the excitation are generally lower than $T_{01}$ for H$_2$ (which should equal the kinetic temperature of the gas traced by H$_2$).
Fits to the CO line profiles indicate that smaller $b$-values (0.3--0.7 km~s$^{-1}$) are needed than for fits to the profiles of \ion{K}{1}, CH, and CN; the CO structure also appears more similar to that of CN (which is concentrated in denser regions).  
The column densities of both $^{12}$CO and $^{13}$CO are most tightly correlated with that of CN, and their abundances, relative to H$_2$, are correlated with the density indicator $N$(CN)/$N$(CH).

These results illustrate the utility of combining information from multiple species sensitive to different combinations of physical parameters in order to probe regions characterized by different properties (along a sight line or within a single ``cloud''). 
They also emphasize the need for developing models capable of incorporating ranges and/or mixtures of physical conditions in order to accurately assess and predict the abundances of various atomic and molecular species in the diffuse ISM. 

\acknowledgments

This work is based in part on data obtained for the Guaranteed Time Team by the NASA-CNES-CSA {\it FUSE} mission operated by the Johns Hopkins University. 
Financial support to U. S. participants has been provided by NASA contract NAS5-32985.
P. Sonnentrucker acknowledges partial support from NASA LTSA grant NAG5-13114 to the Johns Hopkins University. 
D. E. Welty acknowledges support from NASA LSTA grants NAG5-3228 and NAG5-11413 to the University of Chicago. 
We thank E. Jenkins for providing processed STIS spectra of HD~206267 and HD~210839, S. Federman and K. Pan for providing results from their study of the Cep OB2 and OB3 regions in advance of publication, S. Federman for comments on an earlier version of the paper, B. McCall and T. Oka for advice on the analysis of C$_2$, and the anonymous referee for clarifying several points in the discussion.
This work has used the profile fitting procedure Owens.f developed by M. Lemoine and the {\it FUSE} French Team.

Facilities: \facility{FUSE}, \facility{HST (GHRS, STIS)}, \facility{IUE}, \facility{ARC}, \facility{KPNO:coud\'{e} feed}, \facility{McD:2.7m}

\appendix

\section{Component Structures}
\label{sec-comp}

Table~\ref{tab:comp} lists the component structures derived from high-resolution optical spectra of \ion{K}{1}, CH, and CN that have been used to model the profiles of C$_2$ and CO in this paper.  
The \ion{K}{1} structures for HD~206267, HD~207198, and HD~210839 were originally listed by Welty \& Hobbs (2001); they differ slightly from those adopted by Pan et al. (2004). 
The \ion{K}{1} and CH structures for HD~73882, HD~185418, and HD~192639 were used by Snow et al. (2000) and by Sonnentrucker et al. (2002, 2003) in studies of those individual lines of sight.  
Because the total sight line column densities of \ion{K}{1} and CH are generally very tightly correlated, and because the absorption-line profiles of the two species often are strikingly similar in detail (Welty \& Hobbs 2001), we have used the more complex structure discernible in the \ion{K}{1} profiles to model the CH profiles.  
Because CN is likely to be more concentrated in the cooler, denser regions of each sight line, however, we have performed independent fits to the CN profiles.
The CN structures for HD~73882 and HD~210121 are based on analyses of spectra originally reported by Gredel et al. (1991) and Welty \& Fowler (1992), respectively.
A full description of the optical data and profile fits --- for these and other translucent sight lines --- will be given by Welty, Snow, \& Morton (in preparation).

\section{C$_2$ Wavelengths and Line Strengths}
\label{sec-c2bands}

The energy levels of the various rotational levels in the ground vibrational state of the ground $X ^1\Sigma_g^+$ electronic state of C$_2$ were calculated via the usual formula $E_{v=0}^X(J) = B_0^X J (J+1) - D_0^X J^2 (J+1)^2$, using values for the constants $B_0^X$ and $D_0^X$ from Chauville, Maillard, \& Mantz (1977).  
Because C$_2$ is homonuclear, with zero nuclear spin, only even $J$ levels are possible in that ground electronic state.  
The energy levels of the rotational levels of the $v=0$ and $v=1$ vibrational states of the $F ^1\Pi_u$ and $A ^1\Pi_u$ upper electronic states were calculated using the same formula, using constants from Herzberg et al. (1969) and Chauville et al. (1977), respectively.  
The expected wavelengths are then just the reciprocals of the differences between energy levels (in appropriate units); the values for the individual rotational lines in the F-X (1-0) and (0-0) (1314 and 1341 \AA) bands and in the A-X (3-0) and (2-0) (7719 and 8757 \AA) bands are listed in columns (3), (4), (8), and (9) of Table~\ref{tab:c2bands}.  
In the same way, the energy levels of the rotational levels in the $v$ = 0 vibrational state of the $D ^1\Sigma_u^+$ upper electronic state were calculated using constants from Sorkhabi et al. (1998), with $\nu_0$ = 43227.2 cm$^{-1}$ estimated by comparing predicted wave numbers with those measured from their laser-induced fluorescence spectrum.
The corresponding wavelengths for the rotational transitions in the D-X (0-0) band at 2313 \AA\ are listed in column (6) of Table~\ref{tab:c2bands}; these predicted values are in better agreement with the D-X (0-0) lines observed toward HD~24534 than are the measured wave numbers listed by Sorkhabi et al.
The relative intensities of the individual F-X and A-X rotational transitions are given by the usual formulas 
$S_{R(J)} = \frac{J+2}{2(2J+1)}$, $S_{Q(J)} = \frac{1}{2}$, and $S_{P(J)} = \frac{J-1}{2(2J+1)}$; the corresponding relative strengths for the lines in the D-X (0-0) band (which has no Q branch) are given by $S_{R(J)} = \frac{J+1}{2J+1}$ and $S_{P(J)} = \frac{J}{2J+1}$).  

\section{Compilation of CO and Other Molecular Data}
\label{sec-app}

In order to compare the column densities of $^{12}$CO and $^{13}$CO obtained in this paper with those found for other Galactic sight lines, we have collected CO column densities derived from UV absorption-line data from the literature.
Some of the values for sight lines in the {\it Copernicus} survey of Federman et al. (1980) [most with relatively low $N$(CO)] have recently been revised by Crenny \& Federman (2004).  
In the past few years, results from studies of more heavily reddened sight lines [generally with higher $N$(CO)], based on spectra obtained with {\it IUE}, {\it HST}, and/or {\it FUSE}, have become available (e.g., Lambert et al. 1994; Knauth et al. 2001; Sheffer et al. 2002a; Federman et al. 2003, 2004; Pan et al. 2005). 
In this paper, we have analyzed {\it IUE}, {\it FUSE}, and/or {\it HST} spectra of additional translucent sight lines in an attempt to enlarge the sample at higher $N$(CO).
Table~\ref{tab:coldens} lists total column densities for H$_2$, CH, C$_2$, CN, and CO for sight lines where CO has been measured.\footnotemark\
\footnotetext{The column densities in Table~\ref{tab:coldens} are included in a compilation of atomic and molecular column densities maintained at http://astro.uchicago.edu/$\sim$welty/coldens.html.}
This compilation is thus similar to that of Federman et al. (1994a), except that sight lines with no information on CO have not been included.
Moreover, we have generally not included sight lines whose CO column densities were derived from COG analyses of equivalent widths from single {\it IUE} spectra (in view of the low resolution and S/N of the data), particularly if no constraints from higher resolution optical or UV spectra were used or if the CO band structure was not used in computing the theoretical COG.
Of the sight lines in common with the earlier compilation of Federman et al. (1994a), many of those with CO column densities below 10$^{14}$ cm$^{-2}$ have revised values that are significantly smaller; some with column densities above 10$^{14}$ cm$^{-2}$ have revised values different by factors $\ga$ 2.

The C$_2$ column densities in Table~\ref{tab:coldens} have been adjusted for differences in the adopted $f$-values --- e.g., by $-$0.15 dex for references using $f$[A-X(2-0)] = 0.001 (Federman et al. 1994a; Thorburn et al. 2003).
Where the C$_2$ reference is in parentheses, the total column density is based on comparing the $N_{J}$ derived from the equivalent widths in that reference with theoretical C$_2$ rotational distributions (see \S~\ref{sec-nc2} and \S~\ref{sec-c2tex}).

Table~\ref{tab:dq} lists color excesses, visual extinctions, total hydrogen column densities, molecular fractions, excitation temperatures for H$_2$ and C$_2$, as well as several ratios derived from the column densities listed in Table~\ref{tab:coldens}.

\clearpage



\clearpage

\clearpage

\clearpage

\clearpage

\clearpage

\clearpage

\clearpage

\begin{deluxetable}{lllccccrrrrr}
\tabletypesize{\scriptsize}
\tablecolumns{12}
\tablecaption{Sight lines \label{tab:stars}}
\tablewidth{0pt}
 
\tablehead{
\multicolumn{1}{l}{Star}&
\multicolumn{1}{c}{Type}&
\multicolumn{1}{c}{$V$}&
\multicolumn{1}{l}{$E(B-V)$}&
\multicolumn{1}{c}{$A_{\rm V}$}&
\multicolumn{1}{c}{$N$(H~I)}&
\multicolumn{1}{c}{$N$(H$_2$)}&
\multicolumn{1}{c}{$l$}&
\multicolumn{1}{c}{$b$}&
\multicolumn{1}{c}{$d$}&
\multicolumn{1}{c}{$v_{\rm r}$}&
\multicolumn{1}{c}{$vsini$}\\
\multicolumn{5}{c}{ }&
\multicolumn{1}{c}{(cm$^{-2}$)}&
\multicolumn{1}{c}{(cm$^{-2}$)}&
\multicolumn{2}{c}{ }&
\multicolumn{1}{c}{(pc)}&
\multicolumn{1}{c}{(km~s$^{-1}$)}&
\multicolumn{1}{c}{(km~s$^{-1}$)}}
\startdata
\object{HD  24534} & O9.5pe & 6.1  & 0.59 & 2.05 & 20.72 & 20.92 & 163.08 &$-17.14$&  590 &  15.0 & 150     \\
\object{HD  27778} & B3 V   & 6.36 & 0.37 & 1.01 & 20.98 & 20.79 & 172.76 &$-17.39$&  223 &  15.7 & 105     \\
\object{HD  73882} & O8 V   & 7.22 & 0.72 & 2.36 & 21.12 & 21.11 & 260.18 &   0.64 & 1100 &  22.0 & \nodata \\
\object{HD 147888} & B5 V   & 6.74 & 0.47 & 1.91 &\nodata& 20.48 & 353.65 &  17.71 &  136 & $-4.0$& 180     \\
\object{HD 185418} & B0.5 V & 7.45 & 0.51 & 2.03 & 21.12 & 20.76 &  53.60 & $-2.17$&  950 & $-2.0$& \nodata \\
\object{HD 192639} & O8e    & 7.12 & 0.66 & 1.87 & 21.32 & 20.69 &  74.90 &   1.48 & 1100 & $-6.0$& 120     \\
\object{HD 206267} & O6f    & 5.62 & 0.53 & 1.37 & 21.31 & 20.86 &  99.29 &   3.74 & 1000 & $-7.8$&  80     \\
\object{HD 207198} & O9 IIe & 5.95 & 0.62 & 1.36 & 21.34 & 20.83 & 103.14 &   6.99 & 1000 &$-18.4$& \nodata \\
\enddata
\tablecomments{Type, $V$, $E(B-V)$, and $d$ from Thorburn et al. 2003; $A_{\rm V}$, $N$(H~I), and $N$(H$_2$) from Rachford et al. 2002 (and references therein) or Cartledge et al. 2004 (HD~147888); $v_{\rm r}$ and $vsini$ from SIMBAD or Welty \& Fowler 1992 (HD~210121).}
\end{deluxetable} 

\clearpage

\begin{deluxetable}{lccl}
\tabletypesize{\scriptsize}
\tablecolumns{4}
\tablecaption{UV Data Sets \label{tab:uvdata}}
\tablewidth{0pt}
 
\tablehead{
\multicolumn{1}{l}{Star}&
\multicolumn{1}{c}{Mission}&
\multicolumn{1}{c}{Instrument}&
\multicolumn{1}{l}{Data Sets}}
\startdata
HD  24534 & {\it HST} & STIS & o64812010-030, o64813010-020,
                               \dataset[ADS/Sa.HST#o66p01020]{o66p01020}, 
                               \dataset[ADS/Sa.HST#o66p02010]{o66p02010}\\
          & {\it HST} & GHRS & \dataset[ADS/Sa.HST#z2do020at]{z2do020at}\\
HD  27778 & {\it IUE} & SWP  & 21264-66, 21284-85, 21289-92, 21297-98, 
                               \dataset[ADS/Sa.IUE#swp24095]{24095}\\
          & {\it HST} & STIS & \dataset[ADS/Sa.HST#o59s01020]{o59s01020}, o6lb01010-040 \\
          & {\it FUSE}&      & \dataset[ADS/Sa.FUSE#p1160301]{p1160301}\\
HD  73882 & {\it IUE} & SWP  & \dataset[ADS/Sa.IUE#swp25437]{25437}, 30064-65, 30075-77, 
                               \dataset[ADS/Sa.IUE#swp38299]{38299}, 
                               \dataset[ADS/Sa.IUE#swp38311]{38311}\\
          & {\it FUSE}&      & \dataset[ADS/Sa.FUSE#p1161301]{p1161301}, 
                               \dataset[ADS/Sa.FUSE#p1161302]{p1161302}, 
                               \dataset[ADS/Sa.FUSE#x0211301]{x0211301}\\
HD 147888 & {\it HST} & STIS & \dataset[ADS/Sa.HST#o59s05010]{o59s05010}, o6lb04010--040 \\
\enddata
\tablecomments{All the {\it FUSE} data were obtained for programs P116 or X021 (T. Snow, P.I.).
The STIS data were obtained for {\it HST} GO programs 
8273 (HD~27778, HD~147888, HD~207198; D. Meyer, P.I.),
8622 (HD~24534; D. Lambert, P.I.),
8623 (HD~24534; T. Snow, P.I.),
9465 (HD~27778, HD~147888, HD~207198; U. J. Sofia, P.I.),
GTO programs 8043 and 8484 (HD~206267, HD~210839; E. Jenkins, P.I.),
or SNAP program 8241 (HD~185418, HD~192639; J. Lauroesch, P.I.).
The GHRS data were obtained for {\it HST} GO programs 5487 (HD~24534; T. Snow, P.I.) and 5739 (HD~210121; E. van Dishoeck, P.I.).
The {\it IUE} data were obtained for programs IDHTS, IEETS, IMFTS, IMGTS, IMITS, and IMLTS (HD~27778, HD~73882, HD~207198; T. Snow, P.I.) and program IMLLH (HD~210121; L. Hobbs, P.I.).}
\end{deluxetable}

\clearpage

\begin{deluxetable}{lllrccccccc}
\tabletypesize{\scriptsize}
\tablecolumns{11}
\tablecaption{CO Equivalent Widths ({\it FUSE} and STIS) \label{tab:ew1}}
\tablewidth{0pt}

\tablehead{
\colhead{Species} & 
\colhead{Band} &
\colhead{$\lambda$\tablenotemark{a}} & 
\colhead{log $f\lambda$\tablenotemark{a}} &
\colhead{ 27778} &
\colhead{147888} &
\colhead{185418} & 
\colhead{192639} & 
\colhead{206267} &
\colhead{207198} & 
\colhead{210839} } 
\startdata
$^{12}$CO & E-X  (0-0) & 1076.034 &    1.864 & 
     83.9$\pm$7.3 & \nodata      & 64.6$\pm$4.7 & 38.9$\pm$5.5 & 119$\pm$18   & 124$\pm$15   & 84.5$\pm$9.0 \\
          & C-X  (0-0) & 1087.868 &    2.126 & 
     blend        & \nodata      & 96$\pm$20    & 59$\pm$10    & blend        & blend        & blend  \\
          & B-X  (0-0) & 1150.483 &    0.887 & 
     69.9$\pm$5.0 & \nodata      & \nodata      & \nodata      & 65.0$\pm$4.7 & \nodata      & 39.6$\pm$2.6 \\
          & A-X (12-0) & 1246.059 & $-$0.950 & 
     11.9$\pm$3.2 & \nodata      & \nodata      & \nodata      & 11.2$\pm$4.3 & \nodata      &  6.5$\pm$1.9 \\
          & A-X (11-0) & 1263.433 & $-$0.643 & 
     21.1$\pm$3.4 &  4.9$\pm$2.8 & \nodata      & \nodata      & 26.3$\pm$6.5 &  9.1$\pm$2.5 &  6.9$\pm$2.0 \\
          & A-X (10-0) & 1281.866 & $-$0.279 & 
     40.2$\pm$5.0 & 11.3$\pm$2.6 &  3.5$\pm$0.7 & \nodata      & 36.0$\pm$6.3 & 17.3$\pm$2.7 & 14.3$\pm$2.0 \\
          & A-X  (9-0) & 1301.403 &    0.092 & 
     52.5$\pm$5.1 & 14.2$\pm$5.2 &  8.1$\pm$1.9 & \nodata      & 62.0$\pm$7.7 & 29.9$\pm$2.5 & 23.3$\pm$1.9 \\
          & A-X  (8-0) & 1322.150 &    0.423 & 
     74.3$\pm$4.3 & 21.4$\pm$4.6 & 13.9$\pm$2.0 &  3.3$\pm$2.4 & 81.0$\pm$6.2 & 46.5$\pm$2.2 & 30.5$\pm$3.1 \\
\enddata
\tablecomments{Values are equivalent width (m\AA) and 1-$\sigma$ uncertainty.
Equivalent width for the B-X (0-0) band toward HD~73882 is 63.9$\pm$5.5 m\AA.}
\tablenotetext{a}{Wavelengths and $f$-values are from Morton \& Noreau (1994) for the A-X bands and from Federman et al. (2001) for the E-X, C-X, and B-X bands.}
\end{deluxetable}

\clearpage

\begin{deluxetable}{lclrcccccc}
\tabletypesize{\scriptsize}
\tablecolumns{10}
\tablecaption{CO A-X Band Equivalent Widths ({\it IUE} and {\it HST}) \label{tab:ew2}}
\tablewidth{0pt}
 
\tablehead{
\multicolumn{1}{l}{Species}&
\multicolumn{1}{c}{Band}&
\multicolumn{1}{c}{$\lambda$\tablenotemark{a}}&
\multicolumn{1}{c}{log $f\lambda$\tablenotemark{a}}&
\multicolumn{1}{c}{149757\tablenotemark{b}}&
\multicolumn{1}{c}{24534\tablenotemark{c}}&
\multicolumn{1}{c}{27778\tablenotemark{d}}&
\multicolumn{1}{c}{73882\tablenotemark{e}}&
\multicolumn{1}{c}{207198\tablenotemark{f}}&
\multicolumn{1}{c}{210121\tablenotemark{g}}}
\startdata
$^{12}$CO &(12-0) & 1246.059 &$-$0.950 &  \nodata      & 11.9$\pm$1.1 & 11.9$\pm$3.2 & \nodata    & \nodata       & \nodata \\
          &(11-0) & 1263.433 &$-$0.643 &   5.4$\pm$0.7 & 18.6$\pm$1.5 & 21.1$\pm$3.4 & \nodata    &   9.1$\pm$2.5 & \nodata \\
          &(10-0) & 1281.866 &$-$0.279 &  10.2$\pm$0.9 & 28.9$\pm$1.6 & 40.2$\pm$5.0 & \nodata    &  17.3$\pm$2.7 & \nodata \\
          & (9-0) & 1301.403 &   0.092 &  15.7$\pm$1.2 & 40.8$\pm$1.6 & 52.5$\pm$5.1 & \nodata    &  29.9$\pm$2.5 & \nodata \\
          & (8-0) & 1322.150 &   0.423 &  24.6$\pm$1.3 & 55.1$\pm$1.4 &   73$\pm$ 6  & 110$\pm$15 &  50$\pm$ 6    & \nodata \\
          &       &          &         &  \nodata      &\nodata       & 74.3$\pm$4.3 & \nodata    &  46.5$\pm$2.2 & \nodata \\
          & (7-0) & 1344.186 &   0.745 &  33.9$\pm$1.6 & 72.3$\pm$2.1 &   98$\pm$ 9  &  90$\pm$14 &  65$\pm$ 8    & \nodata \\
          &       &          &         &  \nodata      &\nodata       & 97.0$\pm$4.9 & \nodata    &  71.8$\pm$1.9 & \nodata \\
\enddata
\tablecomments{Values for each sight line are equivalent width (m\AA) and 1-$\sigma$ uncertainty (or 3-$\sigma$ limit).  Integer values in the last four columns are from {\it IUE} or GHRS data; decimal values are from STIS data.}
\tablenotetext{a}{Morton \& Noreau 1994 (permitted bands); Sheffer et al. 2002a or Eidelsberg \& Rostas 2003 (intersystem bands)}
\tablenotetext{b}{Federman et al. 1994b; Lambert et al. 1994}
\tablenotetext{c}{STIS data (permitted bands); Sheffer et al. 2002a (intersystem bands)}
\tablenotetext{d}{{\it IUE} data ($\lambda$ $>$ 1320\AA); STIS data ($\lambda$ $<$ 1360\AA); values for stronger $^{13}$CO bands may contain some contribution from $^{12}$C$^{18}$O}
\tablenotetext{e}{{\it IUE} data; values for stronger $^{13}$CO bands may contain some contribution from $^{12}$C$^{18}$O }
\tablenotetext{f}{{\it IUE} data ($\lambda$ $>$ 1320\AA); STIS data ($\lambda$ $<$ 1360\AA)}
\tablenotetext{g}{{\it IUE} data (Welty \& Fowler 1992; first line); 
GHRS data (second line)}
\end{deluxetable}

\clearpage

\begin{deluxetable}{lccccccccc}
\tabletypesize{\scriptsize}
\tablecolumns{10}
\tablecaption{Total $^{12}$CO and $^{13}$CO Column Densities.\label{tab:nco}}
\tablewidth{0pt}

\tablehead{
\multicolumn{1}{l}{Star}& 
\multicolumn{5}{c}{$N$($^{12}$CO)}&
\multicolumn{4}{c}{$N$($^{13}$CO)}\\
\multicolumn{1}{c}{ }& 
\multicolumn{1}{c}{COG\tablenotemark{a}}&
\multicolumn{1}{c}{$b_{\rm eff}$}& 
\multicolumn{1}{c}{AOD}& 
\multicolumn{1}{c}{FIT}&
\multicolumn{1}{c}{Adopt\tablenotemark{b}}&
\multicolumn{1}{c}{COG\tablenotemark{a}}& 
\multicolumn{1}{c}{AOD}& 
\multicolumn{1}{c}{FIT}&
\multicolumn{1}{c}{Adopt\tablenotemark{b}}}
\startdata
HD  24534 & 16.15$\pm$0.15 & 0.8 & $\ga$16.02     & 16.01$\pm$0.09 & 16.15$\pm$0.07 & 14.25$\pm$0.05 & 14.18$\pm$0.04 & 14.29$\pm$0.03 & 14.29$\pm$0.02\\
HD  27778 & 16.10$\pm$0.20 & 1.0 & $\ga$16.03     & 16.08$\pm$0.03 & 16.08$\pm$0.03 & 14.20$\pm$0.10 & 14.16$\pm$0.09 & 14.19$\pm$0.10 & 14.19$\pm$0.10 \\
HD  73882 & 16.6$\pm$0.2   & 0.8 & \nodata        & 16.55$\pm$0.20 & 16.55$\pm$0.20 & 15.2$\pm$0.2   & \nodata        & 15.15$\pm$0.20 & 15.15$\pm$0.20 \\
HD 147888 & 15.30$\pm$0.15 & 0.5 & 15.30$\pm$0.11 & 15.30$\pm$0.08 & 15.30$\pm$0.08 & \nodata        & \nodata        & $<$13.5        & $<$13.5        \\
HD 185418 & 14.70$\pm$0.15 & 1.0 & 14.71$\pm$0.06 & 14.75$\pm$0.03 & 14.75$\pm$0.03 & \nodata        & \nodata        & \nodata        & \nodata        \\
HD 192639 & 14.15$\pm$0.15 & 1.0 & 14.04$\pm$0.12 & 14.13$\pm$0.13 & 14.13$\pm$0.13 & \nodata        & \nodata        & \nodata        & \nodata        \\
HD 206267 & 16.10$\pm$0.25 & 1.0 & 16.06$\pm$0.12 & 16.04$\pm$0.04 & 16.04$\pm$0.04 & 14.40$\pm$0.10 & 14.40$\pm$0.10 & 14.39$\pm$0.03 & 14.39$\pm$0.03 \\
HD 207198 & 15.45$\pm$0.25 & 2.0 & 15.52$\pm$0.08 & 15.52$\pm$0.04 & 15.52$\pm$0.04 & 13.70$\pm$0.10 & 13.78$\pm$0.17 &[13.75$\pm$0.10]&[13.75$\pm$0.10]\\
\enddata
\tablecomments{{Logarithmic} column densities and 1-$\sigma$ uncertainties (or 3-$\sigma$ limits); units are cm$^{-2}$.  
Methods used to derive the column densities are curve of growth (COG), apparent optical depth (AOD), and profile fitting (FIT).
Values in parentheses are for $N_0$ only.
Values for $^{13}$CO in square braces are for $N_0$ + $N_1$, assuming that $N_1$/$N_0$ is the same for $^{13}$CO as for $^{12}$CO.}
\tablenotetext{a}{Curves of growth assume $T_{\rm ex}$ = 3 K (for HD~185418, HD~192639, and HD~207198) or $T_{\rm ex}$ = 5 K (for all others).}
\tablenotetext{b}{Values from Sheffer et al. (2002a, 2002b) are adopted for HD~24534.  FIT values are adopted for all other sight lines.}
\end{deluxetable}


\begin{deluxetable}{lrrrrrrr}
\tabletypesize{\scriptsize}
\tablecolumns{8}
\tablecaption{CO Rotational Level Column Densities \label{tab:conj}}
\tablewidth{0pt}
 
\tablehead{
\multicolumn{1}{l}{Star}&
\multicolumn{4}{c}{$N$($^{12}$CO) $\times$ 10$^{-14}$}&
\multicolumn{3}{c}{$N$($^{13}$CO) $\times$ 10$^{-13}$}\\
\multicolumn{1}{c}{ }&
\multicolumn{1}{c}{$J$ = 0}&
\multicolumn{1}{c}{$J$ = 1}&
\multicolumn{1}{c}{$J$ = 2}&
\multicolumn{1}{c}{$J$ = 3}&
\multicolumn{1}{c}{$J$ = 0}&
\multicolumn{1}{c}{$J$ = 1}&
\multicolumn{1}{c}{$J$ = 2}}
\startdata
HD 24534  & 48.8$\pm$17.8 & 43.8$\pm$9.3  &   9.2$\pm$1.4  &  0.67$\pm$0.30 & 11.9$\pm$1.3 &  6.7$\pm$0.4 & 0.70$\pm$0.08 \\
HD 27778  & 52.1$\pm$5.9  & 53.5$\pm$4.9  &  12.5$\pm$1.4  &  1.17$\pm$0.26 &  8.2$\pm$1.2 &  6.8$\pm$1.3 & \nodata \\
HD 147888 &  5.4$\pm$2.1  &  9.8$\pm$3.2  &   4.0$\pm$0.9  &  0.88$\pm$0.20 &  \nodata     &  \nodata     & \nodata \\
HD 185418 &  3.5$\pm$0.3  &  2.0$\pm$0.2  &  0.21$\pm$0.14 &  \nodata       &  \nodata     &  \nodata     & \nodata \\
HD 192639 & 1.00$\pm$0.27 & 0.35$\pm$0.26 &  \nodata       &  \nodata       &  \nodata     &  \nodata     & \nodata \\
HD 206267 & 43.3$\pm$6.5  & 53.7$\pm$6.3  &  12.3$\pm$1.4  &  1.40$\pm$0.34 & 10.2$\pm$1.2 & 14.1$\pm$1.4 & \nodata \\
HD 207198 & 18.4$\pm$2.6  & 13.1$\pm$1.6  &  1.32$\pm$0.41 &  $<$0.3        &  3.3$\pm$0.7 & (2.3)\tablenotemark{a} & \nodata \\
HD 210121 &   25$\pm$8    &   31$\pm$9    &    12$\pm$4    &  0.31$\pm$0.08 &  2.8$\pm$1.0 &  3.8$\pm$1.1 & \nodata \\
\enddata
\tablenotetext{a}{Assumes same excitation temperature as for $^{12}$CO.}
\end{deluxetable}

\clearpage

\begin{deluxetable}{rccrrrr}
\tabletypesize{\scriptsize}
\tablecolumns{7}
\tablecaption{C$_2$ D-X (0-0) Band Equivalent Widths \label{tab:c2dx}}
\tablewidth{0pt}

\tablehead{
\multicolumn{1}{c}{Line}&
\multicolumn{1}{c}{$\lambda$}&
\multicolumn{1}{c}{S\tablenotemark{a}}&
\multicolumn{1}{c}{HD~24534}&
\multicolumn{1}{c}{HD~27778}&
\multicolumn{1}{c}{HD~147888}&
\multicolumn{1}{c}{HD~207198}}
\startdata
 R(0) & 2313.163 & 1.000 & 3.7$\pm$0.8 & 3.7$\pm$0.5 & 2.6$\pm$0.5 & 4.8$\pm$0.9 \\
 R(2) & 2312.769 & 0.600 & 9.5$\pm$0.8 & 8.4$\pm$0.6 & 3.9$\pm$0.5 &12.1$\pm$0.9 \\
 P(2) & 2313.745 & 0.400 & 7.1$\pm$0.8 & 5.1$\pm$0.5 & 3.0$\pm$0.5 & 7.8$\pm$0.8 \\
 R(4) & 2312.371 & 0.556 & 8.8$\pm$0.8 & 9.2$\pm$0.6 & 4.7$\pm$0.6 &11.5$\pm$0.9 \\
 P(4) & 2314.126 & 0.444 & 8.0$\pm$0.8 & 6.0$\pm$0.5 & 3.6$\pm$0.5 &10.1$\pm$0.9 \\
 R(6) & 2311.967 & 0.538 & 6.9$\pm$0.8 & 6.2$\pm$0.6 & 3.0$\pm$0.5 & 9.4$\pm$0.9 \\
 P(6) & 2314.503 & 0.462 & 6.4$\pm$0.8 & 4.9$\pm$0.5 & 2.8$\pm$0.6 & 8.7$\pm$0.9 \\
 R(8) & 2311.560 & 0.529 & 4.9$\pm$0.7 & 4.0$\pm$0.6 & 2.1$\pm$0.5 & 5.9$\pm$0.8 \\
\enddata
\tablecomments{Equivalent widths are in m\AA.  Uncertainties are 1$\sigma$; limits are 3$\sigma$.}
\tablenotetext{a}{Relative $f$-value for individual line.
Adopted band $f$-value is 0.0545 (Lambert et al. 1995).}
\end{deluxetable}

\clearpage

\setlength{\tabcolsep}{0.03in}

\begin{deluxetable}{lrrcccccccc}
\tabletypesize{\scriptsize}
\tablecolumns{11}
\tablecaption{C$_2$ A-X Band Equivalent Widths \label{tab:c2optw}}
\tablewidth{0pt}

\tablehead{
\multicolumn{1}{c}{Line}&
\multicolumn{1}{c}{$\lambda$}&
\multicolumn{1}{c}{log $f\lambda$}&
\multicolumn{1}{c}{HD 24534}& 
\multicolumn{1}{c}{HD 27778}&
\multicolumn{1}{c}{HD 147888}&
\multicolumn{1}{c}{HD 204827}&
\multicolumn{1}{c}{HD 206267}&
\multicolumn{1}{c}{HD 207198}&
\multicolumn{1}{c}{HD 210121}&
\multicolumn{1}{c}{HD 210839}
}
\startdata
\multicolumn{2}{c}{(2-0) band} & \\
 R(0) & 8757.686 & 1.089 & 1.2$\pm$0.3 & 1.2$\pm$0.3 & 1.0$\pm$0.3 & 16.1$\pm$0.5 & 3.5$\pm$0.4 & 2.6$\pm$0.5 & 3.8$\pm$0.4 & 2.0$\pm$0.4 \\
 R(2) & 8753.949 & 0.690 & 2.9$\pm$0.5 & 2.0$\pm$0.3 & 1.2$\pm$0.3 & 20.1$\pm$0.6 & 3.9$\pm$0.5 & $<$5.0      & 3.6$\pm$0.4 & \nodata     \\
 Q(2) & 8761.194 & 0.788 & 3.2$\pm$0.4 & 3.2$\pm$0.4 & 1.7$\pm$0.4 & 28.2$\pm$0.5 & 7.3$\pm$0.5 & 3.6$\pm$0.6 & 6.0$\pm$0.4 & 1.9$\pm$0.5 \\
 P(2) & 8766.031 & 0.089 & \nodata     & 1.2$\pm$0.3 & \nodata     &  4.7$\pm$0.4 & 2.2$\pm$0.4 & $<$1.5      & \nodata     & \nodata     \\
 R(4) & 8751.685 & 0.611 & \nodata     &(3.1$\pm$0.4)&(0.9$\pm$0.3)&(20.3$\pm$0.6)&(6.3$\pm$0.7)&(3.1$\pm$0.8)&(5.4$\pm$0.8)&(1.6$\pm$0.6)\\
 Q(4) & 8763.751 & 0.788 & 3.4$\pm$0.4 & 3.5$\pm$0.4 & 1.2$\pm$0.4 & 25.3$\pm$0.6 & 7.5$\pm$0.6 & 5.3$\pm$0.5 & 6.0$\pm$0.4 & 3.3$\pm$0.6 \\
 R(6) & 8750.848 & 0.577 & \nodata     & 2.7$\pm$0.4 & 0.7$\pm$0.3 &  9.7$\pm$0.6 & 1.9$\pm$0.5 & 1.5$\pm$0.6 & 3.3$\pm$0.6 & 1.9$\pm$0.4 \\
\enddata
\tablecomments{Equivalent widths are in m\AA; uncertainties are 1$\sigma$; limits are 3$\sigma$.
Values for blended lines [(2-0) R(4)/R(8) and Q(8)/P(4)] are given in parentheses; (2-0) Q(10) is usually blended with a telluric line; (3-0) Q(2) is sometimes blended with a weak diffuse interstellar band.
Assumed $f$ = 0.00140 for the (2-0) band and 0.00065 for the (3-0) band.}
\end{deluxetable}

\clearpage

\setlength{\tabcolsep}{0.05in}

\begin{deluxetable}{llcccccccccccc}
\rotate
\tabletypesize{\scriptsize}
\tablecolumns{14}
\tablecaption{C$_2$ Column Densities \label{tab:nc2}}
\tablewidth{0pt}
 
\tablehead{
\multicolumn{1}{l}{Star}&
\multicolumn{1}{c}{Band}&
\multicolumn{1}{c}{$J$ = 0}&
\multicolumn{1}{c}{2}&
\multicolumn{1}{c}{4}&
\multicolumn{1}{c}{6}&
\multicolumn{1}{c}{8}&
\multicolumn{1}{c}{10}&
\multicolumn{1}{c}{12}&
\multicolumn{1}{c}{14}&
\multicolumn{1}{c}{16}&
\multicolumn{1}{c}{18}&
\multicolumn{1}{c}{0--6}&
\multicolumn{1}{c}{Total\tablenotemark{a}}}
\startdata
HD~24534  & F-X(1-0) & 0.13$\pm$0.03 & 0.71$\pm$0.07 & 0.73$\pm$0.07 & 0.55$\pm$0.05 & 0.37$\pm$0.05 & 0.27$\pm$0.05 & 0.14$\pm$0.05 &
                       \nodata       & \nodata       & \nodata       & 2.12$\pm$0.11 & 3.07$\pm$0.36 \\
          & F-X(0-0) & 0.17$\pm$0.02 & 0.68$\pm$0.05 & 0.77$\pm$0.06 & 0.43$\pm$0.05 & 0.16$\pm$0.03 & 0.05$\pm$0.03 & 0.12$\pm$0.04 &
                       \nodata       & \nodata       & \nodata       & 2.05$\pm$0.09 & 2.97$\pm$0.34 \\
          & F-X(0-0)\tablenotemark{b} & 0.18$\pm$0.02 & 0.65$\pm$0.04 & 0.67$\pm$0.05 & 0.42$\pm$0.04 & 0.12$\pm$0.03 & $<$0.05       & 0.08$\pm$0.03 &
                       0.11$\pm$0.03 & 0.08$\pm$0.03 & 0.05$\pm$0.02 & 1.92$\pm$0.08 & 2.78$\pm$0.31 \\
          & D-X(0-0) & 0.15$\pm$0.02 & 0.67$\pm$0.05 & 0.69$\pm$0.04 & 0.55$\pm$0.04 & 0.37$\pm$0.04 & 0.23$\pm$0.03 & 0.22$\pm$0.03 &
                       0.16$\pm$0.03 & 0.04$\pm$0.03 & \nodata       & 2.06$\pm$0.08 & 2.99$\pm$0.33 \\
          & A-X(3-0) & 0.31$\pm$0.05 & 1.11$\pm$0.10 & 0.98$\pm$0.11 & 0.54$\pm$0.10 & 0.33$\pm$0.11 & \nodata       & \nodata       &
                       \nodata       & \nodata       & \nodata       & 2.94$\pm$0.19 & 4.26$\pm$0.52 \\
          & A-X(2-0) & 0.14$\pm$0.04 & 0.73$\pm$0.08 & 0.77$\pm$0.09 & 0.35$\pm$0.09 & 0.18$\pm$0.08 & \nodata       & \nodata       &
                       \nodata       & \nodata       & \nodata       & 1.99$\pm$0.16 & 2.89$\pm$0.37 \\
          & Average  & 0.16$\pm$0.02 & 0.72$\pm$0.06 & 0.74$\pm$0.06 & 0.51$\pm$0.05 &(0.33$\pm$0.04)&(0.22$\pm$0.03)& 0.18$\pm$0.04 &
                       0.16$\pm$0.03 & 0.04$\pm$0.03 & \nodata       & 2.13$\pm$0.10 & 3.09$\pm$0.35 \\
 & \\
\enddata
\tablecomments{Column densities are in units of 10$^{13}$ cm$^{-2}$.
Uncertainties are 1-$\sigma$, and include contributions from photon noise and uncertainties in continuum placement and $b$-value.
Average values for $J$ = 8 and 10 in parentheses use 2$\times$ the nominal values from the F-X(0-0) band.
Values for $\zeta$ Oph are based on data from Lambert et al. (1995) [D-X(0-0)], van Dishoeck \& Black (1986b) [A-X(3-0)], and Danks \& Lambert (1983) and Hobbs \& Campbell (1982) [(A-X(2-0)].}
\tablenotetext{a}{Estimated total $N$(C$_2$) (including unobserved higher $J$ populations) obtained by multiplying observed sum for levels 0--6 by ratio of theoretical sums for $J$ = 0--20 and $J$ = 0--6, for the best-fit $n$ and $T_{\rm k}$ model.}
\tablenotetext{b}{Column densities $N_J$ from fit to combined (3-epoch) spectrum, with only lines for $J$ $<$ 10 shifted from predicted wavelengths (see Fig. 11); these $N_J$ are not used in the weighted average.}
\end{deluxetable}

\clearpage

\setlength{\tabcolsep}{0.09in}

\begin{deluxetable}{lccccccc}
\tabletypesize{\scriptsize}
\tablecolumns{8}
\tablecaption{Total Column Densities \label{tab:ntot}}
\tablewidth{0pt}
 
\tablehead{
\multicolumn{1}{l}{Star}&
\multicolumn{1}{c}{H$_2$}&
\multicolumn{1}{c}{$^{12}$CO}&
\multicolumn{1}{c}{$^{13}$CO}&
\multicolumn{1}{c}{C$_2$}&
\multicolumn{1}{c}{$^{12}$CO/H$_2$}&
\multicolumn{1}{c}{$^{12}$CO/$^{13}$CO}&
\multicolumn{1}{c}{C$_2$/H$_2$}}
\startdata
HD  24534 & 20.92$\pm$0.04 & 16.15$\pm$0.07 & 14.29$\pm$0.02 & 13.49$\pm$0.05 & $-$4.77$\pm$0.08 &  1.86$\pm$0.07 & $-7.43\pm0.06$ \\
HD  27778 & 20.79$\pm$0.06 & 16.08$\pm$0.03 & 14.18$\pm$0.05 & 13.44$\pm$0.05 & $-$4.71$\pm$0.07 &  1.90$\pm$0.06 & $-7.35\pm0.08$ \\
HD  73882 & 21.11$\pm$0.08 & 16.55$\pm$0.20 & 15.15$\pm$0.20 & 13.60$\pm$0.06 & $-$4.56$\pm$0.25 &  1.40$\pm$0.35 & $-7.51\pm0.10$ \\
HD 147888 & 20.48$\pm$0.08 & 15.30$\pm$0.08 & $<$13.5        & 13.22$\pm$0.05 & $-$5.18$\pm$0.12 & $>$1.8         & $-7.26\pm0.09$ \\
HD 185418 & 20.76$\pm$0.05 & 14.75$\pm$0.03 & \nodata        & $<$12.85       & $-$6.01$\pm$0.06 & \nodata        & $<-7.91$       \\
HD 192639 & 20.69$\pm$0.05 & 14.13$\pm$0.13 & \nodata        & $<$12.85       & $-$6.56$\pm$0.13 & \nodata        & $<-7.84$       \\
HD 206267 & 20.86$\pm$0.04 & 16.04$\pm$0.04 & 14.39$\pm$0.03 & 13.68$\pm$0.05 & $-$4.82$\pm$0.05 &  1.66$\pm$0.05 & $-7.18\pm0.06$ \\
HD 207198 & 20.83$\pm$0.04 & 15.52$\pm$0.04 &[13.75$\pm$0.10]& 13.59$\pm$0.05 & $-$5.31$\pm$0.06 & [1.77$\pm$0.09]& $-7.23\pm0.07$ \\
\enddata
\tablecomments{Column densities are log [$N$ (cm$^{-2}$)]; uncertainties are 1-$\sigma$; limits are 3-$\sigma$.
Values for $^{13}$CO in square braces assume same $N_0$/$N_1$ as found for $^{12}$CO.
$N$(H$_2$) are from Rachford et al. 2002, except for HD~147888 (Cartledge et al. 2004). 
$N$(CO) and $N$(C$_2$) are from this paper, except for CO values toward HD~24534 (Sheffer et al. 2002a).}
\end{deluxetable}

\begin{deluxetable}{llcrrcrc}
\tabletypesize{\scriptsize}
\tablecolumns{8}
\tablecaption{Correlation Coefficients and Fits \label{tab:corr}}
\tablewidth{0pt}

\tablehead{
\multicolumn{1}{c}{Y}&
\multicolumn{1}{c}{X}&
\multicolumn{1}{c}{N$_{\rm d}$\tablenotemark{a}}&
\multicolumn{1}{c}{r\tablenotemark{b}}&
\multicolumn{1}{c}{Slope\tablenotemark{c}}&
\multicolumn{1}{c}{rms\tablenotemark{c}}&
\multicolumn{1}{c}{Slope\tablenotemark{d}}&
\multicolumn{1}{c}{rms\tablenotemark{d}}
}
\startdata
C$_2$               & H$_2$     & 23 &    0.678 &  $1.79\pm0.35$ & 0.11 &  $3.33\pm0.78$ & 0.11 \\
 & \\
$^{12}$CO           & CN        & 31 &    0.900 &  $1.50\pm0.11$ & 0.16 &  $1.51\pm0.11$ & 0.16 \\
$^{12}$CO           & CH        & 44 &    0.854 &  $4.00\pm0.32$ & 0.16 &  $3.81\pm0.30$ & 0.17 \\
$^{12}$CO           & C$_2$     & 21 &    0.450 &  $3.13\pm0.78$ & 0.18 &  $3.70\pm0.82$ & 0.17 \\
$^{12}$CO           & C$_3$     &  7 &    0.586 &  $4.00\pm2.08$ & 0.11 &  $4.76\pm2.72$ & 0.11 \\
$^{12}$CO           & HD        & 19 &    0.905 &  $1.82\pm0.20$ & 0.35 &  $1.68\pm0.17$ & 0.35 \\
$^{12}$CO           & K~I       & 40 &    0.733 &  $3.23\pm0.42$ & 0.21 &  $3.23\pm0.31$ & 0.21 \\
 & \\
$^{12}$CO/H$_2$     & CN/CH     & 28 &    0.815 &  $1.54\pm0.19$ & 0.17 &  $1.56\pm0.16$ & 0.17 \\
$^{13}$CO/H$_2$     & CN/CH     & 11 &    0.868 &  $1.74\pm0.38$ & 0.16 &  $1.65\pm0.28$ & 0.16 \\
C$_2$/H$_2$         & CN/CH     & 26 &    0.208 &  \nodata       &      &  $0.05\pm0.10$ & 0.17 \\
 & \\
$^{12}$CO/$^{13}$CO & $^{12}$CO & 17 & $-$0.647 & $-0.34\pm0.11$ & 0.23 & $-0.37\pm0.10$ & 0.22 \\
$^{12}$CO/$^{13}$CO & CN/CH     & 12 & $-$0.549 &  \nodata       &      & $-0.71\pm0.23$ & 0.22 \\
\enddata
\tablenotetext{a}{Number of detected values}
\tablenotetext{b}{{Linear correlation coefficient}}
\tablenotetext{c}{Weighted fit}
\tablenotetext{d}{Unweighted fit}
\end{deluxetable}

\begin{deluxetable}{llcccccccc}
\tabletypesize{\scriptsize}
\tablecolumns{10}
\tablecaption{CO Excitation Temperatures \label{tab:texco}}
\tablewidth{0pt}
 
\tablehead{
\multicolumn{1}{l}{Star}&
\multicolumn{1}{l}{Name}&
\multicolumn{4}{c}{-~-~-~-~-~-~-~$^{12}$CO~-~-~-~-~-~-~-~-}&
\multicolumn{3}{c}{-~-~-~-~-~-~$^{13}$CO~-~-~-~-~-~-}&
\multicolumn{1}{c}{Ref}\\
\multicolumn{2}{c}{ }&
\multicolumn{1}{c}{$T_{01}$}&
\multicolumn{1}{c}{$T_{12}$}&
\multicolumn{1}{c}{$T_{23}$}&
\multicolumn{1}{c}{$T_{02}$/$T_{03}$}&
\multicolumn{1}{c}{$T_{01}$}&
\multicolumn{1}{c}{$T_{12}$}&
\multicolumn{1}{c}{$T_{23}$}&
\multicolumn{1}{c}{ }}
\startdata
HD  22951 & 40 Per       & 3.0$\pm$0.3 & \nodata     & \nodata     & \nodata & \nodata     & \nodata     & \nodata     & 1 \\
HD  23180 & $o$ Per      & 3.6$\pm$0.3 & \nodata     & \nodata     & \nodata & \nodata     & \nodata     & \nodata     & 1 \\
HD  23625 &              & 2.7$\pm$0.3 & \nodata     & \nodata     & \nodata & \nodata     & \nodata     & \nodata     & 1 \\
HD  24534 & X Per        & 3.3         & \nodata     & \nodata     & \nodata & 3.7         & \nodata     & \nodata     & 2 \\
HD 147933 & $\rho$ Oph A & 2.7$\pm$0.1 & 7.6$\pm$0.5 & 8.4$\pm$0.5 & \nodata &(7.7$\pm$2.5)&(7.7$\pm$2.5)&(7.7$\pm$2.5)& 3 \\
HD 149757 & $\zeta$ Oph  & 3.4$\pm$0.4 & 4.5$\pm$0.2 & 5.9$\pm$0.5 & \nodata & 4.2$\pm$0.4 & 4.4$\pm$0.2 & 5.9$\pm$0.5 & 4 \\
HD 148184 & $\chi$ Oph   & 3.0$\pm$0.3 & 5.2$\pm$0.6 & 7.5$\pm$3.7 & \nodata &(4.4$\pm$2.9)&(4.4$\pm$2.9)&(4.4$\pm$2.9)& 3 \\
HD 203374A&              & 3.2         & 3.5         & 8.7         & \nodata & \nodata     & \nodata     & \nodata     & 5 \\
\enddata
\tablecomments{First set of sight lines is from other references; second set is from this paper; last set gives theoretical values from isothermal and thermal equilibrium (TE) models of Warin et al. 1996.  
CO values in parentheses are from fits to all observed levels.}
\tablerefs{1 = Wannier et al. 1997; 2 = Sheffer et al. 2002b [A-X (8-0)]; 
3 = Federman et al. 2003; 4 = Lambert et al. 1994; 5 = Sheffer et al. 2003; 
6 = this paper; 7 = Warin et al. 1996 (theory)}
\end{deluxetable}

\begin{deluxetable}{llccccccc}
\tabletypesize{\scriptsize}
\tablecolumns{9}
\tablecaption{C$_2$ Excitation: Temperature and Density \label{tab:texc2}}
\tablewidth{0pt}
 
\tablehead{
\multicolumn{1}{l}{Star}&
\multicolumn{1}{l}{Name}&
\multicolumn{1}{c}{H$_2$\tablenotemark{a}}&
\multicolumn{5}{c}{-~-~-~-~-~-~-~-~-~-~-~-~-~-~-~-~-~C$_2$~-~-~-~-~-~-~-~-~-~-~-~-~-~-~-~-~-}\\
\multicolumn{2}{c}{ }&
\multicolumn{1}{c}{$T_{01}$}&
\multicolumn{1}{c}{$T_{02}$}&
\multicolumn{1}{c}{$T_{04}$}&
\multicolumn{1}{c}{$T_{\rm k}$}&
\multicolumn{1}{c}{$n$\tablenotemark{b}}&
\multicolumn{1}{c}{$n_{\rm H}$\tablenotemark{c}}&
\multicolumn{1}{c}{Refs\tablenotemark{d}}}
\startdata
HD  23180 & $o$ Per      &  48        & \nodata   & 48$\pm$37 &(50)       &(200$^{+250}_{-50}$) &(265$^{+335}_{-65}$) & 1 \\
HD  24398 & $\zeta$ Per  &  58        & \nodata   & 84$\pm$53 &(50)       &(150$^{+200}_{-50}$) &(215$^{+285}_{-70}$) & 2 \\
HD  62542 &              &  43$\pm$11 & \nodata   & 43$\pm$16 & 30$\pm$10 & 350$^{+400}_{-100}$ & 500$^{+570}_{-145}$ & 3 \\
HD  80077 &              & \nodata    & 40$\pm$26 & 33$\pm$ 8 & 20$\pm$ 5 & 200$^{+100}_{-25}$  & \nodata             & 4 \\
HD 110432 &              &  68$\pm$ 5 & 13$\pm$ 4 & 27$\pm$ 6 & \nodata   & 100$^{+25}_{-25}$   & 140$^{+35}_{-35}$   & 5 \\
HD 147084 & $o$ Sco      & \nodata    & \nodata   & 57$\pm$29 & 35$\pm$10 & 150$^{+50}_{-25}$   & \nodata             & 5 \\
HD 147889 &              & \nodata    & 52$\pm$22 &116$\pm$28 &(70)       &(350$^{+400}_{-150}$)& \nodata             & 6 \\
HD 148184 & $\chi$ Oph   &  46$\pm$ 6 & 39$\pm$16 & 57$\pm$12 & 50$\pm$15 & 150$^{+100}_{-25}$  & 185$^{+125}_{-30}$  & 6 \\
\enddata
\tablecomments{C$_2$ $T_{\rm k}$ and inferred $n$ and $n_{\rm H}$ values for upper set of sight lines are based on equivalent width data from various references; values for most of the sight lines in the lower set are based on data given in this paper. 
Values in parentheses are estimates based on the assumption that $T_{\rm k}$ $\la$ $T_{01}$ from H$_2$.
All values for $n_{\rm H}$ assume a collisional de-excitation cross-section $\sigma$ = 2 $\times$ 10$^{-16}$ cm$^2$ and the average Galactic radiation field.}
\tablerefs{1 = Hobbs 1981; 2 = Chaffee et al. 1980; 3 = Gredel et al. 1993; 4 = Gredel 1999; 5 = van Dishoeck \& Black 1989; 
6 = van Dishoeck \& de Zeeuw 1984; 7 = Federman et al. 1994a; 8 = Gredel et al. 2001; 9 = Federman et al. 1997b; 10 = Galazutdinov et al. 2006; 
11 = this paper; 12 = Hobbs \& Campbell 1982, Danks \& Lambert 1983, van Dishoeck \& Black 1986b, Lambert et al. 1995}
\tablenotetext{a}{Savage et al. 1977; Rachford et al. 2002}
\tablenotetext{b}{Density of collision partners (H~I + H$_2$).}
\tablenotetext{c}{Density of hydrogen nuclei (H~I + 2H$_2$).
For HD~169454, HD~179406, HD~200775, and HD~204827, $f$(H$_2$) = 0.5 is assumed.}
\tablenotetext{d}{Reference for C$_2$ equivalent widths (temperatures and densities are those calculated in this paper).}
\end{deluxetable}

\begin{deluxetable}{lrcccrcc}
\tabletypesize{\scriptsize}
\tablecolumns{8}
\tablecaption{Component Structures for K~I, CH, and CN \label{tab:comp}}
\tablewidth{0pt}

\tablehead{
\multicolumn{1}{c}{Star}&
\multicolumn{1}{c}{$v$}&
\multicolumn{1}{c}{$b$}&
\multicolumn{1}{c}{$N$(K~I)}&
\multicolumn{1}{c}{$N$(CH)}&
\multicolumn{1}{c}{$v$}&
\multicolumn{1}{c}{$b$}&
\multicolumn{1}{c}{$N$(CN)}\\
\multicolumn{1}{c}{ }&
\multicolumn{1}{c}{(km~s$^{-1}$)}&
\multicolumn{1}{c}{(km~s$^{-1}$)}&
\multicolumn{1}{c}{(10$^{10}$ cm$^{-2}$)}&
\multicolumn{1}{c}{(10$^{12}$ cm$^{-2}$)}&
\multicolumn{1}{c}{(km~s$^{-1}$)}&
\multicolumn{1}{c}{(km~s$^{-1}$)}&
\multicolumn{1}{c}{(10$^{12}$ cm$^{-2}$)}
}
\startdata
HD  24534 &   11.2 & 0.8 &  9.1 &  5.0 &        &     &      \\
          &   12.4 & 0.6 & 46.7 &  6.0 &   12.3 & 0.8 &  0.7 \\
          &   14.5 & 1.0 & 63.5 & 25.9 &   14.7 & 1.0 &  6.6 \\
          &   16.5 & 0.8 &  2.8 &      &        &     &      \\
 & \\
HD  27778 &   12.6 & 0.6 &  2.2 &  0.3 &        &     &      \\
          &   14.1 & 0.4 & 13.6 &  4.0 &        &     &      \\
          &   15.2 & 0.7 & 19.2 & 12.8 &   15.1 & 1.0 &  8.9 \\
\enddata
\tablecomments{CH fitted with K~I component structure; CN fitted independently.
K~I structures for HD~206267, HD~207198, and HD~210839 are from Welty \& Hobbs 2001.
CN structure for HD~73882 is from Gredel et al. 1991 (FWHM $\sim$ 5 km~s$^{-1}$).
CN structure for HD~147888 is from Pan et al. 2004.
CN structure for HD~210121 is from Welty \& Fowler 1992 (FWHM $\sim$ 3 km~s$^{-1}$).
CH structure for HD~210839 is based on new fit to profile from Crane et al. 1995, assumed aligned with strongest components in K~I and CN.}
\end{deluxetable}

\clearpage

\setlength{\tabcolsep}{0.09in}

\begin{deluxetable}{lrcccccccc}
\tabletypesize{\scriptsize}
\tablecolumns{10}
\tablecaption{C$_2$ Band Structures \label{tab:c2bands}}
\tablewidth{0pt}

\tablehead{
\multicolumn{1}{c}{Line}  & 
\multicolumn{1}{c}{E($J$)/k}&
\multicolumn{3}{c}{F-X bands}&
\multicolumn{2}{c}{D-X bands}&
\multicolumn{3}{c}{A-X bands}\\
\multicolumn{2}{c}{ }&
\multicolumn{1}{c}{$\lambda$(1-0)} &
\multicolumn{1}{c}{$\lambda$(0-0)} & 
\multicolumn{1}{c}{$S$\tablenotemark{a}}&
\multicolumn{1}{c}{$\lambda$(0-0)} &
\multicolumn{1}{c}{$S$\tablenotemark{a}}&
\multicolumn{1}{c}{$\lambda$(3-0)} &
\multicolumn{1}{c}{$\lambda$(2-0)} &
\multicolumn{1}{c}{$S$\tablenotemark{a}}\\
\multicolumn{1}{c}{ }&
\multicolumn{1}{c}{(K)}&
\multicolumn{1}{c}{(\AA)}&
\multicolumn{1}{c}{(\AA)}&
\multicolumn{1}{c}{ }&
\multicolumn{1}{c}{(\AA)}&
\multicolumn{1}{c}{ }&
\multicolumn{1}{c}{(\AA)}&
\multicolumn{1}{c}{(\AA)}&
\multicolumn{1}{c}{ }}
\startdata
R(0) &   0.00 & 1314.170 & 1341.631 & 1.000 & 2313.163 & 1.000 & 7719.329 & 8757.686 & 1.000 \\
R(2) &  15.64 & 1314.079 & 1341.532 & 0.400 & 2312.769 & 0.600 & 7716.528 & 8753.949 & 0.400 \\
Q(2) &        & 1314.246 & 1341.708 & 0.500 & \nodata  &\nodata& 7722.095 & 8761.194 & 0.500 \\
P(2) &        & 1314.358 & 1341.826 & 0.100 & 2313.745 & 0.400 & 7725.819 & 8766.031 & 0.100 \\
R(4) &  52.11 & 1314.015 & 1341.458 & 0.333 & 2312.371 & 0.556 & 7714.944 & 8751.685 & 0.333 \\
Q(4) &        & 1314.294 & 1341.752 & 0.500 & \nodata  &\nodata& 7724.219 & 8763.751 & 0.500 \\
P(4) &        & 1314.517 & 1341.988 & 0.167 & 2314.126 & 0.444 & 7731.663 & 8773.430 & 0.167 \\
R(6) & 109.43 & 1313.977 & 1341.411 & 0.308 & 2311.967 & 0.538 & 7714.575 & 8750.848 & 0.308 \\
\enddata
\tablecomments{Wavelengths are adapted from Herzberg et al. 1969 [F-X bands], Sorkhabi et al. 1998 [D-X(0-0)], and van Dishoeck \& de Zeeuw 1984 [A-X(2-0)].}
\tablenotetext{a}{Relative $f$-value for individual line.  Values for higher-$J$ A-X bands differ slightly from those tabulated by van Dishoeck \& de Zeeuw 1984.}
\end{deluxetable}

\setlength{\tabcolsep}{0.03in}

\begin{deluxetable}{llccccccccccccc}
\tabletypesize{\scriptsize}
\tablecolumns{15}
\tablecaption{Column Densities (H$_2$, CH, C$_2$, CN, CO) \label{tab:coldens}}
\tablewidth{0pt}
\tablehead{
\multicolumn{1}{l}{HD}&
\multicolumn{1}{l}{Name}&
\multicolumn{2}{c}{H$_2$\tablenotemark{a}}&
\multicolumn{2}{c}{CH\tablenotemark{b}}&
\multicolumn{2}{c}{C$_2$\tablenotemark{c}}&
\multicolumn{2}{c}{CN\tablenotemark{d}}&
\multicolumn{2}{c}{$^{12}$CO\tablenotemark{e}}&
\multicolumn{2}{c}{$^{13}$CO\tablenotemark{f}}&
}
\startdata
HD  21278   &                    & 
    \nodata        &   & \nodata        &   & \nodata        &   & \nodata        &   & $12.87\pm0.12$ & 1 & \nodata        &   & \\
HD  22951   & 40 Per             & 
    $20.46\pm0.18$ & 1 & $13.08\pm0.04$ & 1 & $12.04\pm0.12$ & 1 & $11.85       $ & 1 & $14.22       $ & 2 & \nodata        &   & \\
HD  23180   & $o        $ Per    & 
    $20.60\pm0.15$ & 1 & $13.33\pm0.03$ & 2 & $13.27\pm0.12$ &(2)& $12.17\pm0.01$ & 2 & $14.93       $ & 2 & \nodata        &   & \\
HD  23408   & 20 Tau             & 
    $19.75\pm0.26$ & 2 & $<11.30$       & 2 & \nodata        &   & $<11.82$       & 3 & $13.73\pm0.05$ & 1 & \nodata        &   & \\
HD  23625   &                    &
    \nodata        &   & $13.08\pm0.15$ & 1 & $12.91\pm0.05$ & 1 & $11.85$        & 1 & $14.26$        & 2 &                &   & \\
HD  23630   & $\eta    $  Tau    & 
    $19.54\pm0.18$ & 2 & $<12.34$       & 3 & \nodata        &   & \nodata        &   & $<12.34$       & 1 & \nodata        &   & \\
HD  24398   & $\zeta   $  Per    & 
    $20.67\pm0.18$ & 1 & $13.34\pm0.03$ & 2 & $13.22\pm0.07$ &(3)& $12.51\pm0.01$ & 4 & $14.86\pm0.17$ & 1 & \nodata        &   & \\
HD  24534   & X Per              & 
    $20.92\pm0.04$ & 3 & $13.57\pm0.02$ & 4 & $13.49\pm0.05$ & 4 & $12.87\pm0.01$ & 5 & $16.15\pm0.07$ & 3 & $14.29\pm0.02$ & 1 & \\
\enddata
\end{deluxetable}

\begin{deluxetable}{llccccccrrr}
\tabletypesize{\scriptsize}
\tablecolumns{11}
\tablecaption{Derived Quantities \label{tab:dq}}
\tablewidth{0pt}
 
\tablehead{
\multicolumn{1}{l}{HD}&
\multicolumn{1}{l}{Name}&
\multicolumn{1}{c}{$E(B-V)$\tablenotemark{a}}&
\multicolumn{1}{c}{A$_{\rm V}$\tablenotemark{a}}&
\multicolumn{1}{c}{$N$(H$_{\rm tot}$)\tablenotemark{b}}&
\multicolumn{1}{c}{log[$f$(H$_2$)]\tablenotemark{b}}&
\multicolumn{1}{c}{$T_{01}$(H$_2$)\tablenotemark{b}}&
\multicolumn{1}{c}{$T_{\rm ex}$(C$_2$)\tablenotemark{c}}&
\multicolumn{1}{c}{CN/CH}&
\multicolumn{1}{l}{($^{12}$CO/H$_2$)}&
\multicolumn{1}{c}{$^{12}$CO/$^{13}$CO\tablenotemark{d}}
}
\startdata
HD  22951  & 40 Per       & 0.27 & 0.66 & 21.23 & $-$0.47& 63$\pm$10 & \nodata    & $-1.23$ & $-6.24$ &   ... \\
HD  23180  & $o$ Per      & 0.31 & 0.84 & 21.20 & $-0.30$& 48        & .../48/... & $-1.16$ & $-5.67$ &   ... \\
HD  24398  & $\zeta$ Per  & 0.31 & 0.83 & 21.20 & $-0.23$& 58        & .../84/... & $-0.83$ & $-5.81$ &   ... \\
HD  24534  & X Per        & 0.59 & 2.05 & 21.34 & $-0.12$& 57$\pm$ 4 & .../72/45  & $-0.70$ & $-4.77$ &  1.86 \\
HD  24912  & $\xi$ Per    & 0.33 & 1.03 & 21.29 & $-0.46$& 61        & \nodata    & $-1.68$ & $-7.04$ &   ... \\
HD  27778  & 62 Tau       & 0.37 & 1.01 & 21.34 & $-0.25$& 55$\pm$ 7 & .../67/50  & $-0.29$ & $-4.71$ &  1.90 \\
HD  30614  & $\alpha$ Cam & 0.3  & 0.88 & 21.09 & $-0.45$& 80        & \nodata    &$<-1.02$ & $-5.85$ &   ... \\
HD  35149  & 23 Ori       & 0.11 &  ... & 20.74 & $-2.14$& 100       & \nodata    &$<-0.97$ &$<-5.30$ &   ... \\
\enddata
\tablenotetext{a}{Thorburn et al. 2003; L. Hobbs, priv. comm.; B. Rachford, priv. comm.}
\tablenotetext{b}{Savage et al. 1977; Rachford et al. 2002; Sheffer et al. 2003; Pan et al. 2005; Cartledge et al. 2004}
\tablenotetext{c}{The three values are $T_{02}$, $T_{04}$, and $T_{\rm k}$; see Table~\ref{tab:texc2}.}
\tablenotetext{d}{Values in square braces assume that $N_0$/$N_1$ is the same for $^{13}$CO as for $^{12}$CO.}
\end{deluxetable}

\end{document}